\documentclass[useAMS,usenatbib]{mn2e}
\usepackage{graphicx}
\usepackage{mathptmx}

\newcommand{\threetotwo}{$J=3\to 2$}
\newcommand{\onetozero}{$J=1\to 0$}
\newcommand{\thirteenco}{$^{13}$CO}
\newcommand{\twelveco}{$^{12}$CO}
\newcommand{\ceighteeno}{C$^{18}$O}
\newcommand{\ammonia}{NH$_3$}
\newcommand{\ntwohplus}{N$_2$H$^+$}

\newcommand{\msun}{M$_\odot$}
\newcommand{\mjybeam}{mJy\,beam$^{-1}$}
\newcommand{\kkms}{K\,km\,s$^{-1}$}
\newcommand{\kms}{km\,s$^{-1}$}
\newcommand{\cmthree}{cm$^{-3}$}
\newcommand{\percmtwo}{cm$^{-2}$}
\newcommand{\apj}{ApJ}
\newcommand{\apjl}{ApJ}
\newcommand{\apjs}{ApJS}
\newcommand{\apjss}{ApJS}
\newcommand{\aap}{A\&A}

\newcommand{\mnras}{MNRAS}
\newcommand{\aj}{AJ}
\newcommand{\araa}{ARA\&A}

\title[The kinematics of the Perseus
  molecular cloud]{A submillimetre survey of the kinematics of the
  Perseus molecular cloud: I. data}
\author[E.~I.~Curtis, J.~S.~Richer and J.~V.~Buckle]{Emily~I.~Curtis$^{1,2}$\thanks{E-mail:e.curtis@mrao.cam.ac.uk}, John~S.~Richer$^{1,2}$ and Jane~V.~Buckle$^{1,2}$\\
$^{1}$~Astrophysics Group, Cavendish Laboratory, J. J. Thomson
Avenue, Cambridge, CB3 0HE\\
$^{2}$~Kavli Institute for Cosmology, c/o Institute of Astronomy, University
of Cambridge, Madingley Road, Cambridge, CB3 0HA}
\begin{document}

\date{Accepted 2009 September 2}

\pagerange{\pageref{firstpage}--\pageref{lastpage}} \pubyear{2009}

\maketitle

\label{firstpage}

\begin{abstract}

We present submillimetre observations of the \threetotwo\ rotational transition of
\twelveco, \thirteenco\ and \ceighteeno\ across over 600\,arcmin$^2$ of the
Perseus molecular cloud, undertaken with HARP, a new array spectrograph
on the James Clerk Maxwell Telescope. The data encompass four regions of the cloud,
containing the largest clusters of dust continuum condensations:
NGC1333, IC348, L1448 and L1455. A new procedure to remove striping
artefacts from the raw HARP data is introduced. We compare the maps to those
of the dust continuum emission mapped with SCUBA
\citep{hatchell05} and the positions of starless and protostellar
cores \citep{hatchell07a}. No straightforward correlation is found between the
masses of each region derived from the HARP CO and SCUBA
data, underlining the care that must be exercised when comparing masses
of the same object derived from different tracers. From the \thirteenco/\ceighteeno\ line
ratio the relative abundance of the two species
($\rmn{[^{13}CO]/[C^{18}O]} \sim 7$) and their opacities (typically
$\tau$ is 0.02--0.22 and 0.15--1.52 for the \ceighteeno\ and
\thirteenco\ gas respectively) are calculated. \ceighteeno\ is
optically thin nearly everywhere, increasing in opacity towards
star-forming cores but not beyond $\tau_{18} \sim 0.9$. Assuming the
\twelveco\ gas is optically thick we compute its excitation
temperature, $T_\rmn{ex}$ (around 8--30\,K), which has little
correlation with estimates of the dust temperature. 

\end{abstract}

\begin{keywords}
submillimetre -- stars: formation -- ISM: kinematics and
dynamics -- ISM: individual: Perseus.
\end{keywords}

\section{Introduction} 

Recent advances in telescope instrumentation have provided an
unprecedented view of the star formation process inside molecular clouds. Near-infrared imaging from
e.g.\ \emph{Spitzer} (e.g.\ \citealp{evans08}) captures the youngest stellar objects,
whilst (sub)millimetre continuum imaging with e.g.\ SCUBA or
MAMBO maps the very earliest stages of star formation, often before a
well-defined central object is established (e.g.\
\citealp*{motte98}; \citealp{johnstone00b};  \citealp{hatchell05,enoch06,difrancesco08}).
Simulations of star formation have also become increasingly
sophisticated (e.g.\ \citealp{klessen00,bate05,bate09a}) with the latest
models including
turbulence, magnetic fields and radiative transfer
\citep{price08,bate09b}. The true, underlying structure in star-forming clouds, essential for
comparison with models, is hard to disentangle from continuum images
alone (\citealp{ballesteros02}; \citealp*{smith_clark08}): density enhancements along
the line of sight may superpose and limited spatial resolution and sensitivity
inevitably blends nearby objects together. Spectral lines from
molecular species can provide crucial information about
the \emph{kinematics} of and physical conditions inside molecular
clouds. They may for
instance allow the separation of multiple objects, moving at distinct
velocities, along a line of sight (e.g.\ \citealp*{kirk07}). Additionally, line emission is arguably
the best discriminator between different support mechanisms for
star-forming cores, i.e.\ whether magnetic fields or turbulence
dominate the required suppression of star formation in the bulk of clouds.  

This paper presents a new survey of the kinematics of star formation in
the Perseus molecular cloud. We observed four fields, totalling
approximately 600\,arcmin$^2$ in the \threetotwo\ rotational
transition of \twelveco, \thirteenco\ and \ceighteeno\ using HARP
(Heterodyne Array Receiver Programme), a new array
spectrograph operating between 325 and 375\,GHz on the
James Clerk Maxwell Telescope (JCMT, see
\citealp{smith03,smith08,hills09}). These data represent some of the
largest and highest-quality maps of the higher-$J$ transitions of
CO and its isotopologues with good angular resolution ($<20$\,arcsec). This paper is the
first in a planned series which will examine the gas structure and core
kinematics (Paper II), molecular outflows (Paper III) and the physical
conditions in the cloud cores (Paper IV). This paper is organized as
follows: the remainder of the
introduction outlines the survey and its aims before
Sections \ref{sec:observations} and \ref{sec:datareduction} detail the
observations and data reduction procedure including an algorithm to
calibrate variable sensitivities across the HARP array for molecular cloud
data. Section \ref{sec:results} presents maps of the integrated intensity of all of the
isotopologues while some initial results including the cloud opacity and
excitation temperature, assuming local thermodynamic equilibrium (LTE)
are presented in Section \ref{sec:discussion}. We summarize this work in Section \ref{sec:summary}.    

\subsection{The survey}

The transitions of CO and its common isotopologues in the 345\,GHz atmospheric
window trace denser and/or warmer regions than their lower $J$
relatives -- compare the temperature of 5\,K and density of
$n_\mathrm{crit}\sim 700$\,cm$^{-3}$ required to
collisionally excite the \twelveco\ $J=1\to 0$ line at 115\,GHz to Table
\ref{table:co_properties}. These conditions are similar to those
inside star-forming SCUBA cores, allowing us to probe the motions of
bulk gas in their vicinity at matched resolution to SCUBA
850\,\micron\ observations. This survey is closely linked to the
JCMT Gould Belt Legacy Survey (GBS, \citealp{gbs}). Our data will eventually
be included in that analysis and we share similar science goals, namely:
\begin{enumerate}
\item Map any high-velocity outflowing gas in the \twelveco\
  \threetotwo\ line to investigate mass-loss and accretion in a large
  sample of sources. The presence of an outflow can differentiate
  between a protostellar or starless continuum core, establishing the
  source age. 
\item Establish the level of non-thermal support inside star-forming
  cores from the \ceighteeno\ transition, which may also be used to
  evaluate the levels of CO depletion. 
\item Investigate the turbulent structure of the molecular gas.  
\end{enumerate} 

\begin{table}
\caption{The transitions of
  CO and its isotopologues observed.} 
\begin{tabular}{lcccc}
\hline
Molecule & $\Delta J$~$^\mathrm{a}$ & $\nu_\mathrm{trans}$~$^\mathrm{b}$ &
$T_\mathrm{trans}$~$^\mathrm{c}$ & $n_\mathrm{crit}$~$^\mathrm{d}$ \\
 &  & (GHz) & (K) & (cm$^{-3}$) \\
\hline
CO & $3\to 2$ & 345.796 & 33.2 & $2.54\times 10^{4}$ \\
$^{13}$CO & $3\to 2$ & 330.558 & 31.7 & $2.22\times 10^{4}$ \\
C$^{18}$O &  $3\to 2$ & 329.331 & 31.6 & $2.19\times 10^{4}$ \\
\hline
\end{tabular}
\label{table:co_properties}

$^\mathrm{a}$~Transition quantum numbers. \\
$^\mathrm{b}$~Rest frequency. \\
$^\mathrm{c}$~Height of the upper $J$ level above ground. \\
$^\mathrm{d}$~Estimated critical density, assuming a collision
cross-section $\sigma = 10^{-16}$\,\percmtwo\ and velocity 1\,\kms\ so
$n_\rmn{crit}= A\times 10^{10}$\,\percmtwo. The Einstein $A$ values were
taken from the LAMDA database \citep{schoeier05}.
\end{table}

\begin{figure*}
\hbox{\hspace{2.4cm} \includegraphics[width=13.6cm]{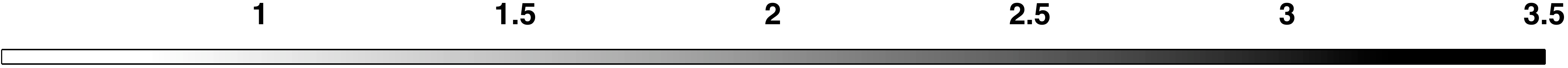}}
\includegraphics[width=0.8\textwidth]{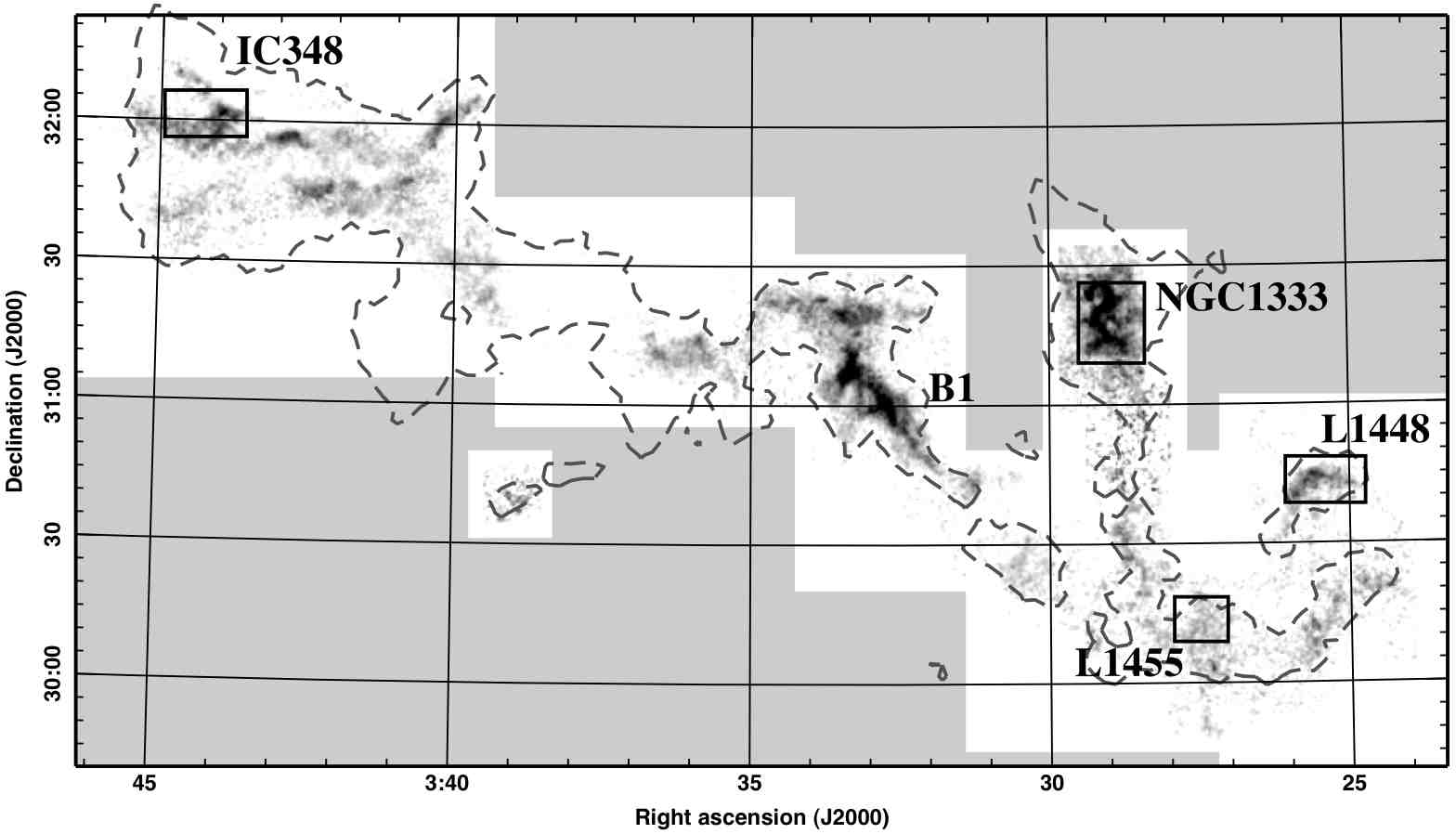}
\caption{Overview of the molecular gas in the Perseus molecular
  cloud. The grey-scale is \ceighteeno\ \onetozero\ integrated
  intensity in \kkms\ from
  the FCRAO 14-m telescope \citep{hatchell05}. The dotted contour
  encloses regions where the visual extinction is $>3.5$, from the map
  of the COMPLETE team \citep{complete}. The four fields observed in
  this survey are enclosed in boxes: NGC1333, IC348, L1448 and
  L1455. Also marked is the B1 molecular ridge. B5 the other large
  cluster of SCUBA cores lies roughly 45\,arcmin to the northeast.}
\label{fig:perseus}
\end{figure*}

Perseus (see Fig. \ref{fig:perseus}) is an intermediate star-forming environment between low-mass,
quiescent Taurus and high-mass, turbulent Orion (e.g.\ \citealp*{ladd93}). The complex consists of a series of dark clouds at
approximately $3^\mathrm{h}30^\mathrm{m}$, $+31^\circ$ with an
angular extent of about $(1.5 \times 5)$\,deg, totalling $\sim 1.7\times
10^4$\,M$_\odot$ \citep{bachiller86a}. The cloud is associated
with Per OB2, the second closest OB association to the Sun, with an
age $<15$\,Myr (see \citealp{bally08}). The association has blown a
20\,deg diameter shell of atomic hydrogen into the
interstellar medium with the Perseus molecular cloud embedded in its
western rim. The region is
well-studied both as a whole and in individual sub-sections with a wealth
of ancillary data available, particularly from the COMPLETE project\footnote{the COordinated Molecular Probe Line
  Extinction Thermal Emission survey of star-forming regions, see \url{http://www.cfa.harvard.edu/COMPLETE}} \citep{complete}. 

Notably, large-area surveys of
the dust continuum emission in Perseus have been completed with SCUBA at 850\,\micron\ (\citealp{hatchell05}; \citealp*{kirk06}) and Bolocam at 1.1\,mm
\citep{enoch06}. The 3\,deg$^2$ SCUBA survey \citep{hatchell05} identified 92
submillimetre cores above their completeness limit of
0.4\,\msun\ (in a 14\,arcsec beam). 80~per cent of these cores were
grouped in six clusters:
NGC1333, IC348, L1448, L1455, B1 and B5. The larger and more
sensitive Bolocam survey (7.5\,deg$^2$ down to 0.18\,\msun;
\citealp{enoch06}) identified 122 compact cores. However, only 5--10
cores were in
areas not observed with SCUBA by \citeauthor{hatchell05},
illustrating that most of Perseus is devoid of active star
formation. In this survey we target the clusters of star-forming
cores (see Section \ref{sec:observations}). \emph{Spitzer} has
provided a census of deeply embedded young stellar objects (YSOs) in many
nearby 
molecular clouds (e.g.\ \citealp{evans08}) and this
population has been associated with
continuum cores identified by SCUBA \citep{jorgensen07,jorgensen08} or Bolocam
\citep{enoch08,enoch08b}. In this survey, we use the catalogue of
\citet{hatchell07a}, who identified the SCUBA cores from
\citet{hatchell05} as starless, Class
0 or Class I protostars on the basis of their spectral energy
distributions (SEDs), which incorporated \emph{Spitzer} fluxes from the
IRAC wavebands.\footnote{IRAC (Infrared Array Camera) on
  \emph{Spitzer} has four channels at 3.6, 4.5, 5.8, and 8\,\micron.}     

Estimates of the distance to Perseus vary from 220\,pc \citep{cernis90} to
350\,pc \citep{herbig83}. Larger distances are often based
on the Perseus OB2 association which has an established
distance of $\sim320$\,pc from \emph{Hipparcos} \citep{dezeeuw99}. However, there is some evidence that Per OB2 may
lie behind the molecular clouds of interest which are probably at
closer to 250\,pc \citep{cernis93}. In fact, it may not be appropriate to use a
single distance for the entire cloud and extinction studies suggest
increasing distances from 220 to 260\,pc from west to east (previous
references and \citealp{cernis2003}). Furthermore many authors have
suggested that Perseus is a superposition of a least two smaller clouds -- the closer is thought to be an extension of Taurus
with the more distant a shell-like structure
(e.g.\ \citealp{ridge06}). The latest and possibly most reliable
measure of the distance to the H$_2$O maser in
NGC1333 SVS13 found $(235\pm 18)$\,pc using its parallax at radio
frequencies \citep{hirota08}. In this survey we assume
Perseus to be a single entity at a distance of 250\,pc for consistency
with the majority of recent studies e.g.\ the \emph{Spitzer} Cores to
Disks team, c2d 
\citep{evans03,evans08} and the Bolocam Perseus survey \citep{enoch06}.  

\section{Observations} \label{sec:observations}

We selected the four largest clusters
of continuum cores -- NGC1333, IC348, L1448 and L1455 -- to map in the \twelveco, \thirteenco\ and \ceighteeno\ $J=3\to 2$
lines (see Fig. \ref{fig:perseus}; the exact areas are detailed in Table
\ref{table:regions}). All the regions were observed in the three
tracers except for NGC1333 where only \thirteenco\ and \ceighteeno\ data
were taken since it had already been observed in
\twelveco\ \threetotwo\ with HARP (J. Swift, personal communication).  

\begin{table}
\caption{The regions observed.} 
\begin{tabular}{lccccc}
\hline
Region & \multicolumn{2}{c}{Centre~$^{\rmn{a}}$} &
\multicolumn{2}{c}{Dimensions} & Area \\
 & RA & Dec & Width & Height \\
 & (h~m~s) & ($^\circ$~$'$~$''$) & (arcsec) & (arcsec) &  (arcmin$^2$)  \\
\hline
NGC1333 & 03:28:56 & +31:17:30 & 760 & 900 & 190 \\
IC348   & 03:44:14 & +31:49:49 & 1100 & 650 & 199\\
L1448   & 03:25:30 & +30:43:45 & 800 & 500 & 111\\
L1455   & 03:27:27 & +30:14:30 & 880 & 520 & 127\\
\hline
\end{tabular}

$^\mathrm{a}$~Position of the map centre in J2000 coordinates. \\
\label{table:regions}
\end{table} 

The data were taken as part of the Guaranteed Time programme for the
HARP instrument team, over a period of
nine nights between 17th December 2007 and 12th January 2008 with one map of NGC1333 taken on 28th July 2007. The data comprise $\sim$43\,hrs total
observing time, approximately 25\,hrs on sky translating to an observing
efficiency of 58~per cent. The weather conditions were good to excellent throughout with
median receiver and system temperatures of $T_\rmn{rx}=180$\,K and
$T_\mathrm{sys}=325$\,K for $^{12}$CO compared to $T_\rmn{rx}=179$\,K
and $T_\mathrm{sys}=415$\,K for $^{13}$CO/C$^{18}$O. 

The HARP imaging array comprises 16 SIS detectors, arranged on a
$4\times 4$ grid, separated by 30\,arcsec. This under-samples the focal plane with
respect to the Nyquist criterion ($\lambda/2D$ is 6.0 and
6.2\,arcsec at 345 and 330\,GHz respectively), therefore standard HARP
observing strategies take data points in between the nominal detector
positions to produce a fully-sampled map (see \citealp{hills09}). The
observations for this survey were taken in the raster mode at the default
sample spacing of 7.3\,arcsec. The
$^{13}$CO and C$^{18}$O observations were taken simultaneously,
utilizing the capability of the new back-end correlator, ACSIS
(Auto-Correlation Spectral Imaging System, \citealp{hills09}), to split its bandpass into two
separate sub-bands. Both sub-bands provide $\sim$250\,MHz of
bandwidth on the two lines with a channel spacing of 61\,kHz corresponding to
approximately 0.05\,km\,s$^{-1}$. The $^{12}$CO observations were also
taken using a sub-band mode, with both centred at the same frequency: (i) for
high-velocity gas with a bandwidth of $\sim$1\,GHz, broken into
channels 977\,kHz wide (0.8\,km\,s$^{-1}$) and (ii) to trace the gas in detail
with $\sim$250\,MHz bandwidth at a channel spacing of 61\,kHz (0.05\,km\,s$^{-1}$). 

\subsection{Scan strategy} \label{sec:scanstrategy}

During raster or ``on-the-fly'' mapping, the telescope scans continuously
along a set direction (usually parallel to the longest side of the
map), dumping the data gathered at discrete intervals -- as little as
100\,ms apart with HARP/ACSIS. Once the telescope has scanned the length of the map, it
steps perpendicular to the scan direction and begins a new row. The HARP array is angled at
$\arctan(1/4)\sim 14$\,deg with respect to the scan direction to
produce a fully-sampled image on a 7.3\,arcsec grid \citep{hills09}. 

\begin{figure}
\includegraphics[angle=270,width=0.47\textwidth]{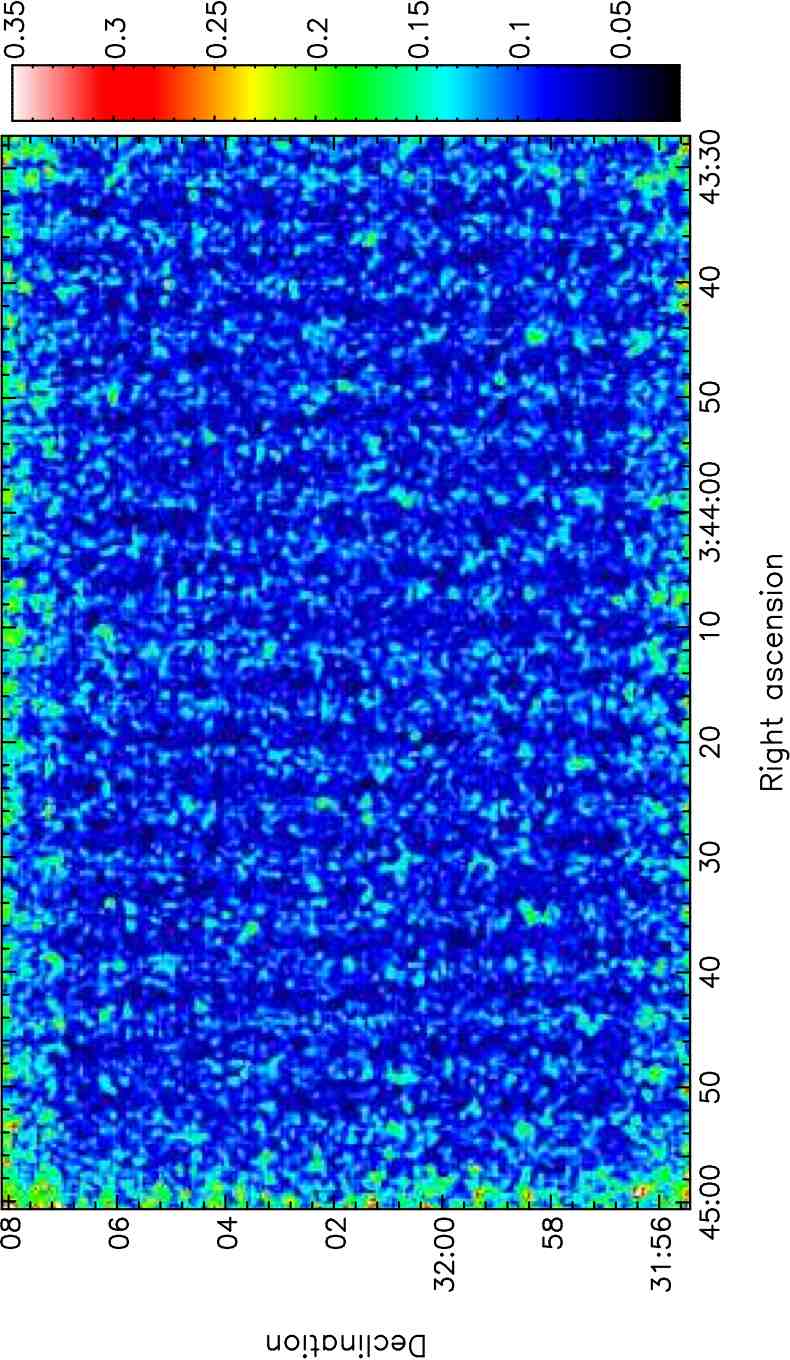}

\vspace{0.2cm}
\includegraphics[angle=270,width=0.47\textwidth]{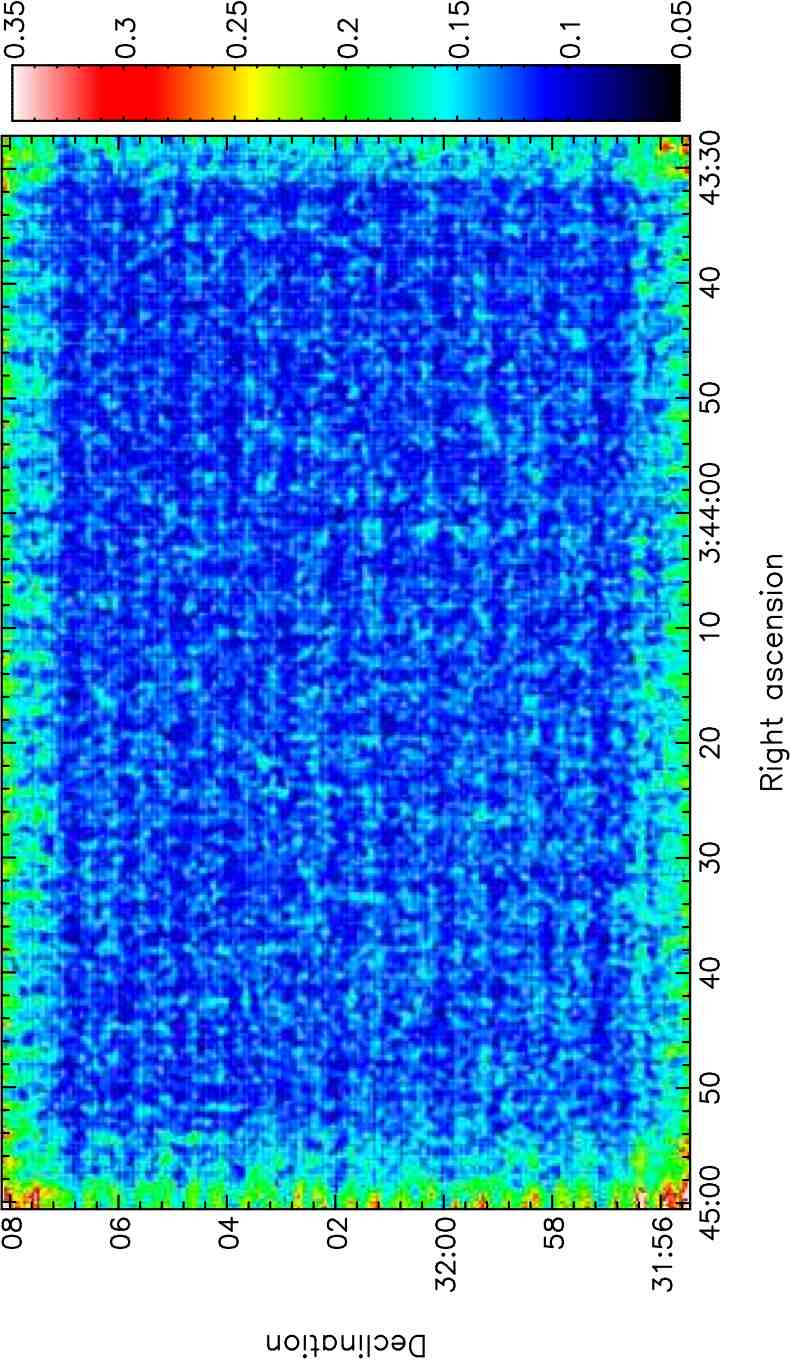}

\vspace{0.2cm}
\includegraphics[angle=270,width=0.47\textwidth]{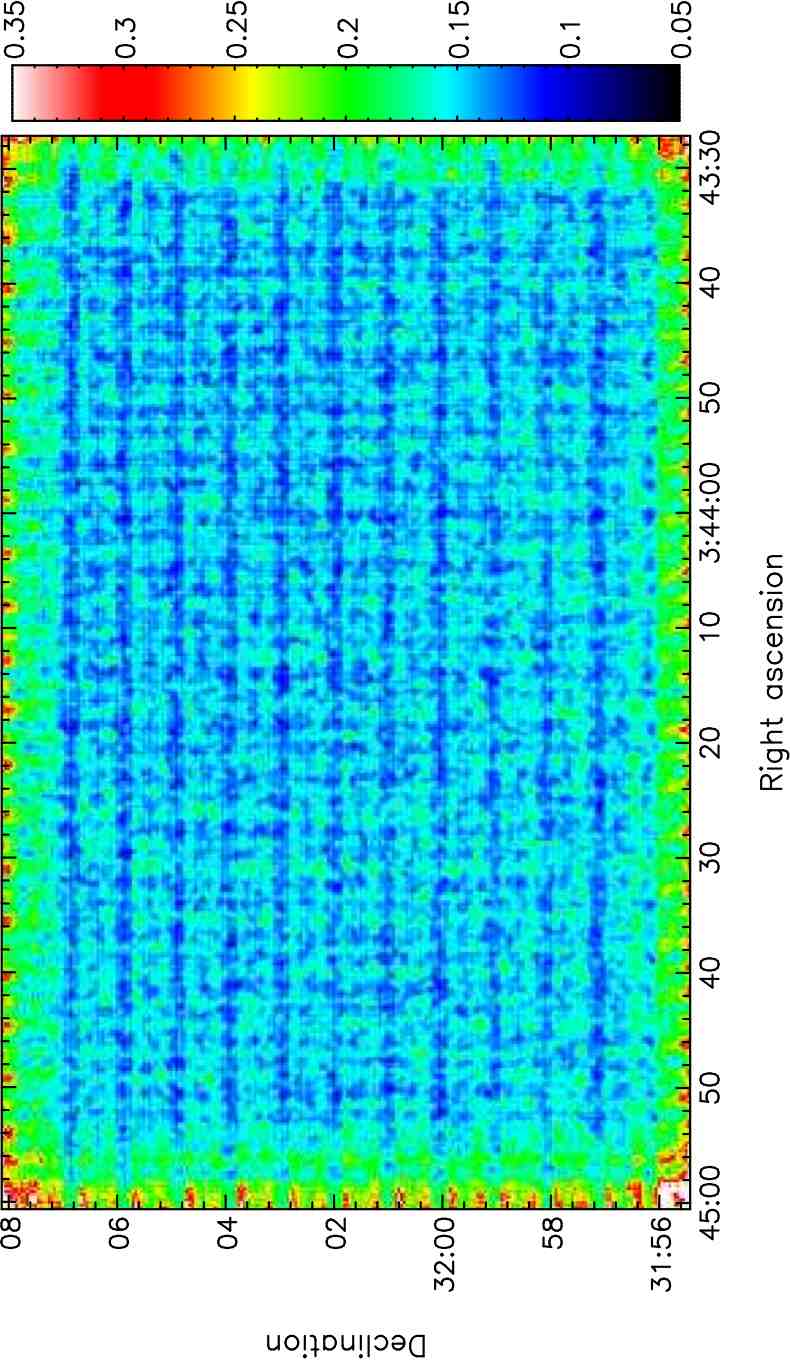}
\caption{Representative maps of the RMS noise in K towards IC348 in
  the three CO isotopologues. Top: \twelveco\ \threetotwo\
  noise in 1\,\kms\ channels. Middle and bottom:
  \thirteenco\ and \ceighteeno\ noise in 0.15\,\kms\ channels respectively.}
\label{fig:rms_noise}
\end{figure}

The detectors in HARP have varying responses across the
array. Additionally, in any one observation the data from a number of
detectors can be unusable with poor baselines. These in combination can lead to maps with variable noise or
entirely blank strips. Our observing strategy employed three factors
to reduce these effects:   
\begin{enumerate}
\item The perpendicular spacing between scan rows
  was set to be half the width of the tilted array. Thus, in
  any row, half of the detectors are going over previously
  observed positions and half are scanning new parts of the sky. 
\item The most effective way to
  `fill-in' entirely missing rows is to `basket-weave' i.e.\ re-observe the map with the scan
  direction perpendicular to the original.
\item A small offset between map repeats, of one pixel's length
  perpendicular to the scan direction, will cause the new spectra from one
  detector to coincide with those from a different one in the previous
  repeat.
\end{enumerate}
Each area was observed at least four times, with one repeat offset
from the original and the other two basket-weaved variants of the
original and offset scans. Maps of the resulting root mean
square (RMS) noise for the three transitions towards IC348 are shown
in Fig. \ref{fig:rms_noise}. 

\subsection{References and calibrations}

Finding reference positions has proved considerably harder for HARP than 
single-pixel receivers, as all 16 detectors must point at
line-free regions, requiring a `blank' piece of sky approximately
2\,arcmin square. Three such positions were used for our Perseus survey (see Table \ref{table:offs}). Separate one minute
`stare' observations were undertaken at $^{12}$CO $J=3\to 2$ towards these
positions, using references even further from the cloud. All were
found to have no emission above the noise, re-binned in
1\,km\,s$^{-1}$ velocity channels, in every working detector.  

\begin{table}
\caption{Reference positions used.} 
\begin{tabular}{ccc}
\hline
RA (J2000) & Dec (J2000) & Sub-Region used \\
(h~m~s) & ($^\circ$~$'$~$''$) \\
\hline
03:29:00 & +31:52:30 & NGC1333 \\
03:48:38 & +31:49:39 & IC348 \\
03:33:11 & +31:52:02 & L1448 and L1455 \\
\hline
\end{tabular}
\label{table:offs}
\end{table} 

All the standard telescope observing procedures were
followed. Calibration spectra were taken frequently towards CRL 618 of the $^{12}$CO or $^{13}$CO
$J=3\to 2$ lines as appropriate. The intensity in the reference detector was then compared to previously recorded standards and only
if they matched within a calibration tolerance,
were any subsequent observations allowed to continue. All the data
products and images we present are on the antenna temperature scale
($T_\rmn{A}^*$, \citealt{kutner81}), which can be converted to
main beam brightness temperature, $T_\rmn{mb}$, using
$T_\rmn{mb}=T_\rmn{A}^*/\eta_\rmn{mb}$. The efficiency we use,
$\eta_\rmn{mb}=0.66$, was measured during the commissioning of HARP.

\section{Data reduction} \label{sec:datareduction}

We used the Starlink software
collection\footnote{Now maintained and developed by the Joint
  Astronomy Centre, see \url{http://starlink.jach.hawaii.edu}} for the
analysis and reduction of HARP data. First, bad or extremely noisy
data were flagged in the supplied HARP time-series format. The
resultant spectra were placed on to a spatial grid using the {\sc
  smurf} reduction package \citep{jenness08}. Our final data products are sampled
on a 3\,arcsec grid, using a 9\,arcsec full-width half-maximum (FWHM) Gaussian
gridding kernel, resulting in an equivalent FWHM
beam size of 17.7 and 16.8\,arcsec for $^{13}$CO/C$^{18}$O and $^{12}$CO respectively. 
After gridding, a linear baseline was removed from each spectrum by
fitting line-free portions using the Starlink
{\sc kappa} applications. The spectra were
 also re-binned spectrally to resolutions of 0.15\,km\,s$^{-1}$ for
$^{13}$CO/C$^{18}$O and 1\,km\,s$^{-1}$ for the low-resolution $^{12}$CO
data. 

\subsection{The HARP flatfield} \label{sec:flatfield}

\begin{figure}
\begin{center}
\includegraphics[width=0.37\textwidth]{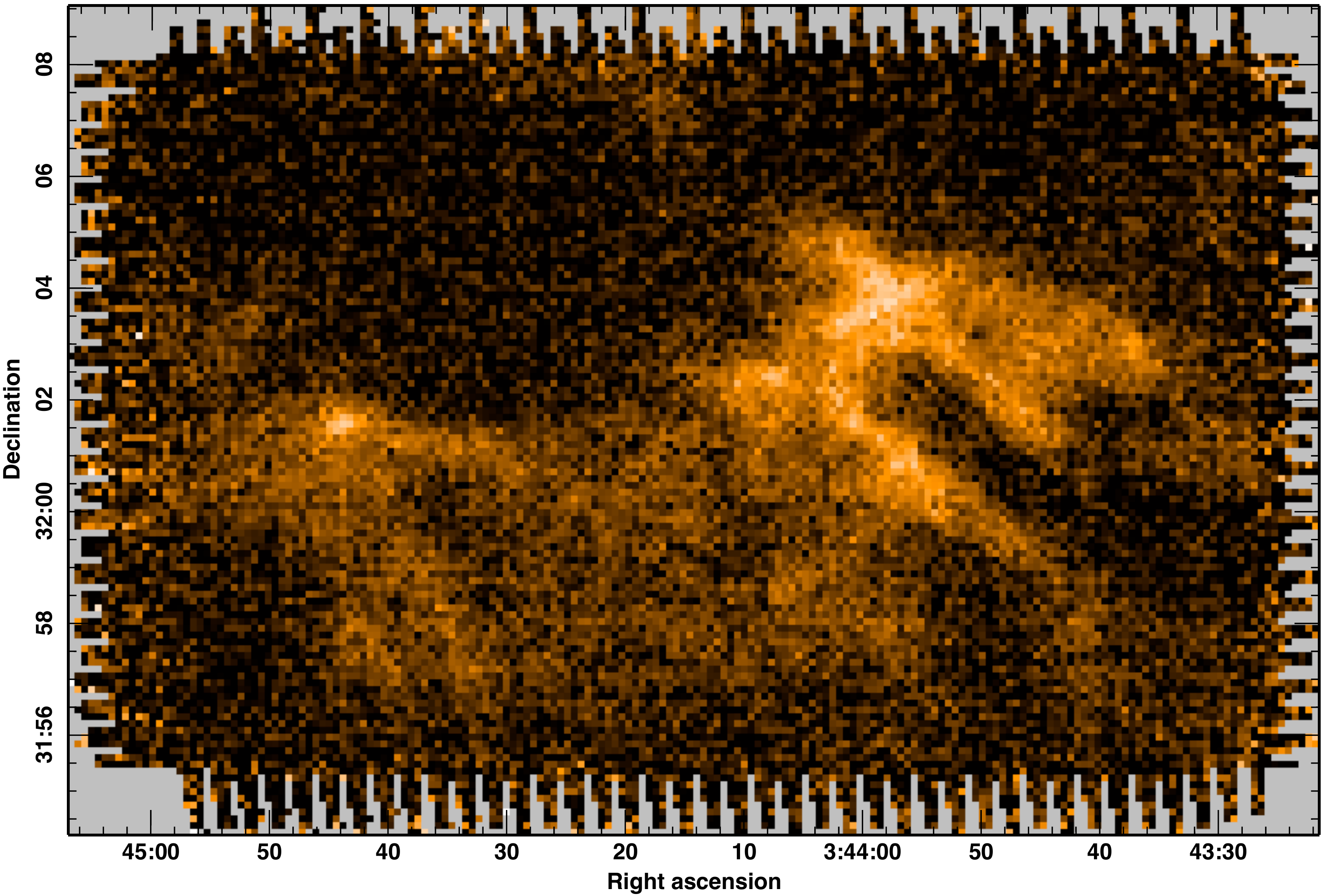}

\vspace{0.2cm}
\includegraphics[width=0.37\textwidth]{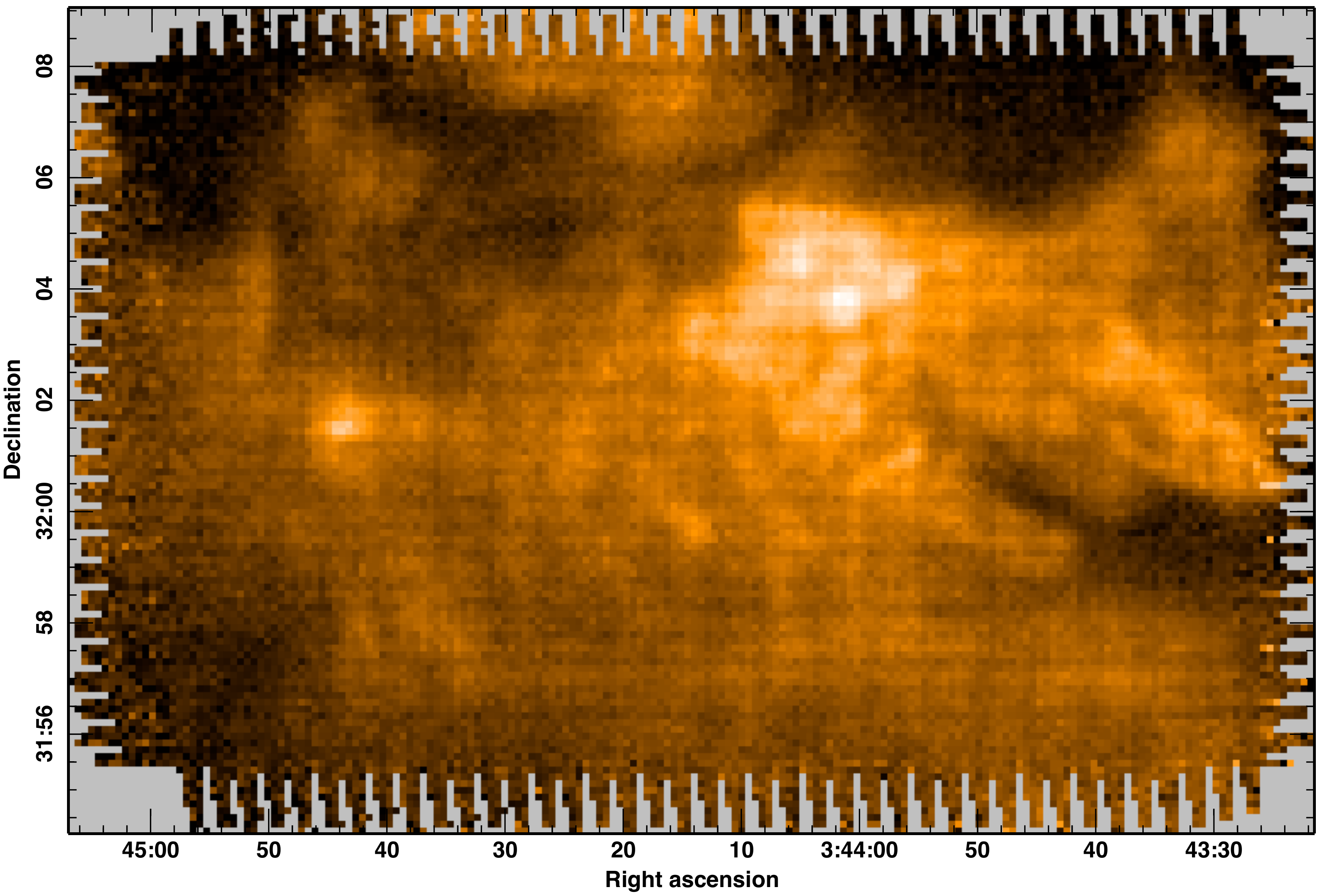}

\vspace{0.2cm}
\includegraphics[width=0.37\textwidth]{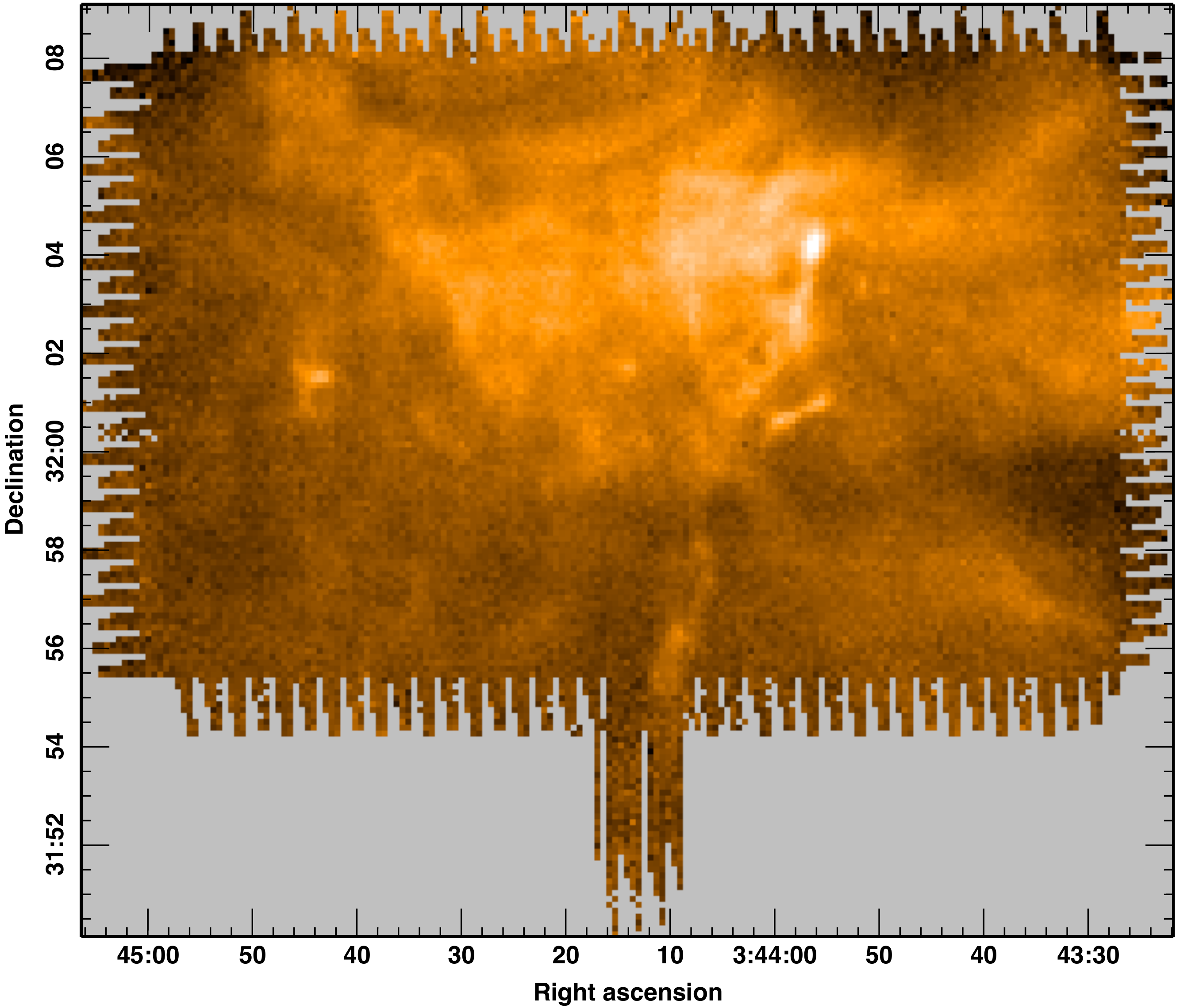}

\vspace{0.2cm}
\includegraphics[width=0.37\textwidth]{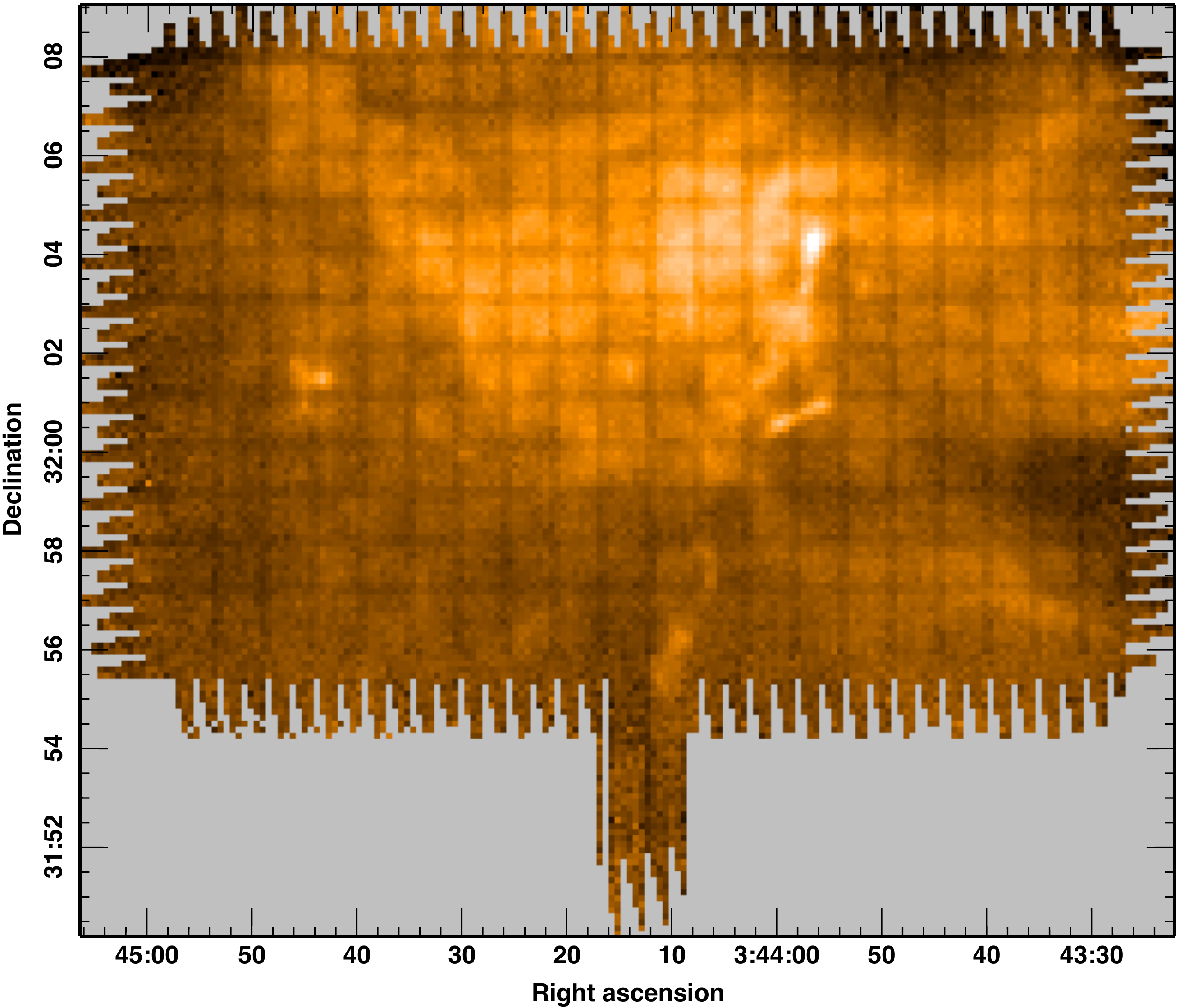}
\caption{Integrated
 intensity images for the different \threetotwo\ transitions towards IC348
 displaying the striping artefacts from our basic data reduction. The
 spectra are distributed on a 7.3\,arcsec grid using nearest-neighbour
 allocation to emphasize detector based variations. From top to bottom
 are \ceighteeno, \thirteenco, \twelveco\ low-resolution and
 \twelveco\ high-resolution maps. The $^{12}$CO data
 from a few telescope scan rows appear to have slipped, this is a
 known problem where the telescope does not return to the top of the
 map before commencing integrating.} 
\label{fig:all_stripes}
\end{center}
\end{figure}

After the basic data reduction, 
distinctive stripes were apparent in the integrated intensity
images. Near the peak of the line intensity, rows of high and low
pixel values lay parallel to the scan direction,
implying certain detectors were systematically higher or
lower by as much as a factor of two. The worst affected data were from the $^{13}$CO and
high-resolution $^{12}$CO scans (see Fig. \ref{fig:all_stripes}). A similar pattern was clear in \emph{every}
individual scan.  

The origin of such systematic calibration
differences between detectors is unknown but is likely to be in the
intermediate frequency system. However, for the purposes of this
paper we require only a pragmatic method to eliminate the striping
artefacts from molecular cloud data. Therefore, we derive a set of temperature conversion factors (TCFs), by which we can multiply the spectra from each
detector to get them on to a common intensity scale. This procedure is
similar to the `flatfield' procedure which accounted for the flux conversion
factors with SCUBA \citep{holland99}, therefore we refer to the analogous
flatfield for HARP.

\begin{figure*}
\includegraphics[width=0.3\textwidth]{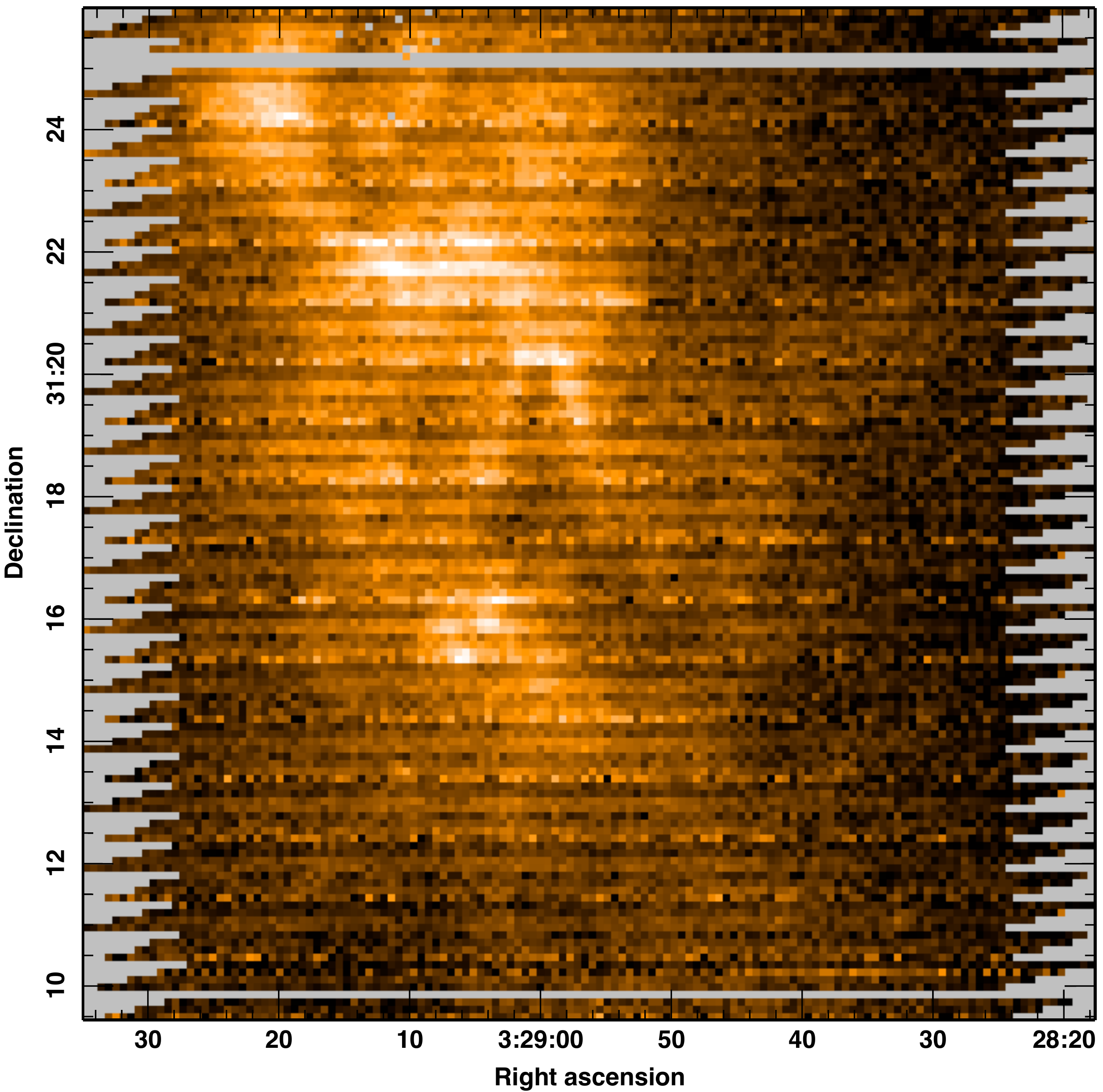}
\includegraphics[width=0.3\textwidth]{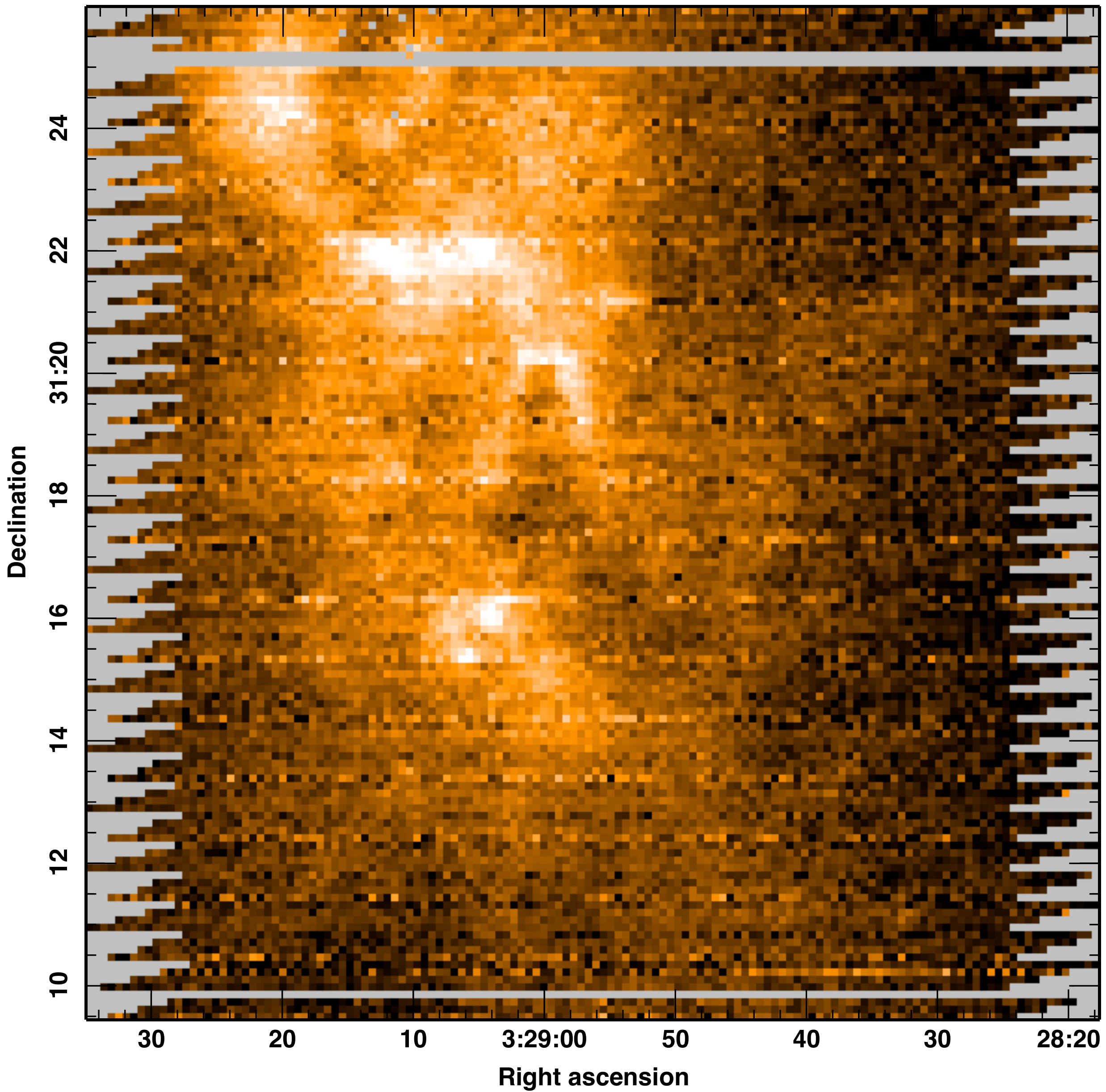}\\

\vspace{0.2cm}
\includegraphics[width=0.3\textwidth]{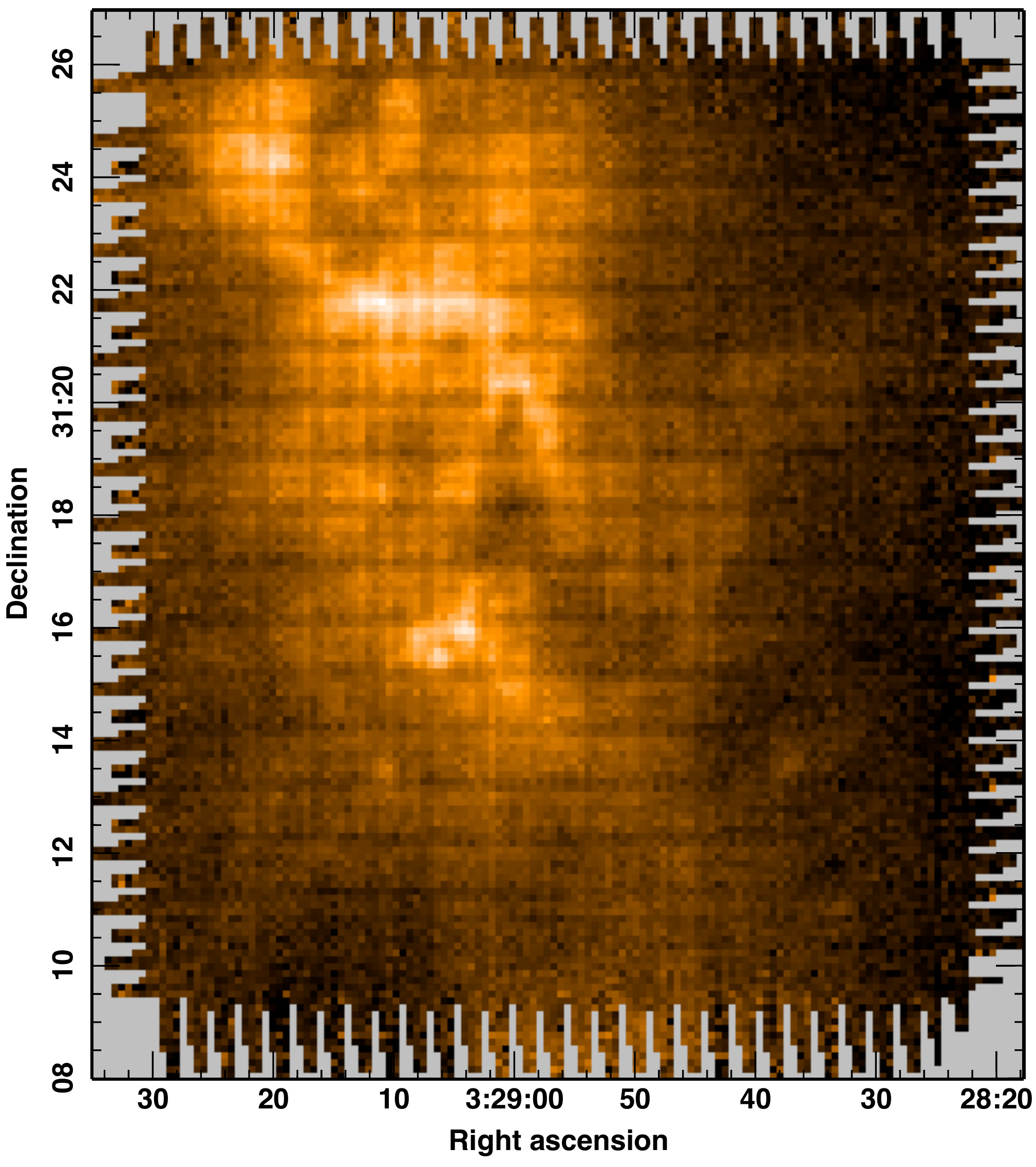}
\includegraphics[width=0.3\textwidth]{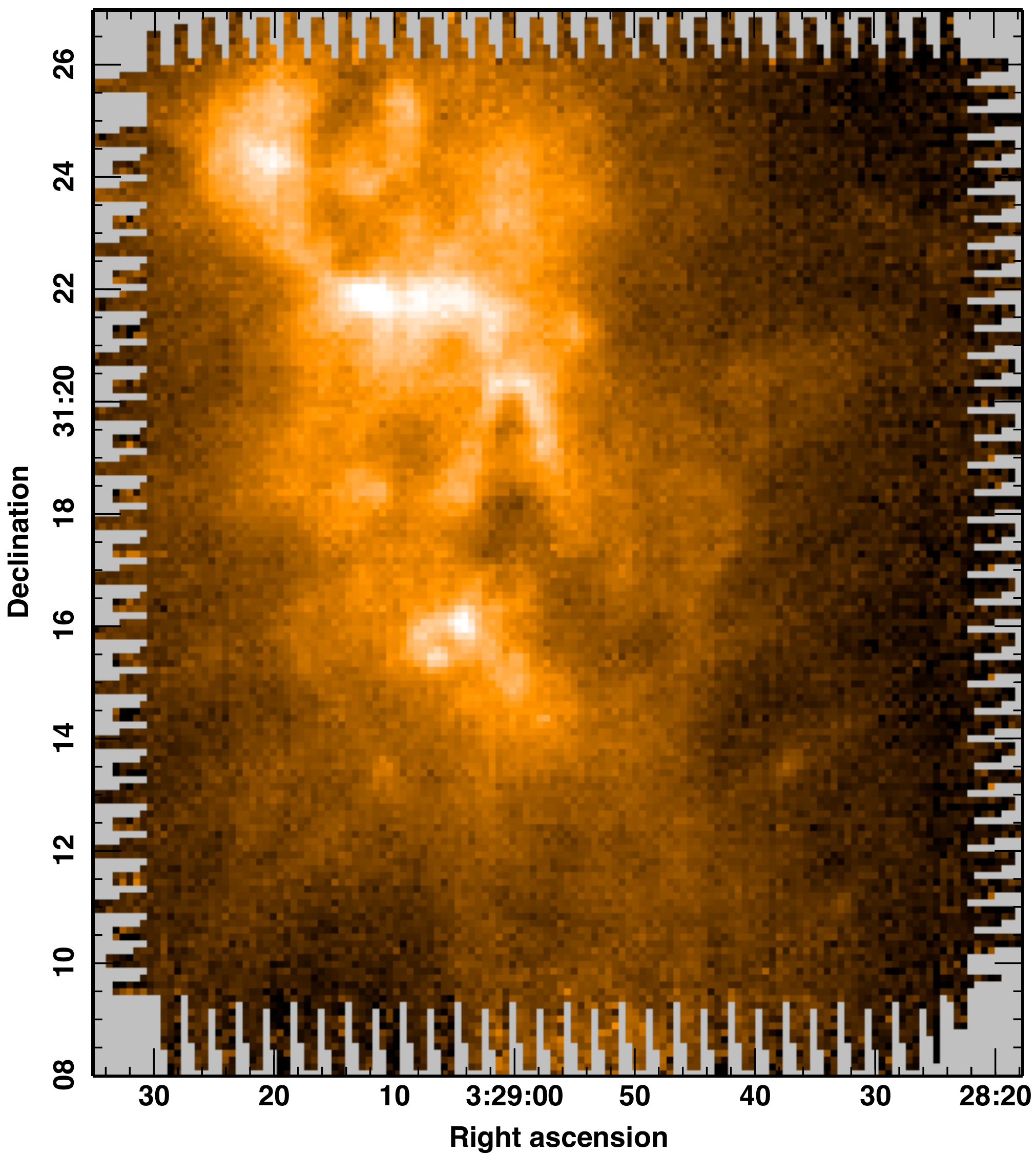}
\caption{Action of the flatfield procedure on $^{13}$CO integrated
  intensity images towards NGC1333. The panels are from top to bottom
  and left to right: data
  from a single scan, unaltered from the basic reduction; the same
  data as in the previous panel with the flatfield applied; data from all the
  NGC1333 scans, unaltered from the basic reduction; and the same data
  as in the previous panel but with the flatfield applied.} 
\label{fig:flatfield}
\end{figure*}

Ideally, to derive these factors from observations, one would observe
a small source, providing a fully-sampled map for each detector, so
that the total flux can be compared between them. However, we must derive
the factors from the  scientific observations
themselves, comparing the intensity detector-by-detector for an entire
map. First, datacubes
were produced for each detector \emph{individually} and the spectra
integrated over the line and summed over the map. We divided the sum from
the reference detector (R11, one of the central four in the array, see \citealp{hills09} for an explanation
of how the JCMT points and tracks with HARP) by those from each of the other
working detectors in turn (up to 15 in number), to calculate the 
respective TCF. The reference detector was taken as the intensity
standard as it is not masked out in any of our
observations. Additionally, all the calibrations are centred
on it, so we have a robust determination of its
performance. Once the factors are known, the output from each detector
is multiplied by its respective TCF. The procedure worked very
effectively for most of the data as shown in Fig. \ref{fig:flatfield}
and all the subsequent images in this paper have been multiplied by
TCFs, except the \ceighteeno\ data towards L1455 which had an
insufficient signal-to-noise ratio (SNR)and all the low-resolution
\twelveco\ datasets, which did not display the artefacts. 

It is perhaps surprising that this procedure worked so effectively and
undoubtedly the nature of the targets contributed to its success. The
method depends on each detector receiving exactly the same amount of
emission. Offset map repeats and basket-weaving helped each
detector to sample a larger proportion of the target field. However,
it is the uniformity of a molecular cloud's emission that means, in general,
there is little difference in intensity between the different
detectors. If there were many compact sources or intense emission
towards the edges of these maps, which are only observed by a few detectors, then this
technique would be ineffective. Other datasets taken in the same period show similar striping artefacts, notably those of the JCMT GBS, who have implemented our algorithm across a number of their
fields. 

\section{Results} \label{sec:results}

\subsection{NGC1333}

NGC1333 is a young stellar cluster in the west of Perseus, associated
with the reflection nebula of the same name and the dark cloud L1450. It
has been widely studied (see \citealp{walawender08} for a review) and
is the most active region of star formation in the Perseus
complex. The stellar cluster is very young ($<1$\,Myr) and highly obscured,
containing about 150 stars totalling 79\,M$_\odot$
(\citealp*{lada96}; \citealp{wilking04}). The latest infrared (IR) survey with \emph{Spitzer}
\citep{jorgensen06,rebull07,gutermuth08} identified 137 objects in the
cluster: 98
pre-main sequence stars and 39 protostars. These protostars are
correlated with the position of dense molecular material and dust. In
total, there is approximately 450\,\msun\ of gas in the region
\citep{warin96}. Much of this gas lies in dense filaments, surrounding
cavities \citep{lefloch98,quillen05}. The YSOs in NGC1333 also drive a
large number of overlapping outflows, providing one of the clearest examples of the
self-regulation of star formation \citep{knee00}.

\begin{figure*}
\begin{minipage}{0.48\textwidth}
\hbox{\hspace{0.62cm}\includegraphics[width=7.85cm]{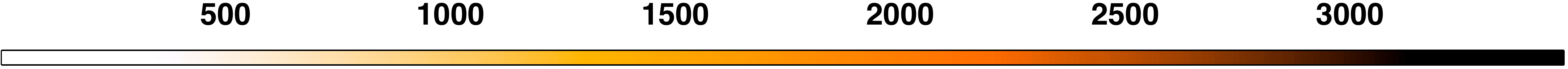}}
\includegraphics[width=\textwidth]{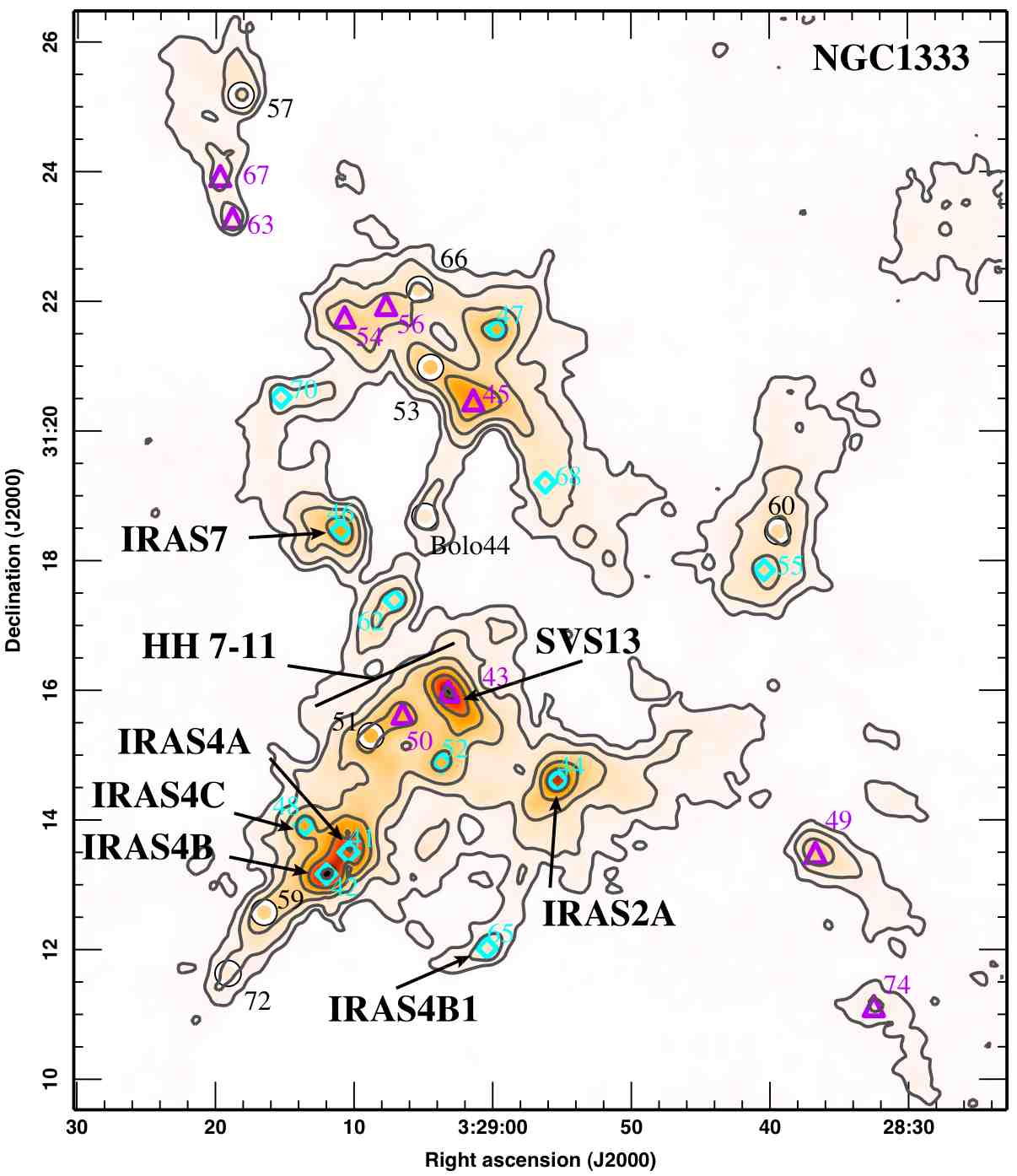}\\
\end{minipage}
\begin{minipage}{0.48\textwidth}
\vspace*{-0.4cm}
\hbox{\hspace{0.62cm}\includegraphics[width=7.9cm]{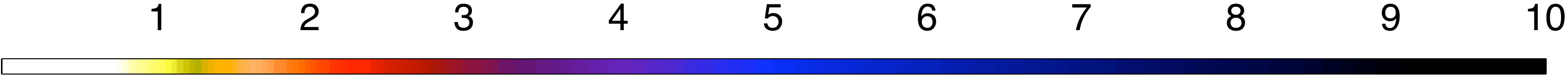}}
\includegraphics[width=\textwidth]{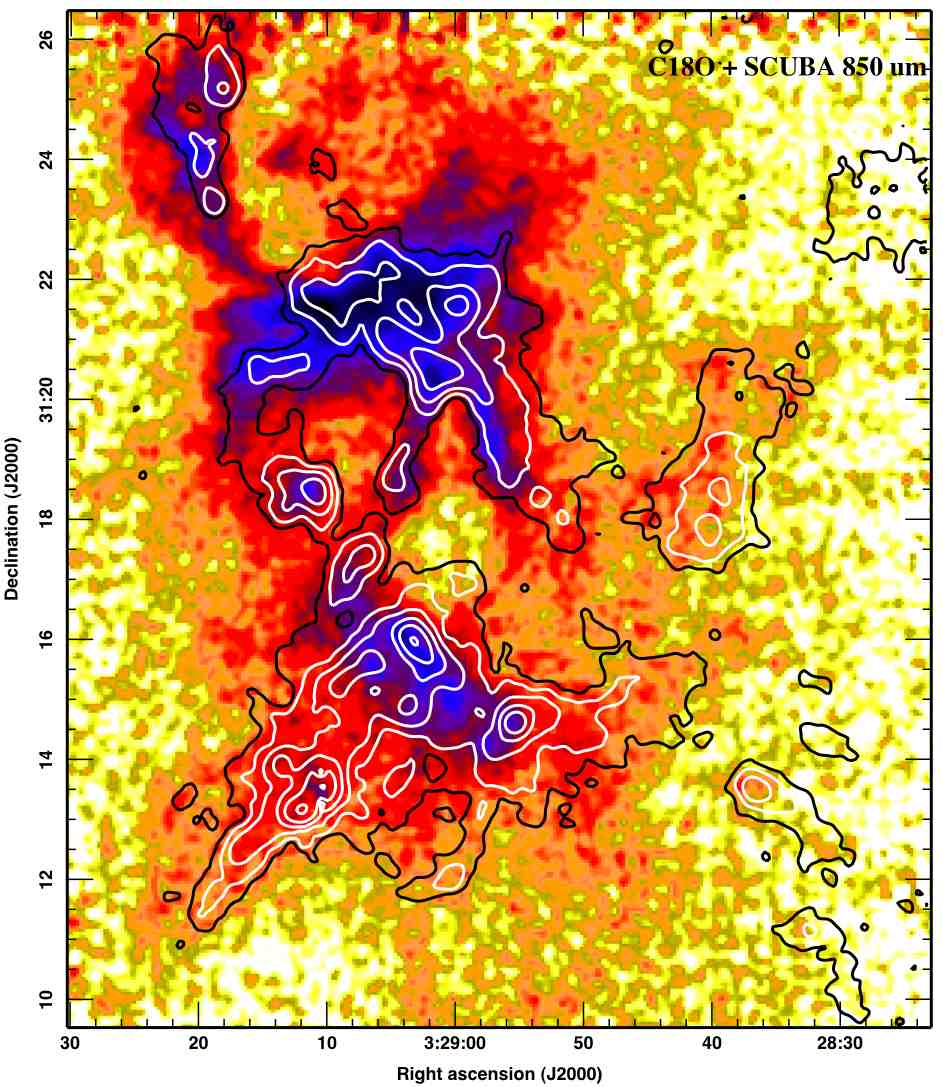}
\end{minipage}
\begin{minipage}{0.48\textwidth}
\vspace{-0.4cm}
\hbox{\hspace{0.62cm}\includegraphics[width=7.9cm]{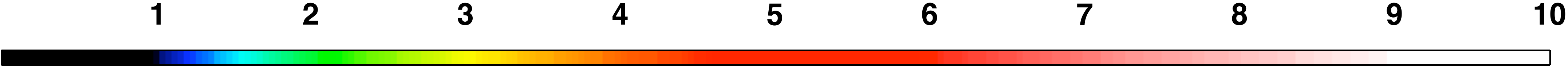}}
\includegraphics[width=\textwidth]{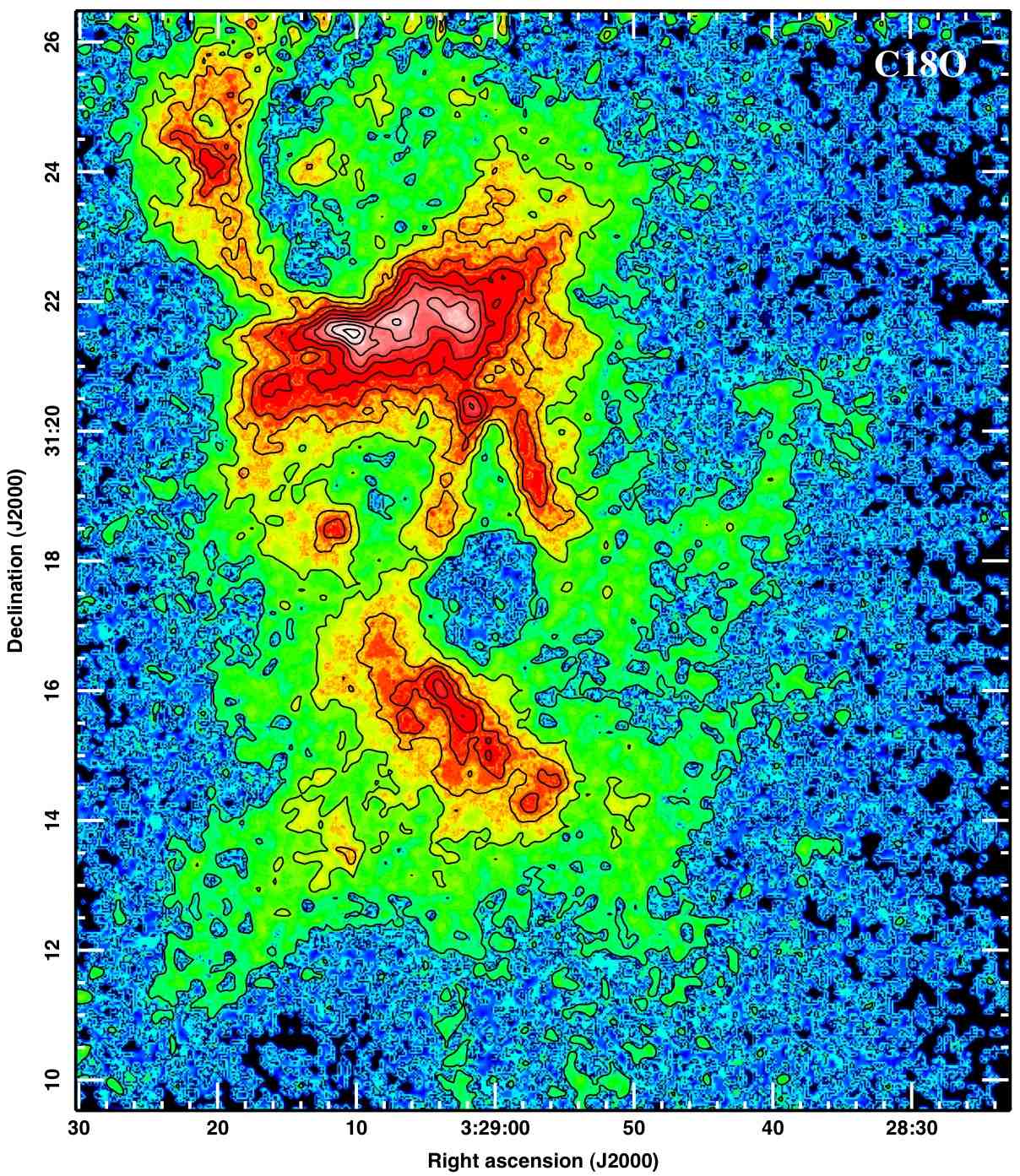}
\end{minipage}
\begin{minipage}{0.48\textwidth}
\vspace{-0.4cm}
\hbox{\hspace{0.65cm}\includegraphics[width=7.8cm]{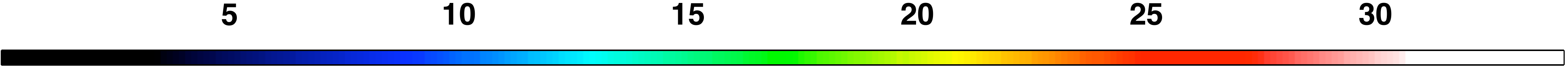}}
\includegraphics[width=\textwidth]{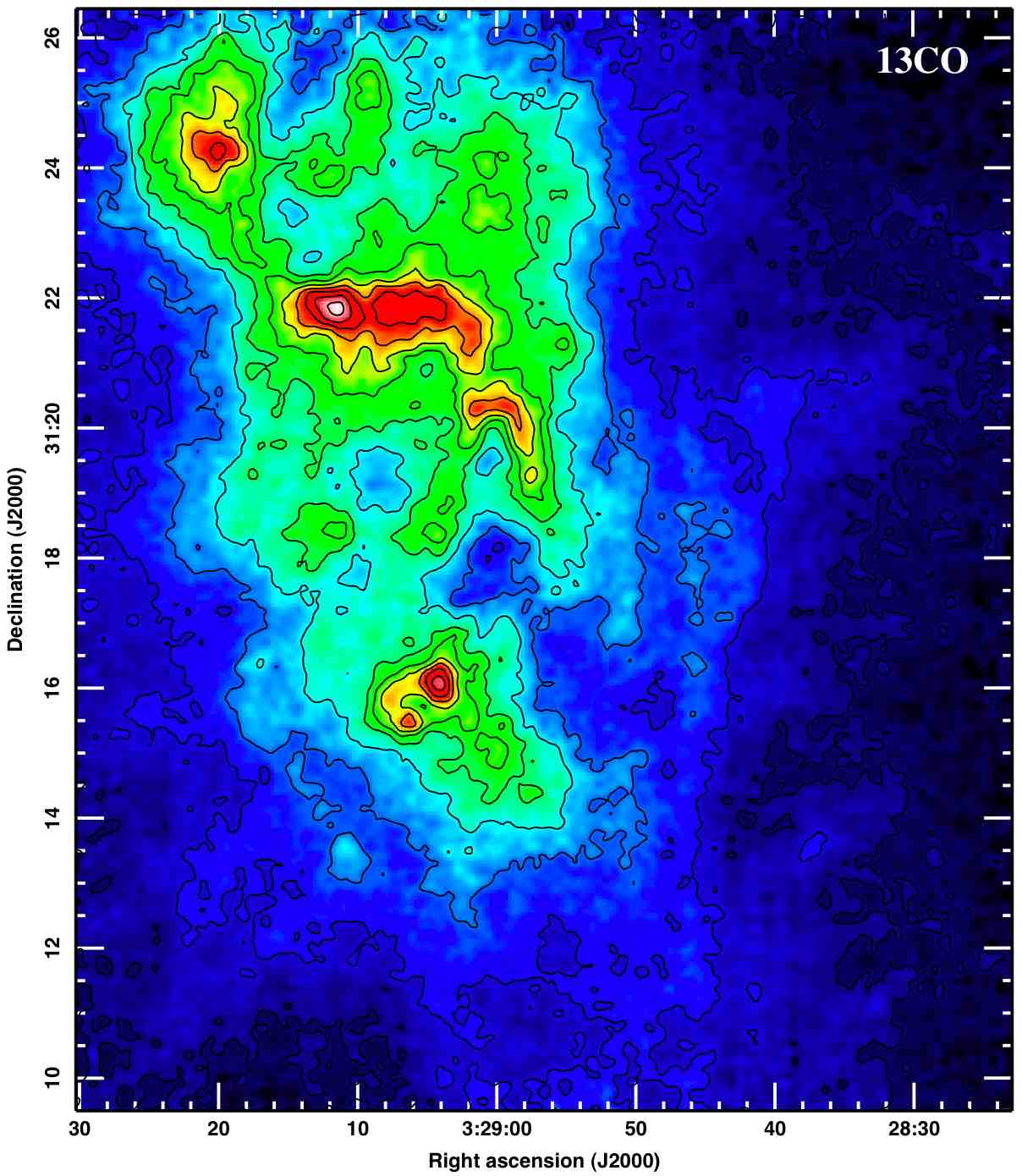}
\end{minipage}
\caption{Overview of the data towards NGC1333. Top left: SCUBA
  850\,\micron\ emission with contours at 100, 200, 400, 800, 1600 and
  3200\,\mjybeam\ \citep{hatchell05} with a colour-scale as shown in \mjybeam. The peak of submillimetre cores
  identified and classified by \citet{hatchell07a} are labelled with
  their \citeauthor{hatchell07a} source number and marked with:
  black and white circle (starless cores), light blue diamonds (Class
  0 protostars) or purple triangles (Class I
  protostars). \citet{sandell01} originally presented this data and
  located many submillimetre sources. Additionally
labelled are prominent sources in the field, NGC1333-IRAS2A
(\citealp{jennings87,blake95}; \citealp*{looney00}), NGC1333-IRAS4A to C
(\citealp{jennings87}; \citealp*{lay95,rodriguez99}; \citealp{looney00}), SVS13 (\citealp*{strom76}; \citealp{haschick80,snell81,grossman87,hirota08}),
HH7-11 \citep{herbig74} and NGC1333-IRAS7
(\citealp{lightfoot86,jennings87}; \citealp*{cohen91};
\citealp{jorgensen07b}). Top right: Contours of
SCUBA 850\,\micron\ emission (as previously)
overlaid on the \ceighteeno\ \threetotwo\ integrated intensity as
displayed on the bottom left. Bottom
  left: Integrated \ceighteeno\ \threetotwo\ intensity, $\int
T_\rmn{A}^* \rmn{d}v$, from 5 to 10\,\kms\ with contours from 1 to
10\,\kkms\ in steps of 1\,\kkms. Bottom right: Integrated \thirteenco\ \threetotwo\ intensity, $\int
T_\rmn{A}^* \rmn{d}v$, from 2 to 17\,\kms\ with contours from 3 to
33\,\kkms\ in steps of 3\,\kkms.}
\label{fig:ngc1333_data}
\end{figure*}

We present maps of the \thirteenco\ and \ceighteeno\ HARP data in
Fig. \ref{fig:ngc1333_data}, alongside the SCUBA 850\,\micron\
emission over the observed region, originally from observations by
\citet{sandell01} but later analysed as part of the survey of
\citet{hatchell05}. Qualitatively, the \ceighteeno\ and SCUBA maps
closely resemble each other. The central cavity, just north of SVS13
is clear in all the tracers, as is the horseshoe of emission
enclosing its northern boundary. This would suggest that similar
material is being explored in both the \ceighteeno\ and SCUBA data. However, there are clear differences in the structures as
well. IRAS4 and IRAS2 \citep{jennings87}, are two of the brightest
sources in the SCUBA field, yet they have much weaker
\ceighteeno\ emission and can scarcely be discerned in the
\thirteenco\ data. The string of cores to the north of the central
cavity, (sources 54, 56 and 66\footnote{The source numberings we refer
  to in the rest of this paper are the catalogue number from \citet{hatchell07a}}) have quite
weak SCUBA emission but are the brightest objects in the \ceighteeno\
maps.   

\subsection{The IC348 molecular ridge}

Our second field is in the vicinity of another young IR cluster
($\sim$2\,Myr old), IC348 (see \citealp{herbst08} for a review), in
the east of Perseus. It comprises several hundred members totalling about 160\,M$_\odot$
\citep{luhman03}. The low stellar disc fraction and lack of
outflow activity may indicate that the cluster is coming to the end of its
star-forming phase \citep{luhman98}. In fact the region we refer to as
IC348 hereafter (Fig. \ref{fig:ic348_data}) is really a bright molecular ridge some 10\,arcmin southwest of
the cluster, sometimes associated with the
``Flying Ghost Nebula'' \citep{boulard95}. This area, in contrast to
IC348, \emph{is} currently undergoing star formation with many embedded
objects and outflows, possibly triggered by
the nearby cluster (e.g.\ \citealp{bally08}). The best-known feature
in the area is HH211, a highly-collimated bipolar outflow driven by a Class 0
protostar discovered by \citet*{mccaughrean94}, which has been
the target of many interferometric studies subsequently (e.g.\ \citealp{gueth99,chandler01}). 

\begin{figure}
\hbox{\hspace{0.4cm}\includegraphics[width=7.9cm]{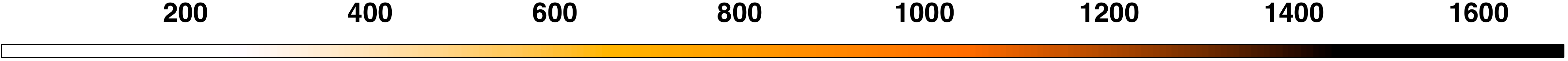}}
\includegraphics[width=0.47\textwidth]{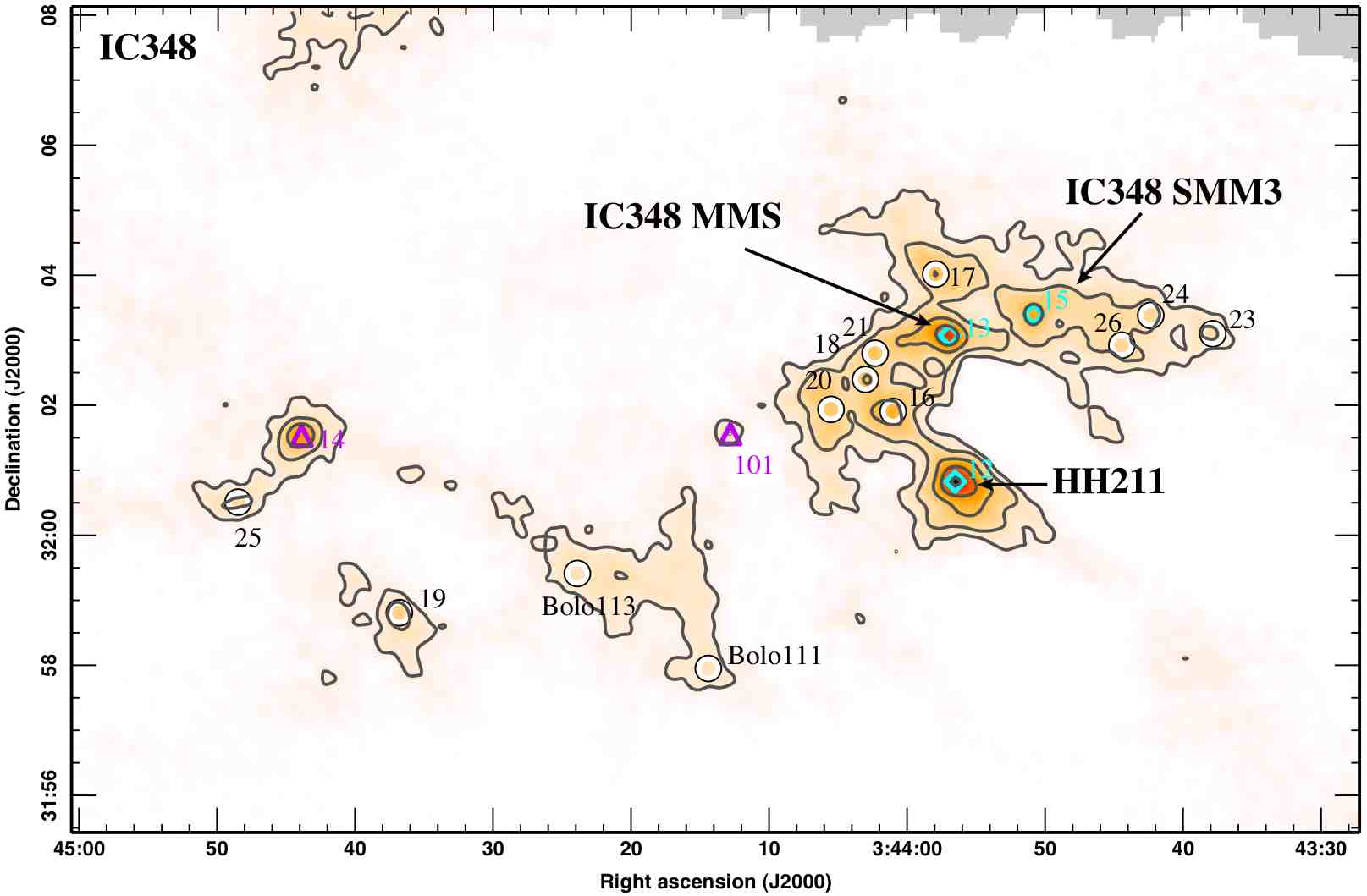}
\vspace{0.2cm}

\hbox{\hspace{0.4cm}\includegraphics[width=7.9cm]{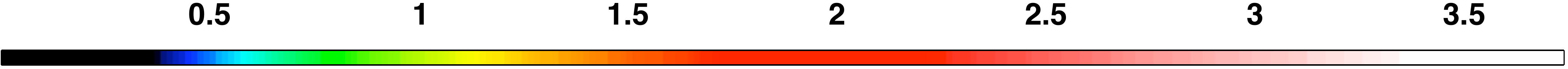}}
\includegraphics[width=0.47\textwidth]{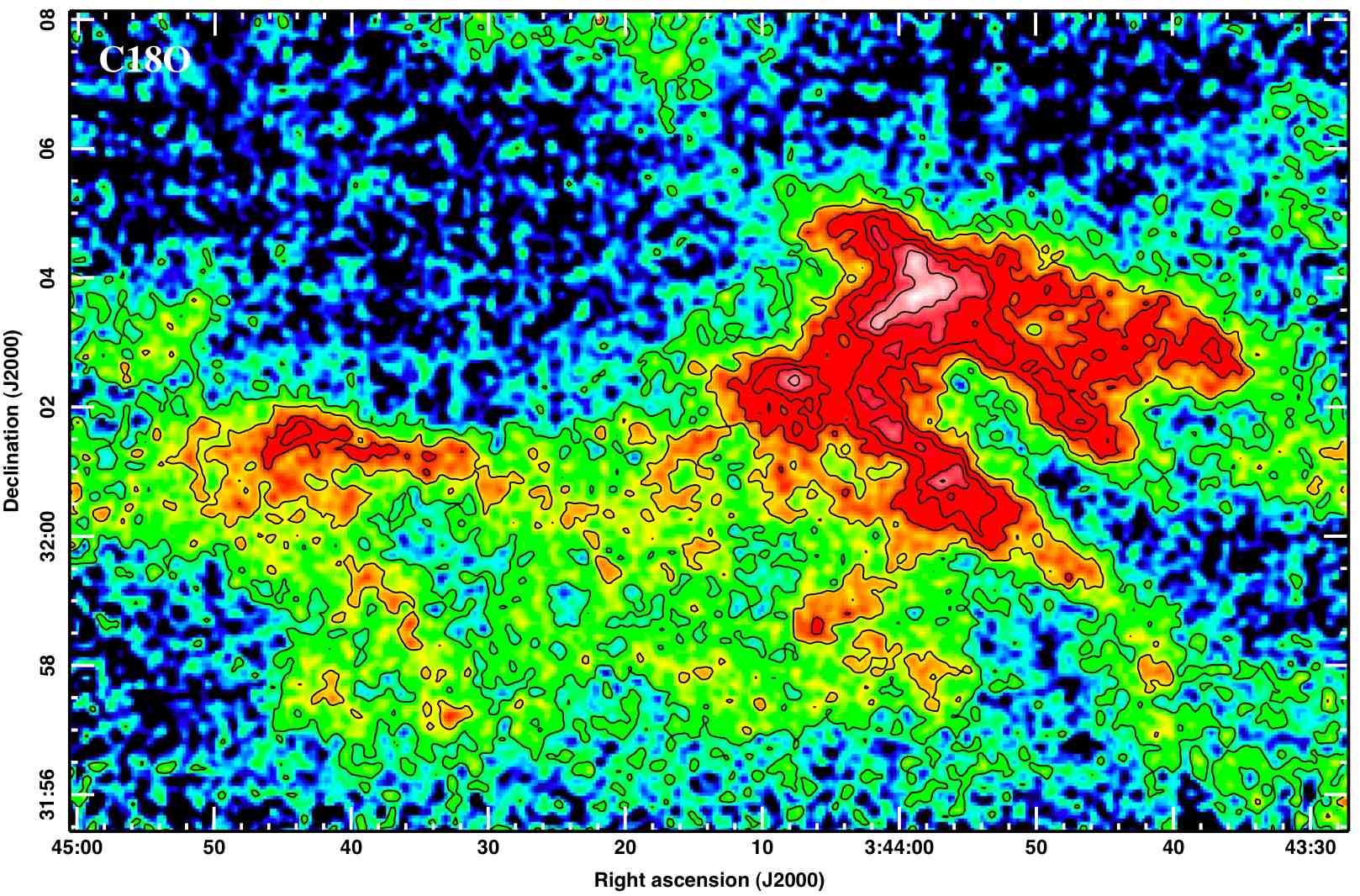}
\vspace{0.2cm}

\hbox{\hspace{0.4cm}\includegraphics[width=7.9cm]{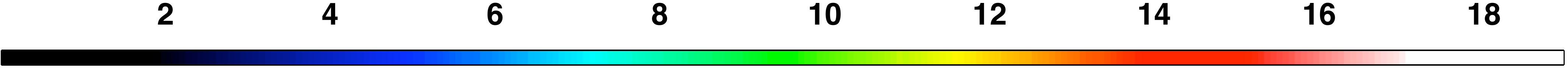}}
\includegraphics[width=0.47\textwidth]{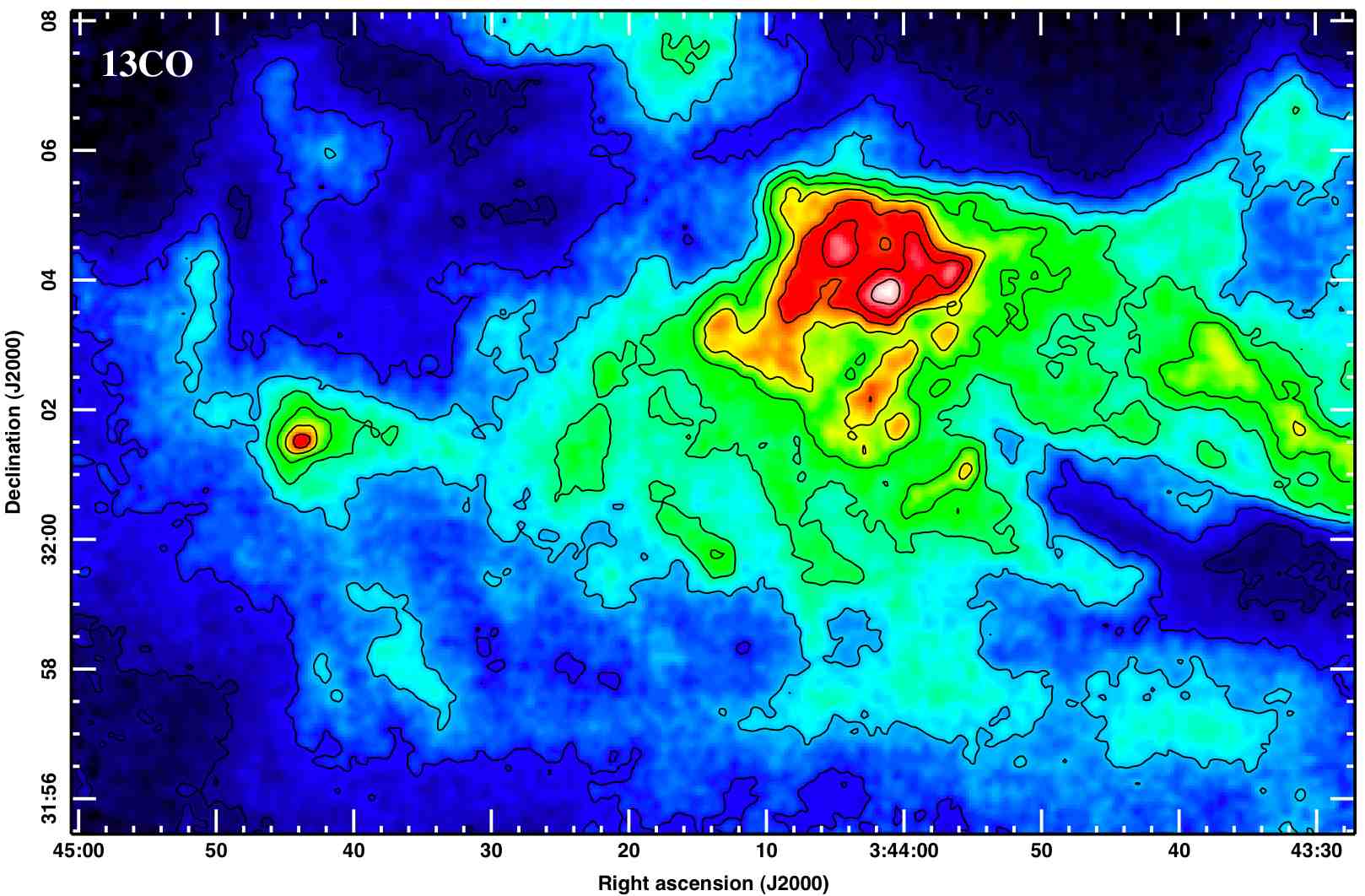}
\caption{Overview of the data towards IC348 as for NGC1333 in
  Fig. \ref{fig:ngc1333_data}. Top: SCUBA 850\,\micron\ emission. Prominent sources in the field are labelled: HH211
\citep{mccaughrean94,gueth99,chandler01}, IC348-MMS
\citep{eisloffel03,tafalla06,walawender06} and IC348-SMM3
\citep{tafalla06}. Middle: Integrated \ceighteeno\ \threetotwo\
intensity from 7 to 10\,\kms\ with contours from 0.5 to 3.0\,\kkms\ in
0.5\,\kkms\ steps. Bottom: Integrated \thirteenco\ \threetotwo\
intensity from 5 to 12\,\kms\ with contours from 2 to 18\,\kkms\ in
2\,\kkms\ steps.}
\label{fig:ic348_data}
\end{figure}

\begin{figure}
\hbox{\hspace{0.35cm}\includegraphics[width=8.05cm]{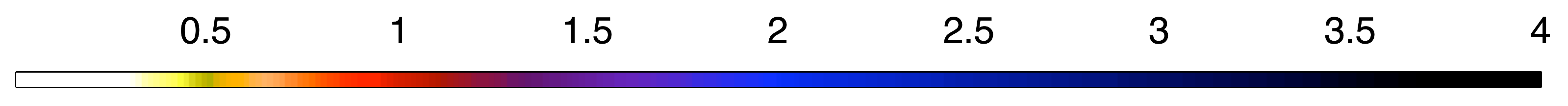}}
\includegraphics[width=0.47\textwidth]{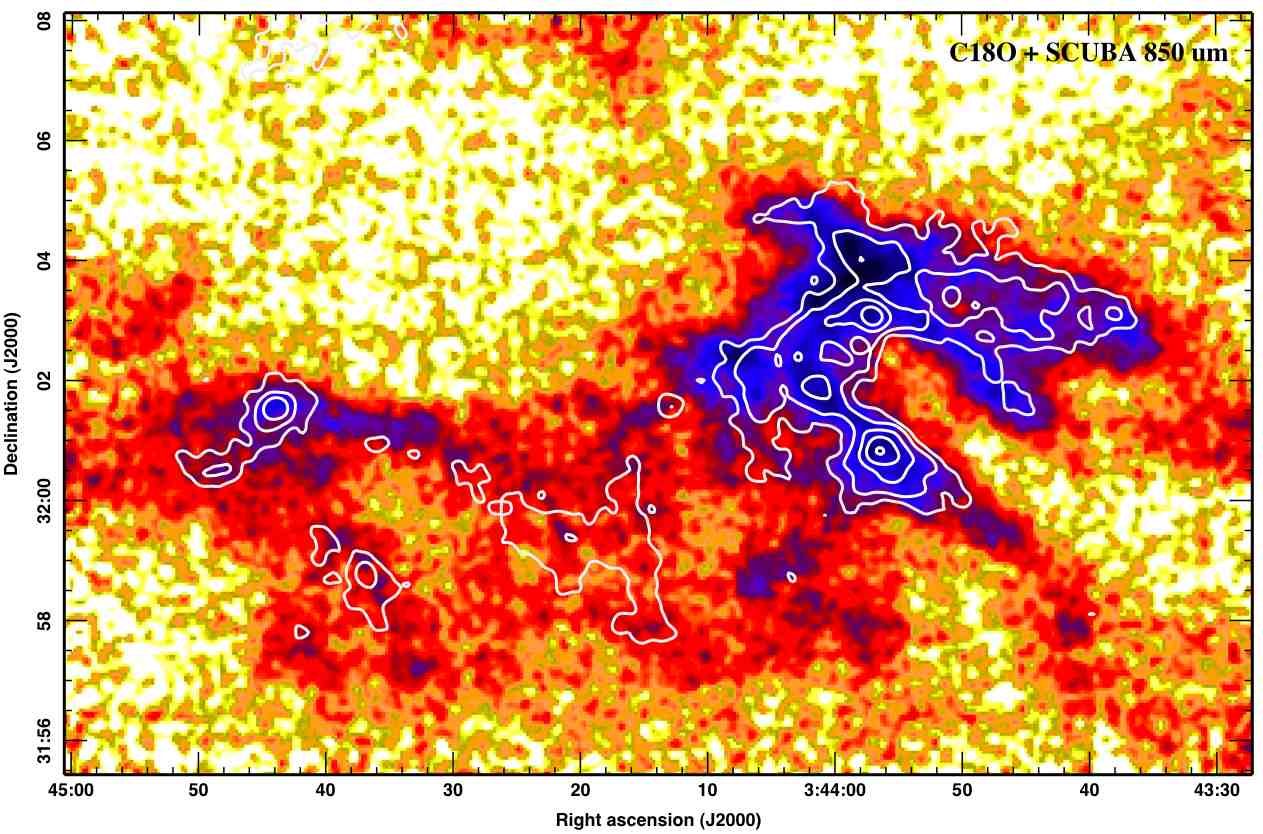}
\hbox{\hspace{0.4cm}\includegraphics[width=7.9cm]{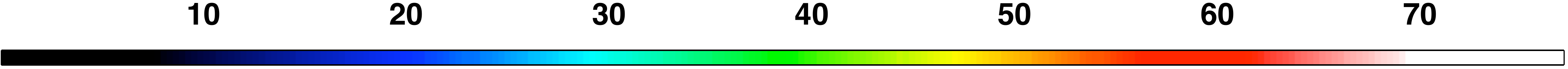}}
\includegraphics[width=0.47\textwidth]{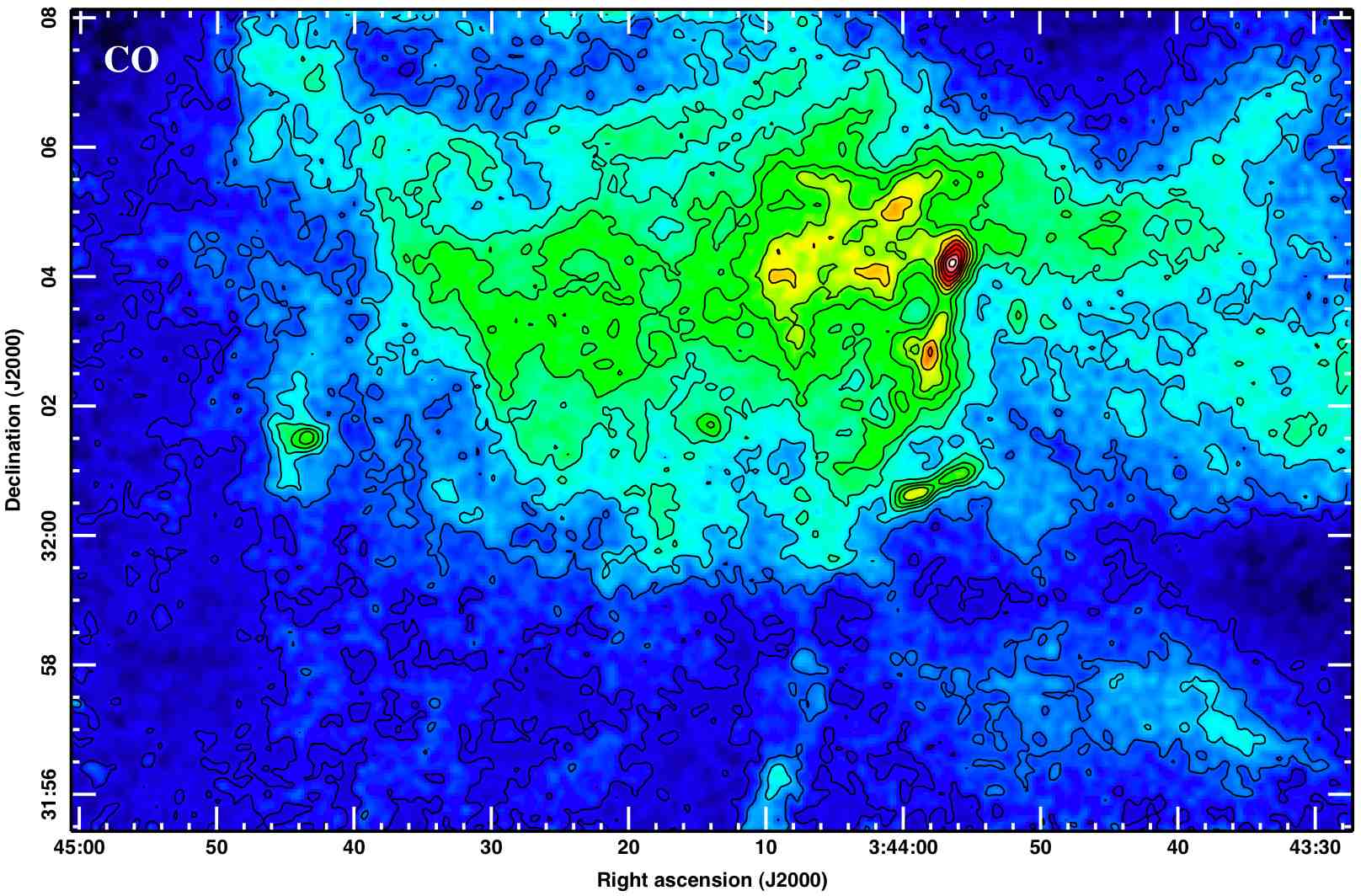}
\vspace{0.2cm}

\hbox{\hspace{0.4cm}\includegraphics[width=7.9cm]{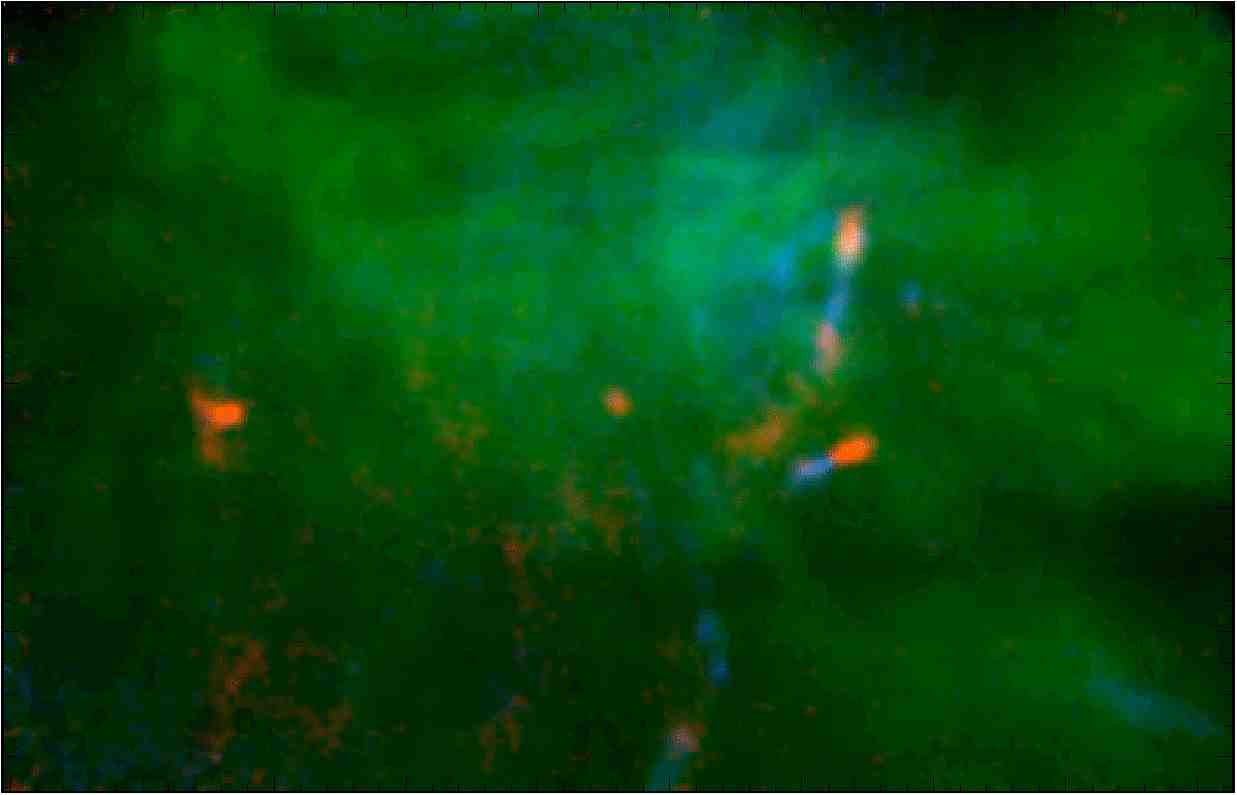}}
\contcaption{Top: Contours of
SCUBA 850\,\micron\ emission (levels as previously)
overlaid on the \ceighteeno\ \threetotwo\ integrated intensity as
before. Middle: Integrated \twelveco\ intensity from 0 to
  20\,\kms\ with contours from 15 to 75\,\kkms\ in steps of
  5\,\kkms. Bottom: Red-green-blue colour composite image of the mean \twelveco\ data value in various velocity ranges: from 3.5
to 5.5\,\kms\ (blue), 7.0 to 9.0\,\kms\ (green) and 12.0 to
14.0\,\kms\ (red).}
\end{figure}

In our \ceighteeno\ data, depicted in Fig. \ref{fig:ic348_data}, the
close correspondence with the SCUBA emission is striking. Again
there is not a simple scaling between SCUBA flux and \ceighteeno\
integrated intensity, for example the brightest \ceighteeno\ core is
the starless source 17, while the driving source of
HH211 (source 12), the brightest SCUBA core, is less prominent in
\ceighteeno. The \twelveco\ data looks very different, as it is
probably optically thick (thus tracing only outer layers of the cloud)
and has very bright outflow lobes. Two bipolar outflow
structures can be seen in the \twelveco\ integrated intensity and are more obvious in the
false colour red-green-blue (RGB) image of the same gas. The first
highly symmetric outflow (driven by
source 12) is HH211, while the second is the more confused north-south
flow from source 13 -- known also as IC348-MMS \citep{eisloffel03} or IC348-SMM2 (\citealp*{tafalla06};
\citealp{walawender06}). The bipolar
outflow discovered by \citet{tafalla06} from IC348-SMM3 (source 15), slightly west of IC348-MMS, is
also faintly discernible. Two Class I protostars (sources
14 and 101) are very red in the RGB image. This may indicate either an
outflow where only one lobe is visible due to e.g.\ an inhomogeneous
environment or that the \twelveco\ emission comes from the protostars themselves
and they are moving with respect to the ambient gas. 

\subsection{L1448}

The dark Lynds cloud L1448 (see Fig. \ref{fig:l1448_data}), the most
westerly of our targets, is a region dominated by outflow activity. Half a dozen or so YSOs reside in its
dual-core molecular structure of $\sim$100\,M$_\odot$
(\citealp{bachiller86b}; \citealp*{wolf-chase00}). Its outflows are well
studied, particularly the highly-collimated, symmetric flow originating
from L1448-C, one of the youngest known at the time (e.g
\citealp{bachiller90}; \citealp*{bally93}). Given that the energy in the outflows
exceeds the gravitational binding energy of the cloud, L1448 is likely
to be dispersed by its outflow activity \citep{wolf-chase00}.

\begin{figure}
\hbox{\hspace{0.35cm}\includegraphics[width=7.9cm]{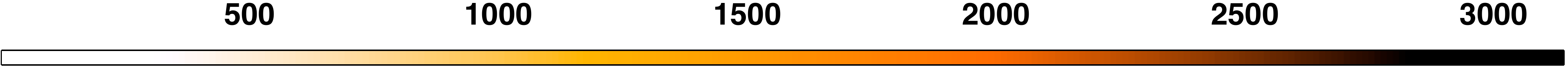}}
\includegraphics[width=0.47\textwidth]{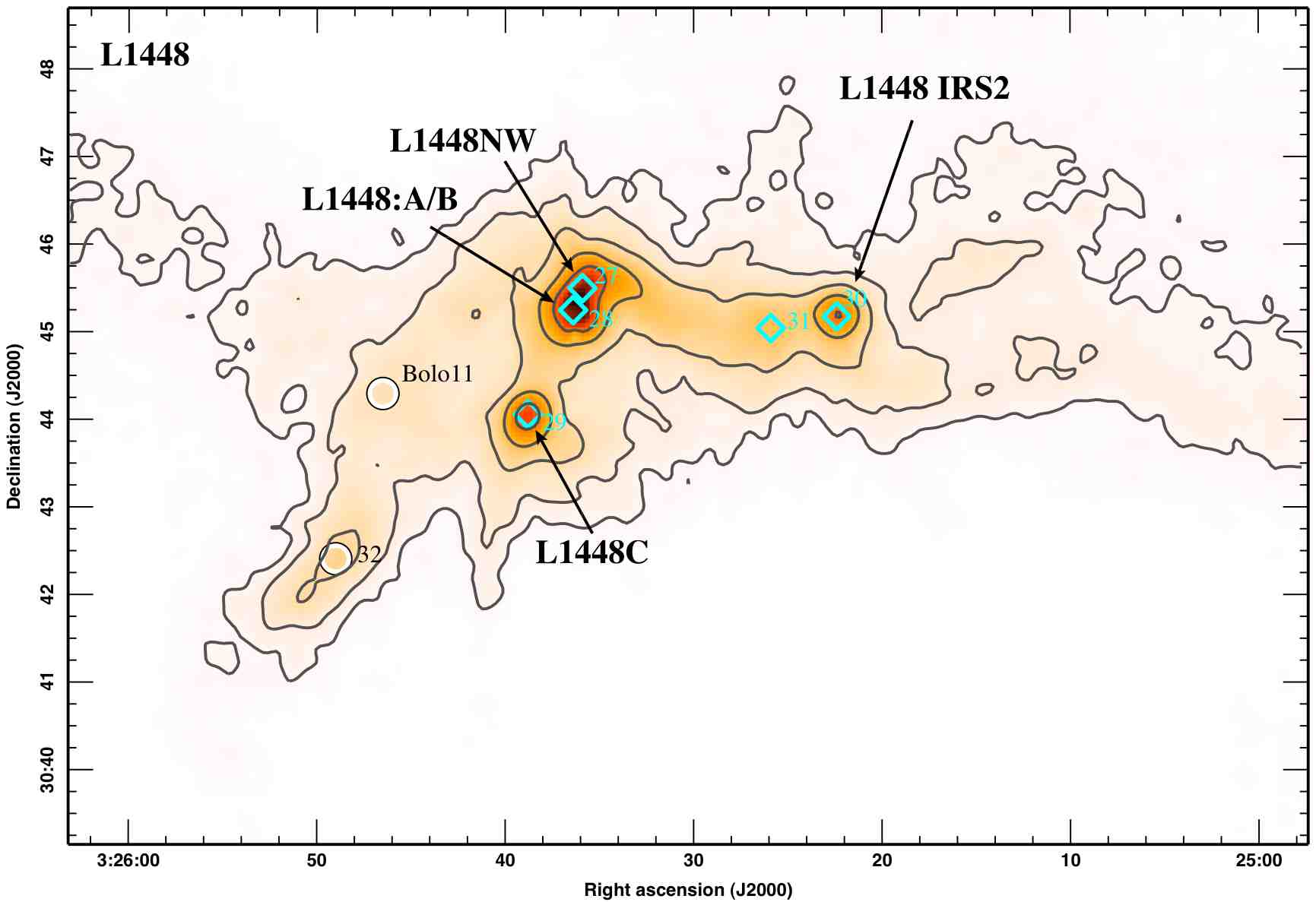}
\vspace{0.2cm}

\hbox{\hspace{0.4cm}\includegraphics[width=7.9cm]{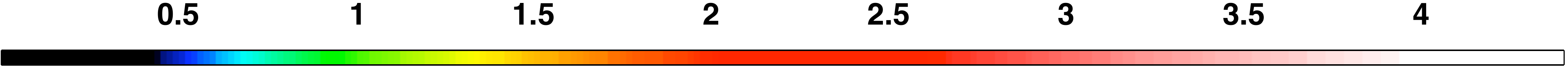}}
\includegraphics[width=0.47\textwidth]{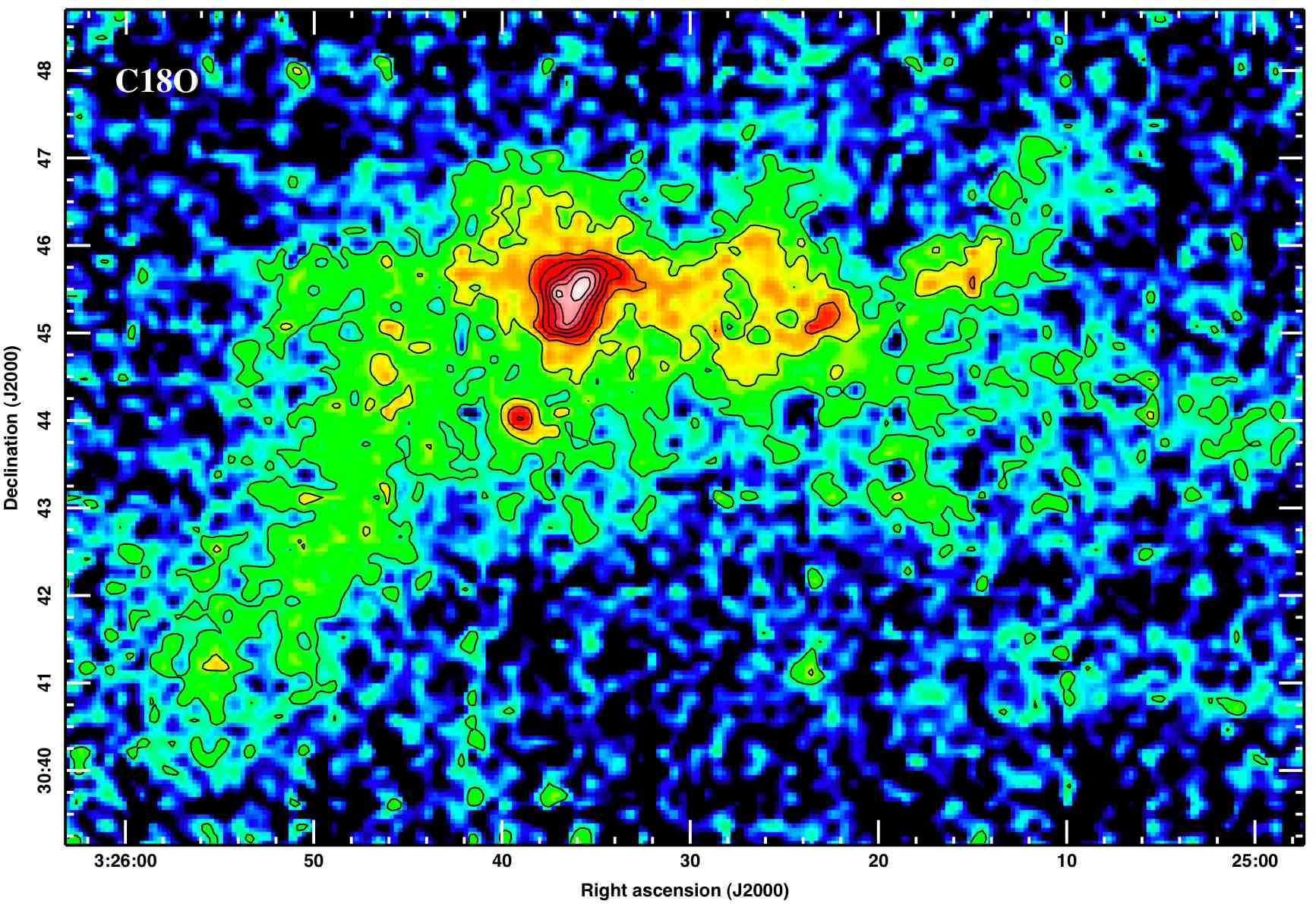}
\vspace{0.2cm}

\hbox{\hspace{0.4cm}\includegraphics[width=7.9cm]{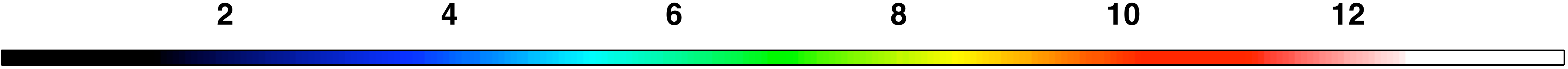}}
\includegraphics[width=0.47\textwidth]{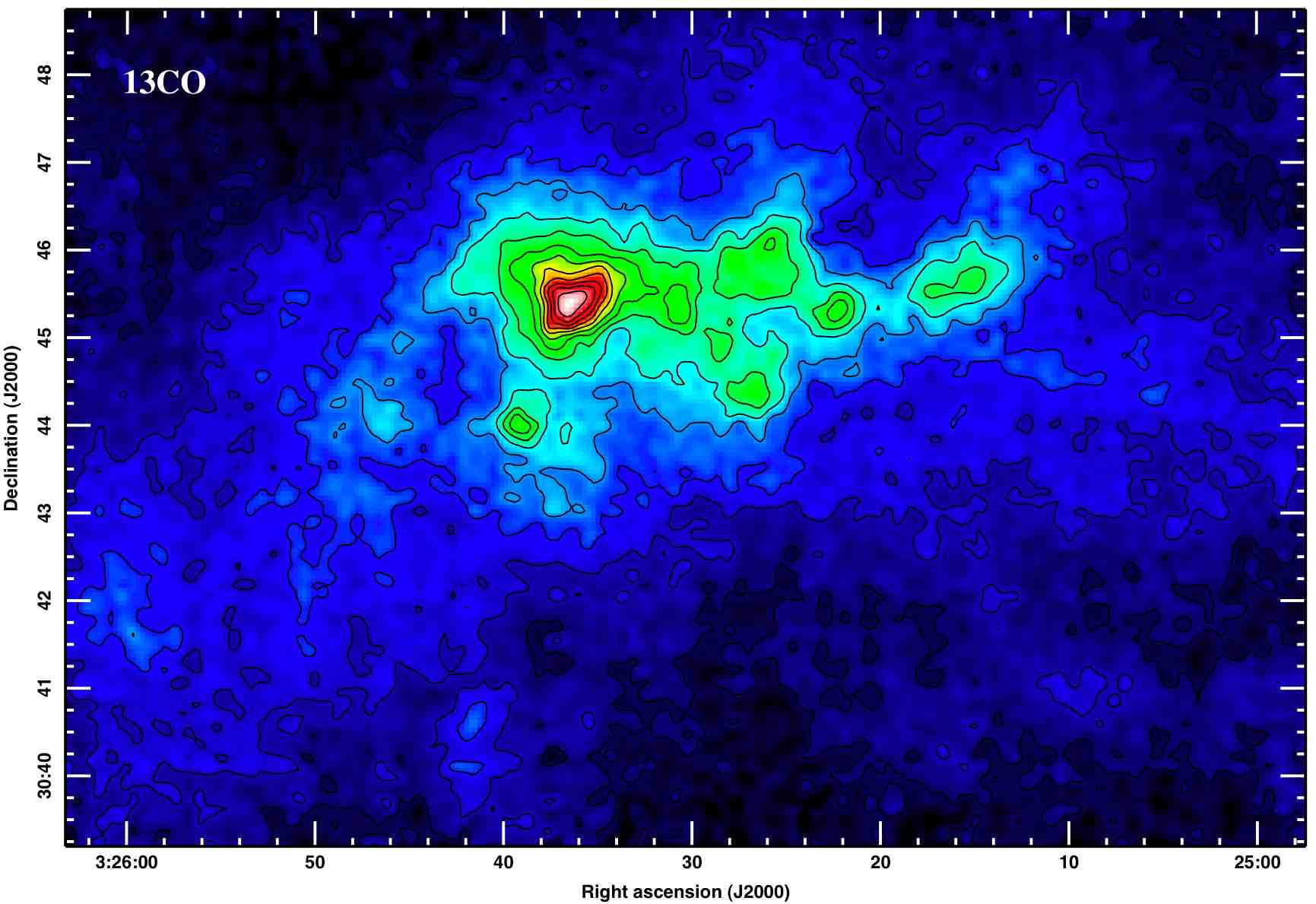}
\caption{Overview of the data towards L1448 as for IC348 in
  Fig. \ref{fig:ic348_data}. Top: SCUBA 850\,\micron\ emission. Prominent sources in the field are labelled: L1448NW, L1448:A/B
\citep{curiel90,curiel99}, L1448C \citep{bachiller90,bachiller95} and L1448-IRS2 \citep{wolf-chase00}.}
\label{fig:l1448_data}
\end{figure}

\begin{figure}
\hbox{\hspace{0.35cm}\includegraphics[width=8.05cm]{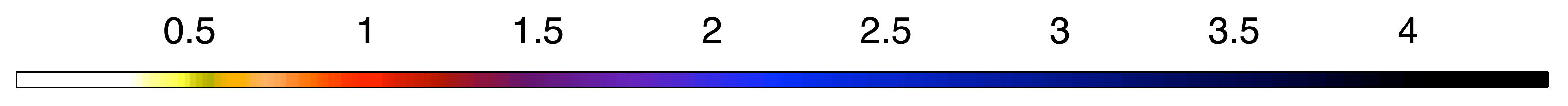}}
\includegraphics[width=0.47\textwidth]{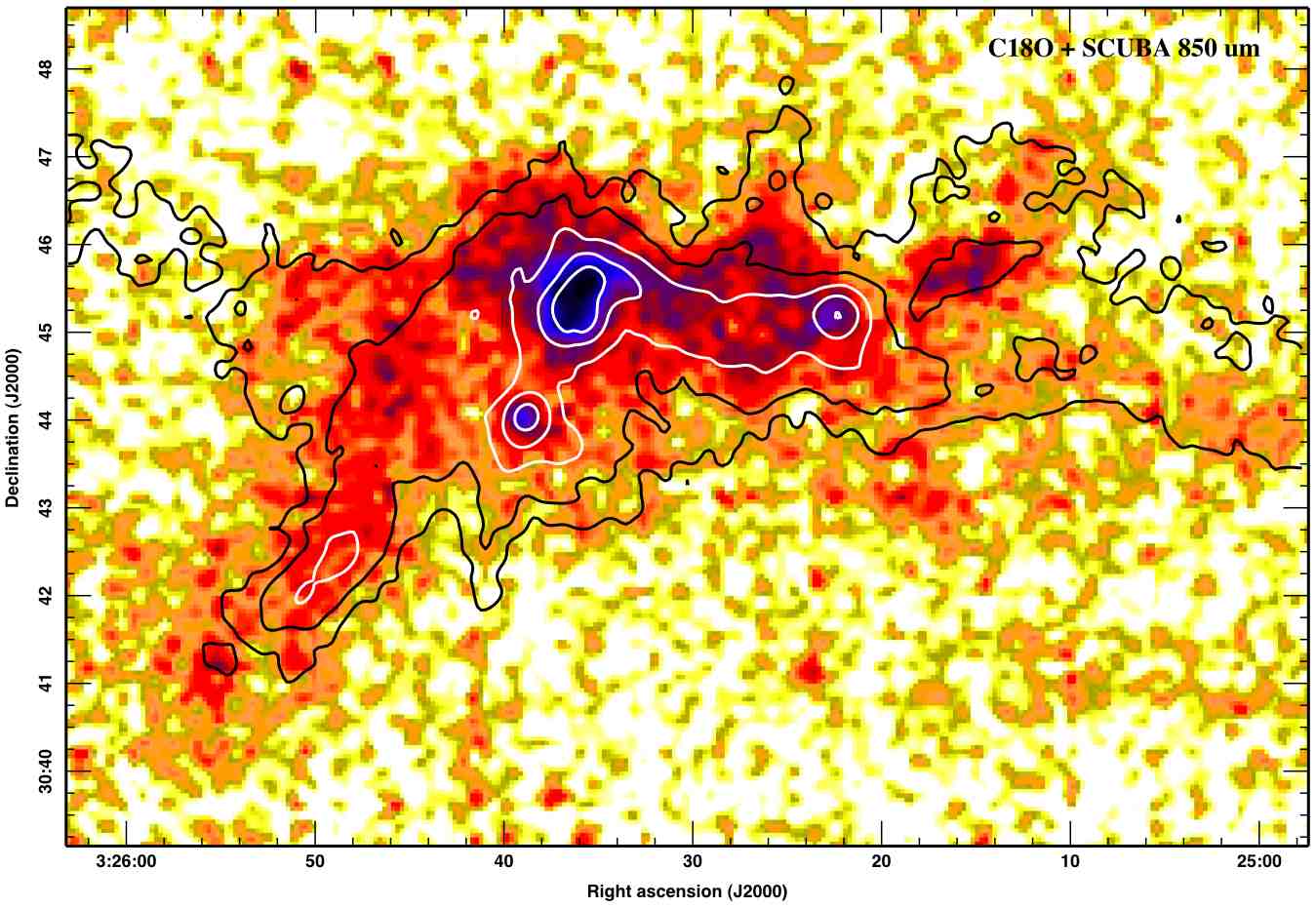}
\hbox{\hspace{0.4cm}\includegraphics[width=7.9cm]{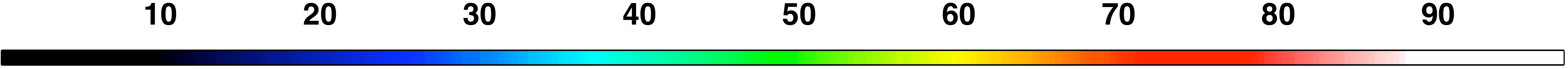}}
\includegraphics[width=0.47\textwidth]{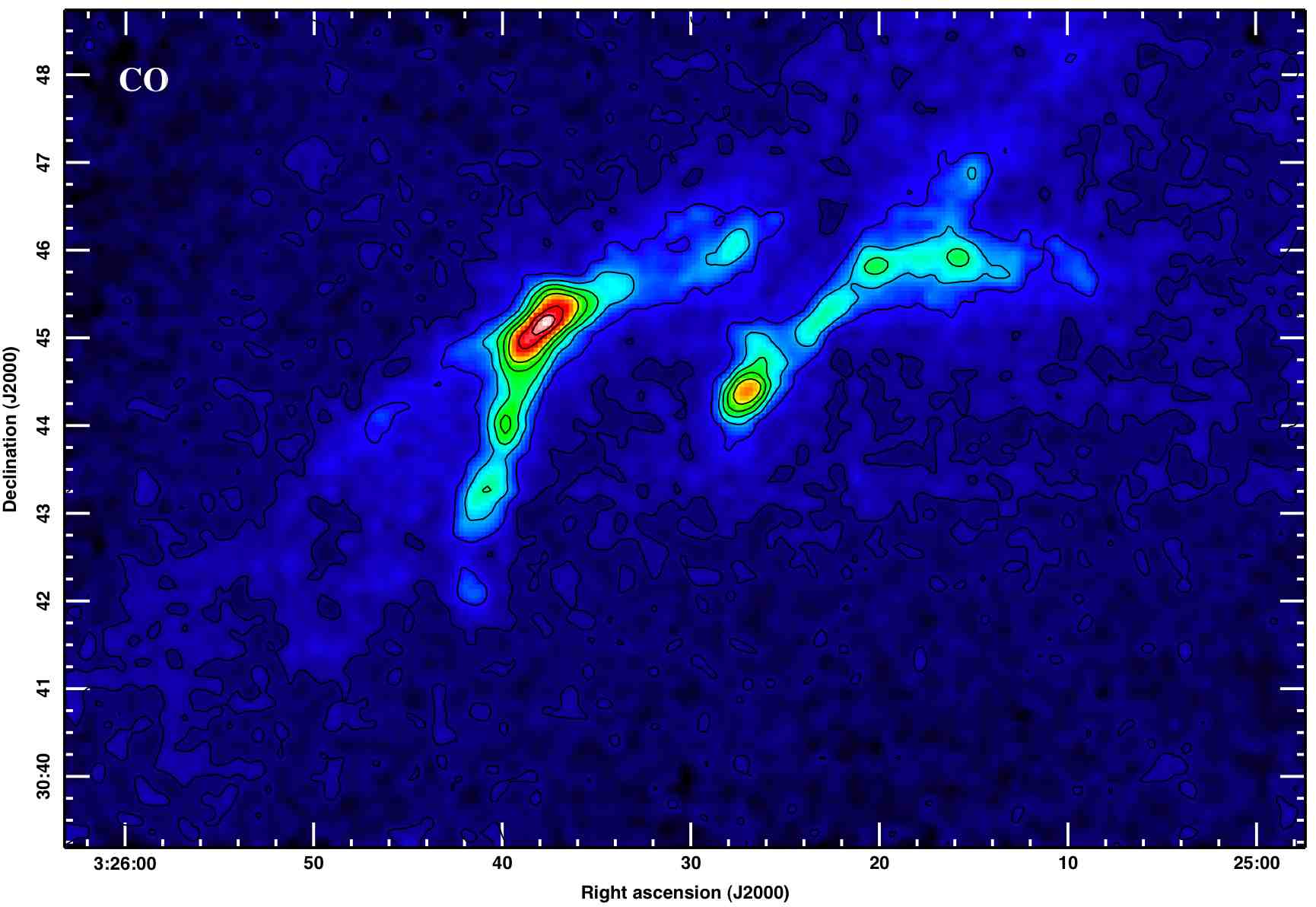}
\vspace{0.2cm}

\hbox{\hspace{0.4cm}\includegraphics[width=7.9cm]{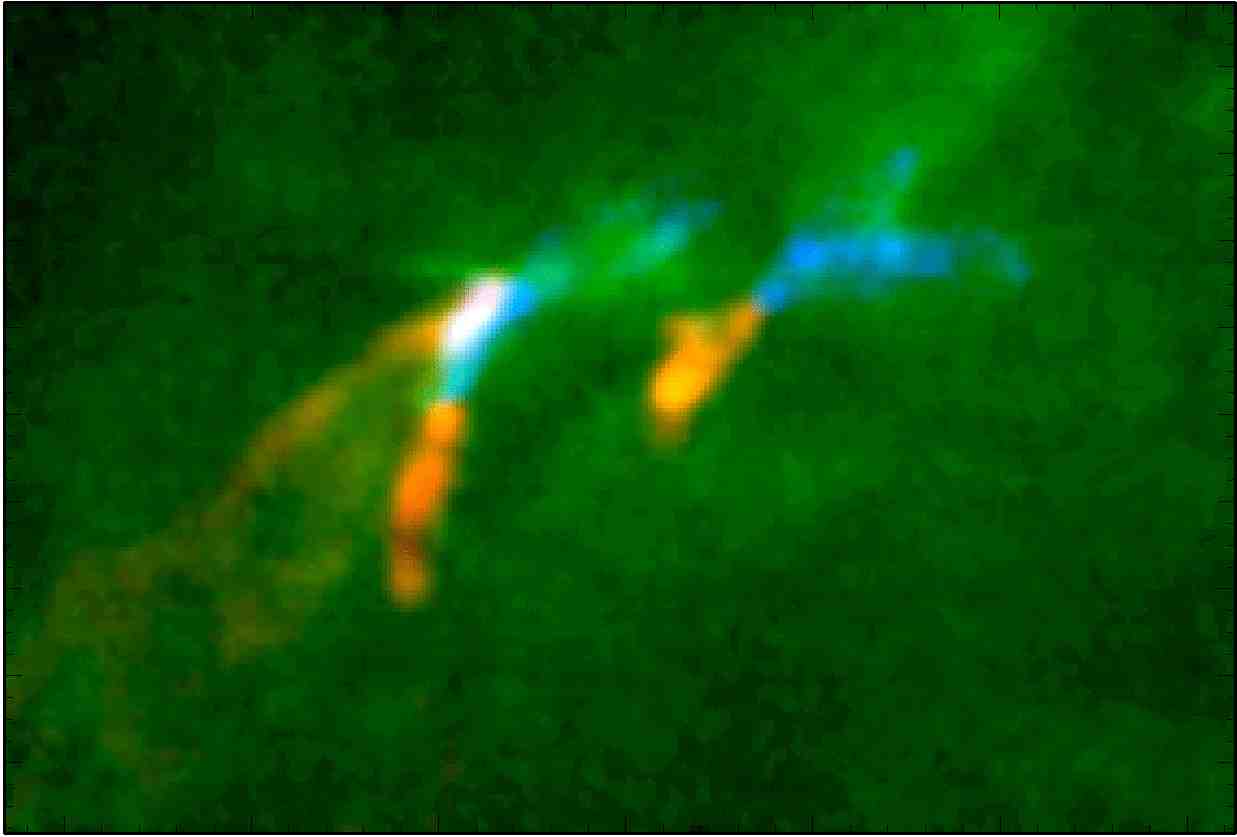}}
\contcaption{Top: Contours of
SCUBA 850\,\micron\ emission (levels as previously)
overlaid on the \ceighteeno\ \threetotwo\ integrated intensity as
before. Middle: Integrated \twelveco\ intensity from $-25$ to
  25\,\kms\ with contours from 10 to 90\,\kkms\ in steps of
  10\,\kkms. Bottom: Ranges: from $-25$
to 0\,\kms\ (blue), 2 to 6\,\kms\ (green) and 7 to
14\,\kms\ (red).}
\end{figure}

The HARP \ceighteeno\ data seems to reflect straightforwardly the
SCUBA emission in Fig. \ref{fig:l1448_data}, with the brightest cores at 850\,\micron\ also
having the greatest \ceighteeno\ integrated intensity. Most of the
SCUBA cores also appear as clumps of \ceighteeno\ emission. The same
structure is clear in \thirteenco\ as well, even though there is some
hint that the emission is lined up along the main southeast-northwest
outflow axis. The outflows themselves dominate the \twelveco\ data with
perhaps three to four flows overlapping. L1448C (source 29) drives a
highly collimated outflow discovered by \citet{bachiller90}, which was
one of the highest-velocity and youngest found at the time
\citep{bachiller90,bachiller95}. Its northern, blue-shifted lobe
intersects an outflow from a cluster of Class 0 objects: L1448N:A/B
and L1448NW. Finally, west of this cluster there are perhaps two
outflows emanating from the vicinity of L1448-IRS2 (sources 30 and 31).  

\subsection{L1455}

Our final field is towards another Lynds cloud, L1455 (Fig.
\ref{fig:l1455_data}) in the south-west of the complex with a
mass of some 40--50\,M$_\odot$ \citep{bachiller86a}. Although the
smallest and faintest of our targets, it has some of the most
interesting outflow structure. The \ceighteeno\ emission is weak and
only detectable towards the handful of protostellar and starless cores
in the southeast of the field. The \thirteenco\ data also mainly pick
out the compact sources with associated SCUBA emission. One of the brightest \thirteenco\ clumps (also detected in
\ceighteeno\ at 3$^\rmn{h}$27$^\rmn{m}$33$^\rmn{s}$,
30$^\circ$12$'$45$''$) has no compact SCUBA emission and appears
coincident with collimated blue-shifted emission from RNO15-FIR
(L1455-FIR, source 35), a
Class I source associated with the red reflection nebula of the same
name. The \twelveco\ maps are more intriguing, with a
prominent northwest-southeast CO outflow
\citep{goldsmith84,levreault88} some distance from the cluster of
protostars. Either this is driven by an unknown low-luminosity source
or one of the known protostars. The latter explanation is promoted by
\citet{davis08} although the structure of the outflow might suggest
the former is more likely.  

\begin{figure}
\hbox{\hspace{0.4cm}\includegraphics[width=7.9cm]{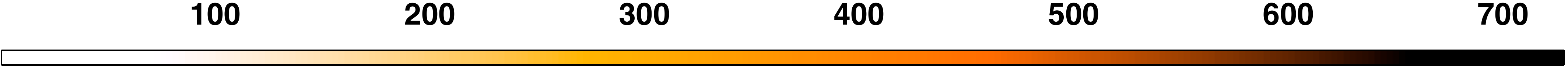}}
\includegraphics[width=0.47\textwidth]{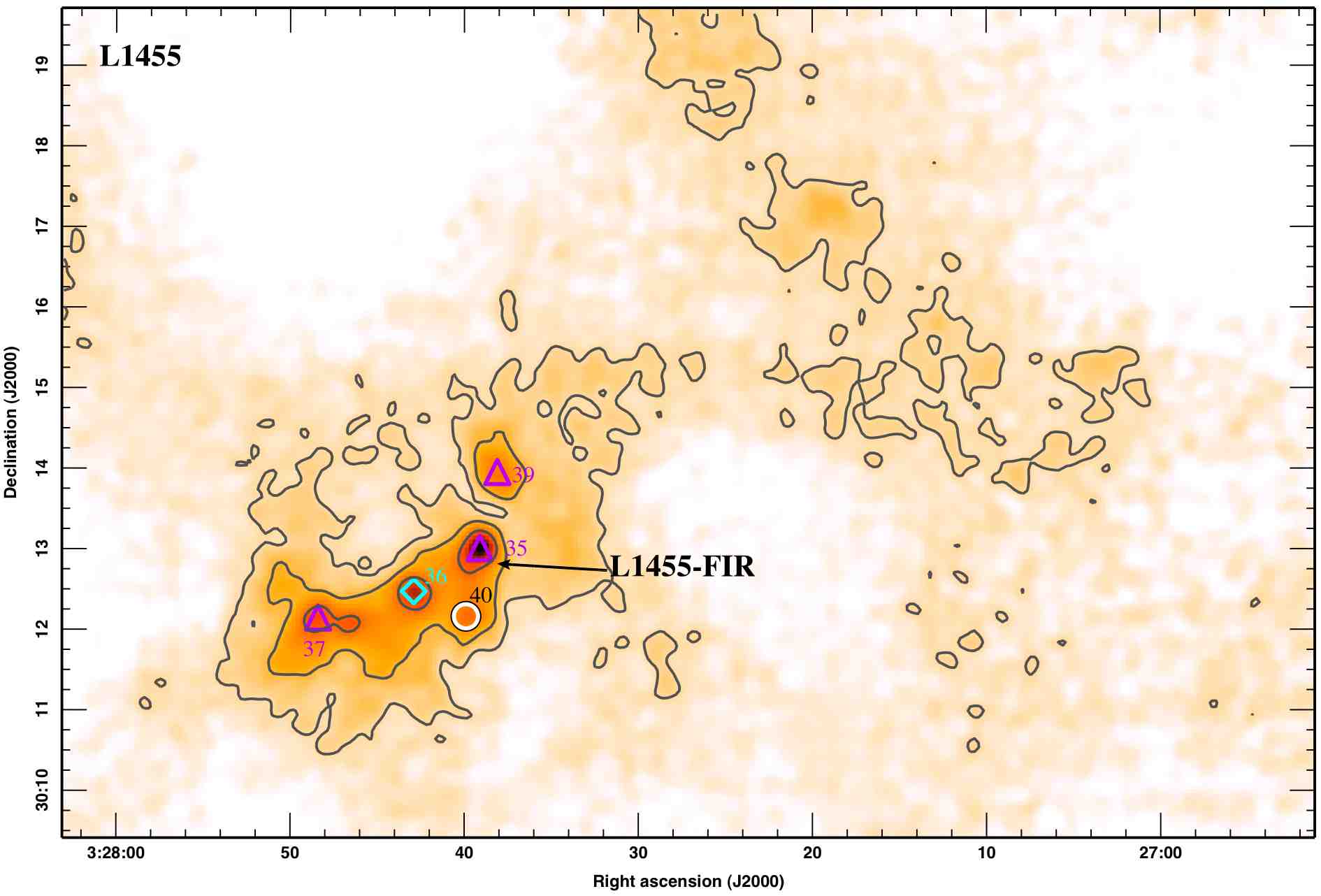}
\vspace{0.2cm}

\hbox{\hspace{0.4cm}\includegraphics[width=8cm]{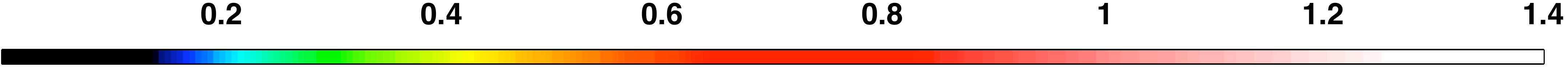}}
\includegraphics[width=0.47\textwidth]{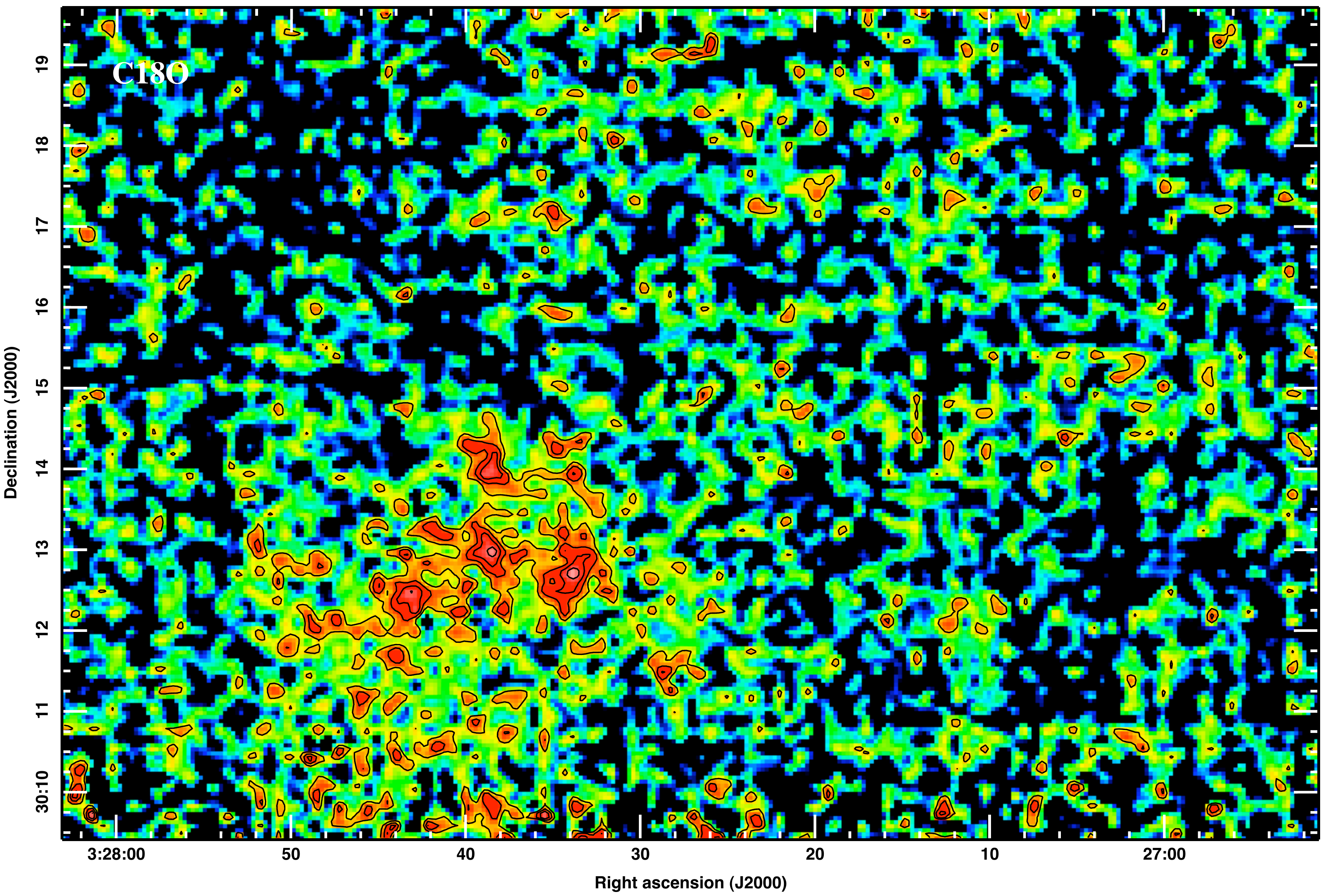}
\vspace{0.2cm}

\hbox{\hspace{0.4cm}\includegraphics[width=7.9cm]{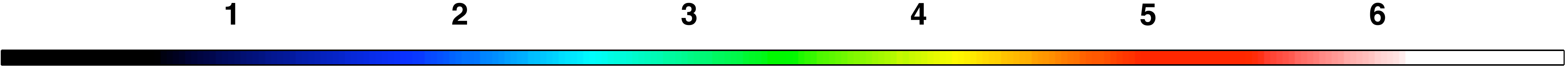}}
\includegraphics[width=0.47\textwidth]{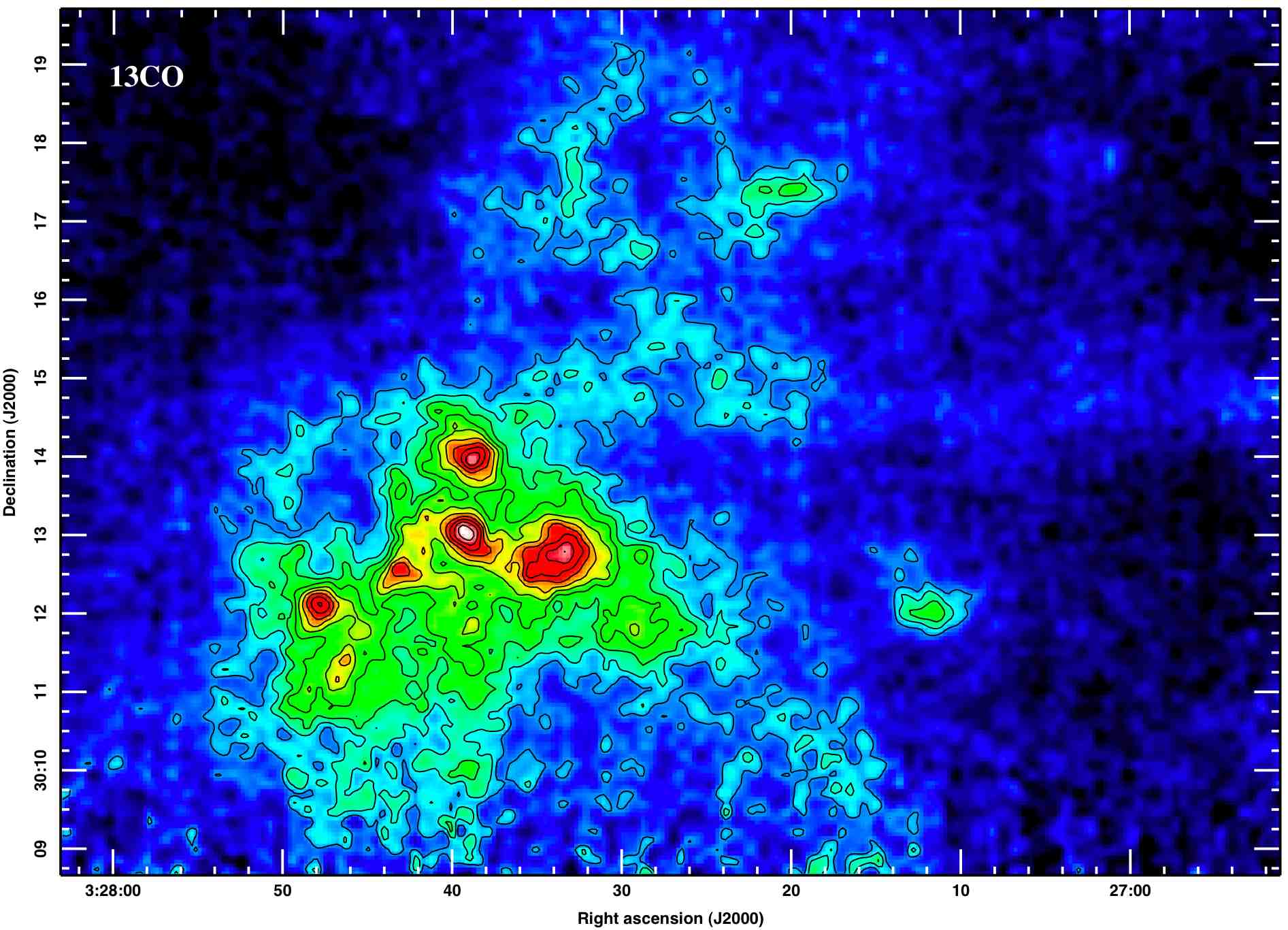}

\caption{Overview of the data towards L1455 as for IC348 in Fig. \ref{fig:ic348_data}. Top: SCUBA
  850\,\micron\ emission. Additionally
labelled is L1455-FIR \citep{davis97}.}
\label{fig:l1455_data}
\end{figure}

\begin{figure}
\hbox{\hspace{0.39cm}\includegraphics[width=7.9cm]{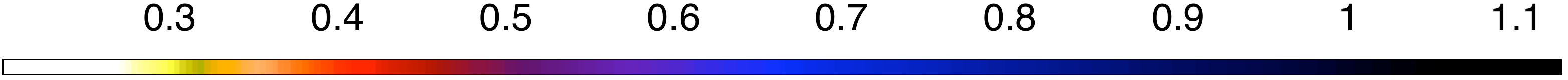}}
\includegraphics[width=0.47\textwidth]{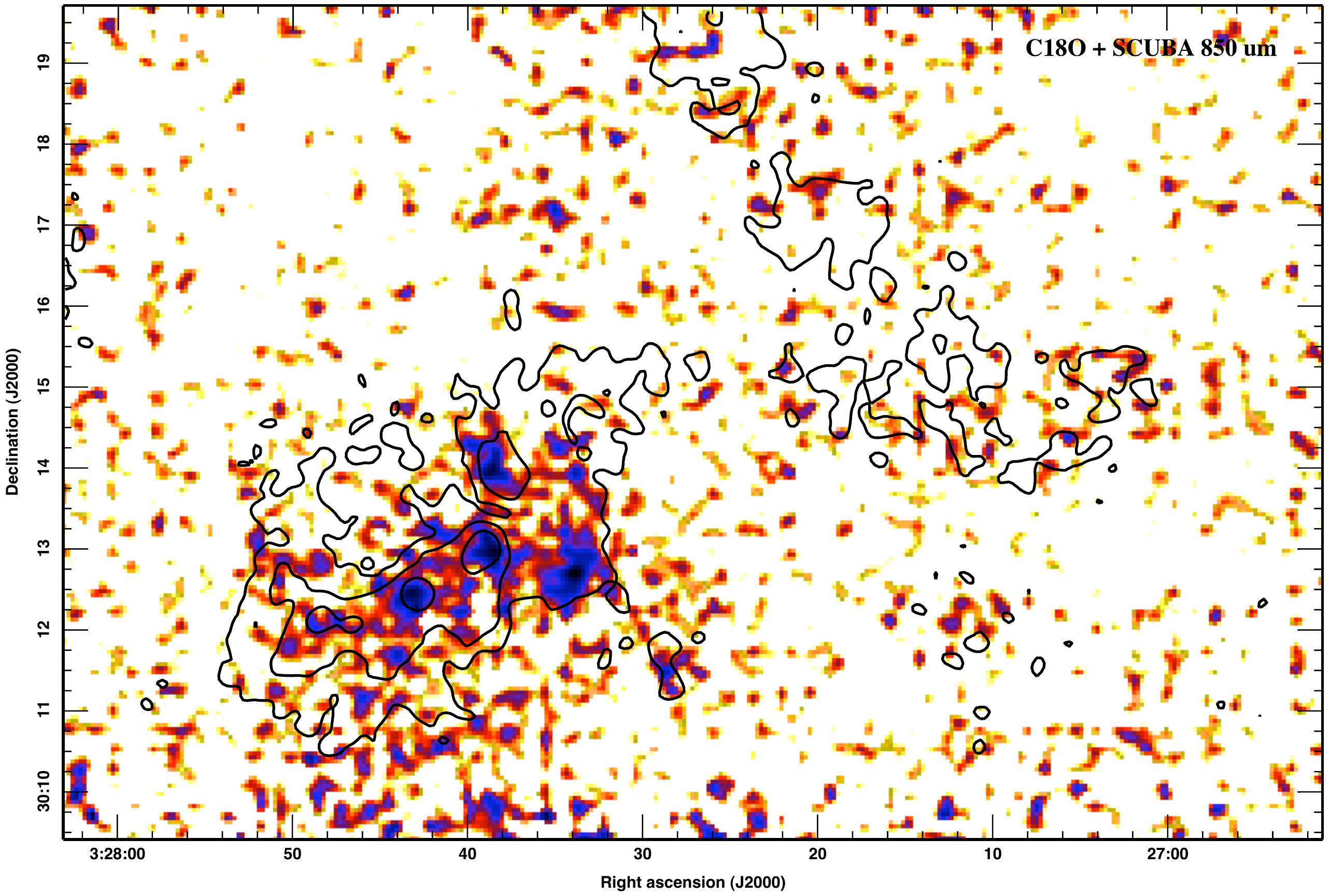}
\hbox{\hspace{0.4cm}\includegraphics[width=7.9cm]{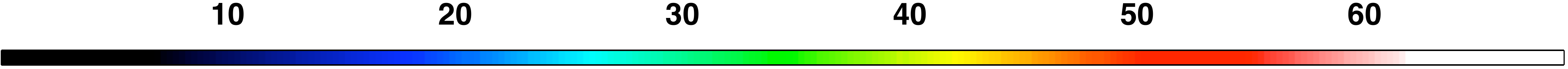}}
\includegraphics[width=0.47\textwidth]{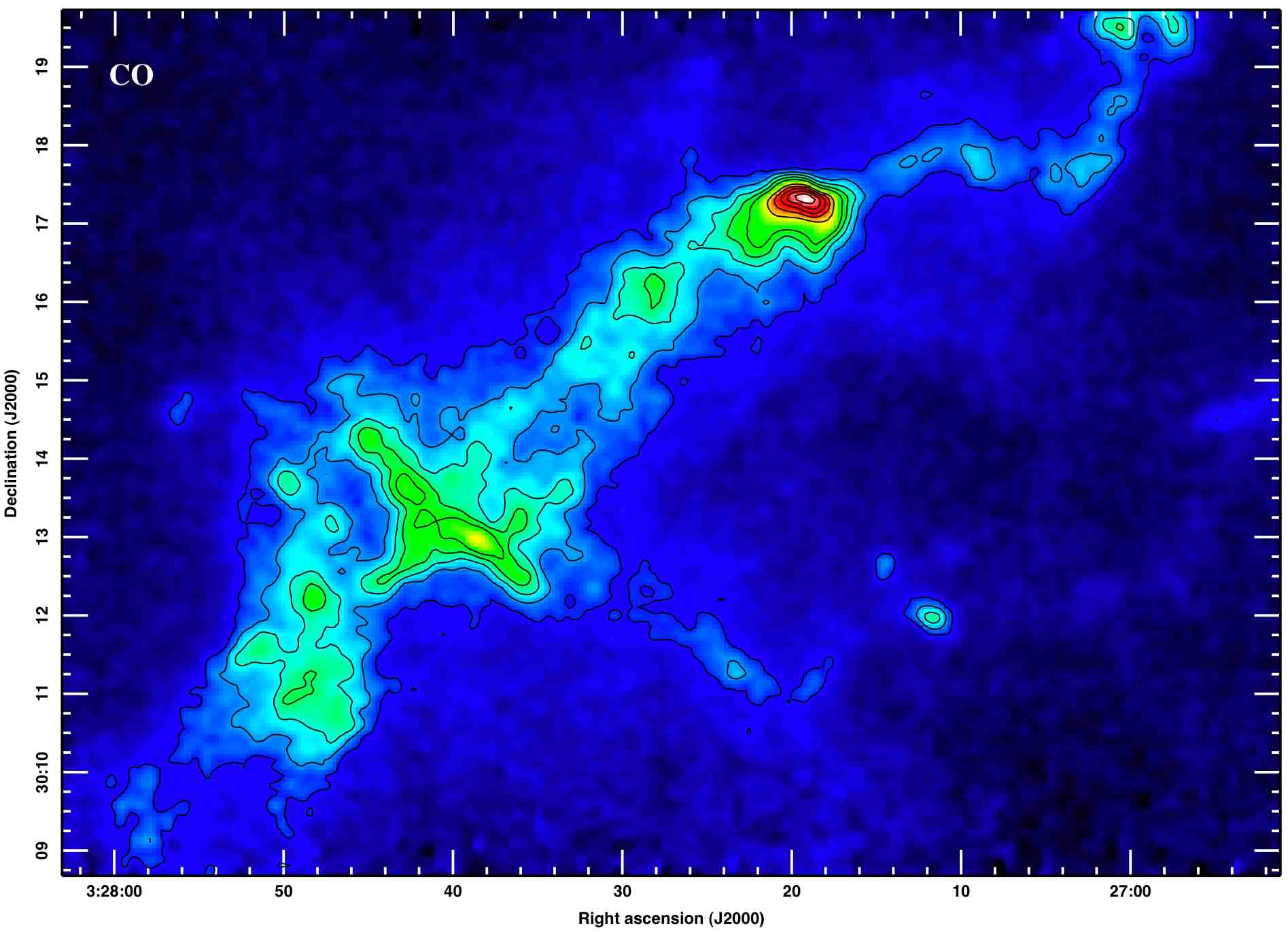}
\vspace{0.2cm}

\hbox{\hspace{0.4cm}\includegraphics[width=7.9cm]{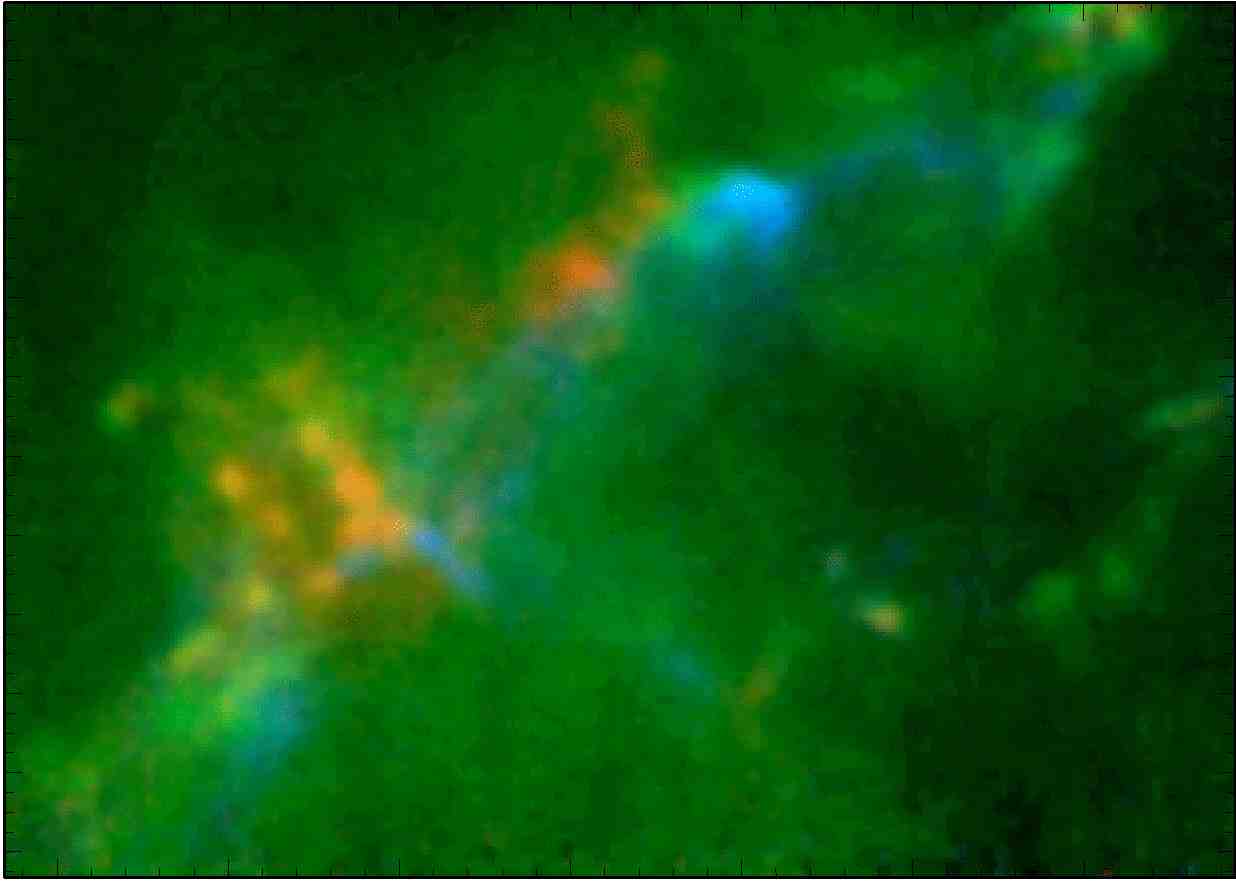}}
\contcaption{Top: Contours of
SCUBA 850\,\micron\ emission (levels as previously)
overlaid on the \ceighteeno\ \threetotwo\ integrated intensity as
before. Middle: Integrated \twelveco\ intensity from $-5$ to
  15\,\kms\ with contours from 15 to 65\,\kkms\ in steps of
  5\,\kkms. Bottom: Ranges: from $-5.0$
to 3.5\,\kms\ (blue), 5.0 to 6.5\,\kms\ (green) and 8.0 to
14.0\,\kms\ (red).}
\end{figure}

\section{Discussion} \label{sec:discussion}

\subsection{Average region properties} \label{sec:averageregionproperties}

In Table \ref{table:region_masses}, we list the masses of the regions derived
from the CO isotopologues and SCUBA 850\,\micron\ data. For the
continuum data, we follow \citet*{visser01} in calculating the region
masses, assuming a constant dust temperature and opacity of 12\,K and
$\kappa_{850}=0.012$\,cm$^2$g$^{-1}$ respectively. The gas masses
assume each spectral line is optically thin, in LTE and its emitting gas
has an excitation temperature, $T_\rmn{ex}=12$\,K. We use a standard method to derive the molecular
column densities (e.g.\ \citealp*{minchin93}), which for a CO isotopologue, X,
yields: \begin{equation} N(\rmn{X}) = 5\times 10^{12} ~ T_\rmn{ex}
  \exp\left( \frac{T_\rmn{trans}}{T_\rmn{ex}}\right) \int T_\rmn{mb}
  \rmn{d}v \quad \rmn{cm^{-2},} \end{equation} where $T_\rmn{trans}$
is 33.7, 31.8 and 31.6\,K for the \threetotwo\ transition of
\twelveco, \thirteenco\ and \ceighteeno\ respectively and the integral
is in \kkms. To
calculate total masses from the derived column densities we assume
molecular abundances of $10^{-7}$, $10^{-6}$ and $10^{-4}$ for
\ceighteeno, \thirteenco\ and \twelveco\ respectively relative to
neutral hydrogen. A main-beam efficiency of $\eta_\rmn{mb}=0.66$ was
used for all the isotopologues as measured during HARP's commissioning. In each map, only
positions with peak values greater than three times the average
noise, $\langle \sigma_\mathrm{RMS} \rangle$, have been included. 

The pattern of the different masses is intriguing and \emph{not} purely
what is expected given the different optical depths of the CO
isotopologues. NGC1333 clearly shows the naively-expected trend; as the
CO lines become increasingly optically thick
(\ceighteeno$\to$\thirteenco$\to$\twelveco), only the cloud surface layers
are probed, the total emission falls and the mass decreases. The SCUBA 850\,\micron\
emission, although optically thin almost everywhere, is less sensitive
than the CO observations to large-scale structure ($\ga 2$\,arcmin), due to the spatial chopping in the
observing technique. Therefore, it potentially misses much of the cloud mass. This simple prescription is not followed in
any of the other regions, except perhaps L1455. However, here the \ceighteeno\
emission is particularly weak so most of the spectra have peaks $<3
\sigma_\rmn{RMS}$ and are therefore not included in the mass estimates
 of Table \ref{table:region_masses}, which probably relate to three or
 four cores. This is a potential reason why
 the \ceighteeno\ mass is so much smaller than that from SCUBA. IC348 and L1448 make an interesting contrast: the
\ceighteeno\ and \thirteenco\ masses are not very different, with the \thirteenco\ mass marginally larger in both cases,
implying the \thirteenco\ gas is probably optically thin and more
widespread than the \ceighteeno. However, L1448 has a SCUBA mass some
3--4 times larger than these gas estimates in contrast to IC348 whose
SCUBA mass is 2--3 times smaller. \ceighteeno\ excitation effects
may explain the large \ceighteeno\ mass in IC348, if there is a
significant amount of subthermal excitation, the \ceighteeno\ emission
for a given quantity of gas would be boosted relative to LTE causing
an over-estimate of the \ceighteeno\ mass. In L1448, the small
\ceighteeno\ mass is unlikely to be caused by subthermal
excitation and probably arises from a different gas-to-dust ratio or
if we underestimate the dust temperature (which in any case will vary from our
constant assumption of 12\,K). In Section \ref{sec:12co_temp}, we show the
dust temperature in L1448 is always higher than the excitation
temperature of the gas. If we underestimate the dust temperature then
we overestimate the corresponding dust mass, which may account for
some of the discrepancy between the tracers.
 
\begin{table}
\caption{Region masses from the various HARP \threetotwo\ integrated
  intensities and SCUBA 850\,\micron\ emission. All spectral-line estimates
  assume LTE, $\tau \ll 1$ and $T_\mathrm{ex}=12$\,K, while the SCUBA
  estimate assumes a constant dust temperature of 12\,K.  Only points with peaks $>3\langle \sigma_\mathrm{RMS} \rangle$ have been included.}
\begin{tabular}{lcccc}
\hline
Region & \multicolumn{4}{c}{Masses (\msun)} \\
 & 850\,\micron\ & \ceighteeno\ & \thirteenco\ & \twelveco\ \\
\hline
NGC1333 & 250 & 412 & 301 & 13 \\
IC348 & 71 & 160 & 212 & 35\\
L1448 & 153 & 38 & 45 & 9 \\
L1455 & 45 & 7 & 35 & 12\\
\hline
\end{tabular}
\label{table:region_masses}
\end{table}

\begin{figure}
\begin{center}
\includegraphics[angle=270,width=0.235\textwidth]{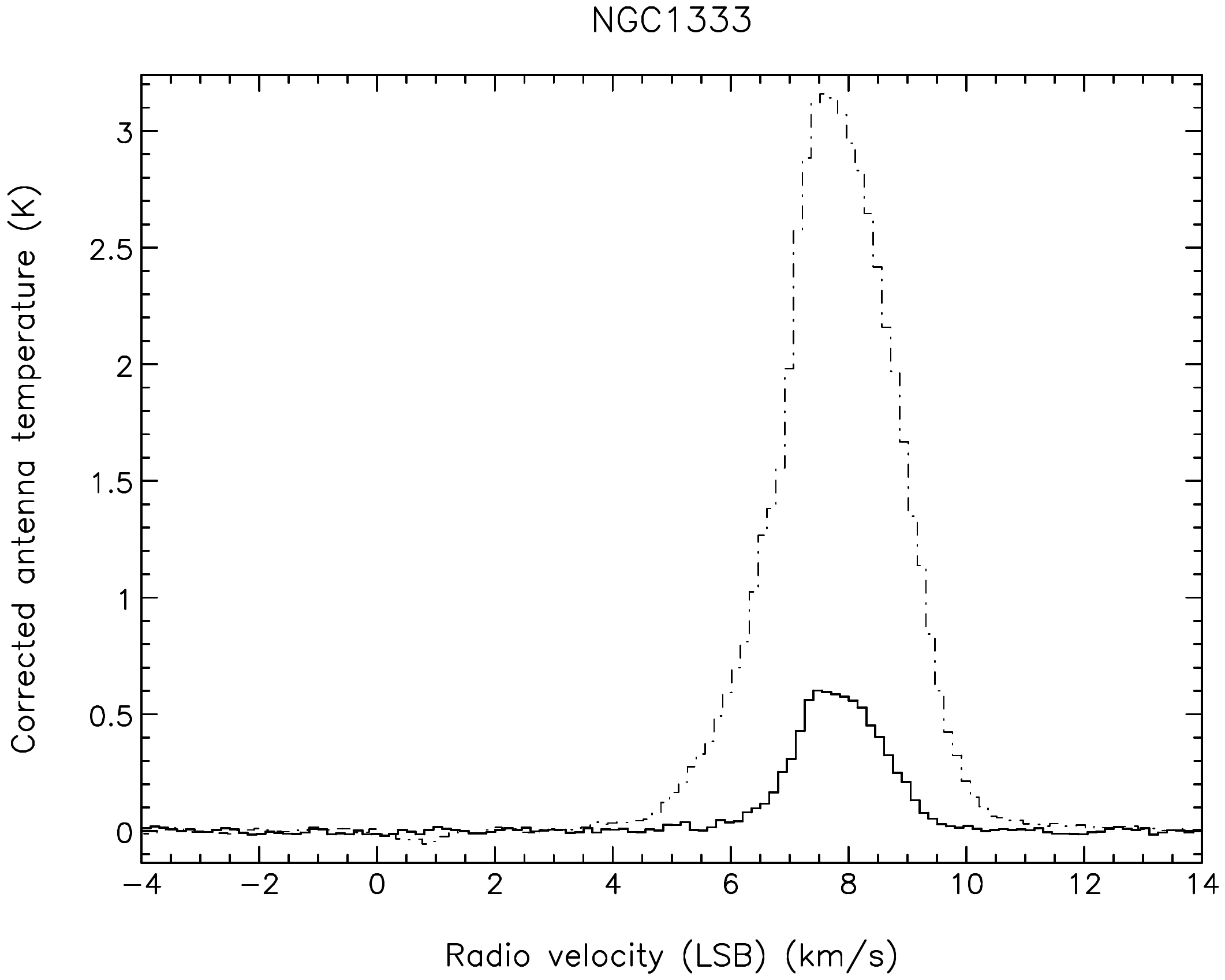}
\includegraphics[angle=270,width=0.235\textwidth]{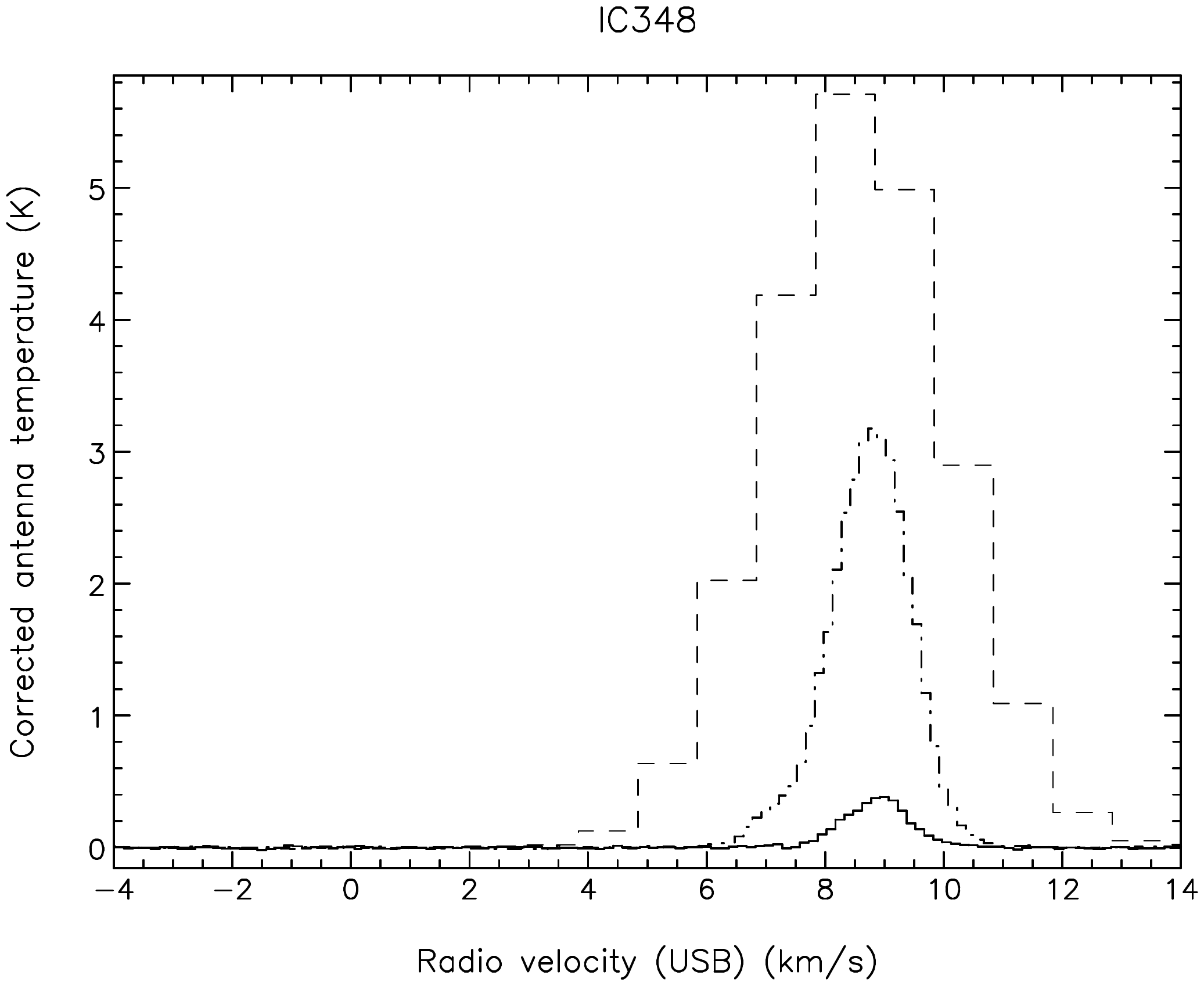}
\includegraphics[angle=270,width=0.235\textwidth]{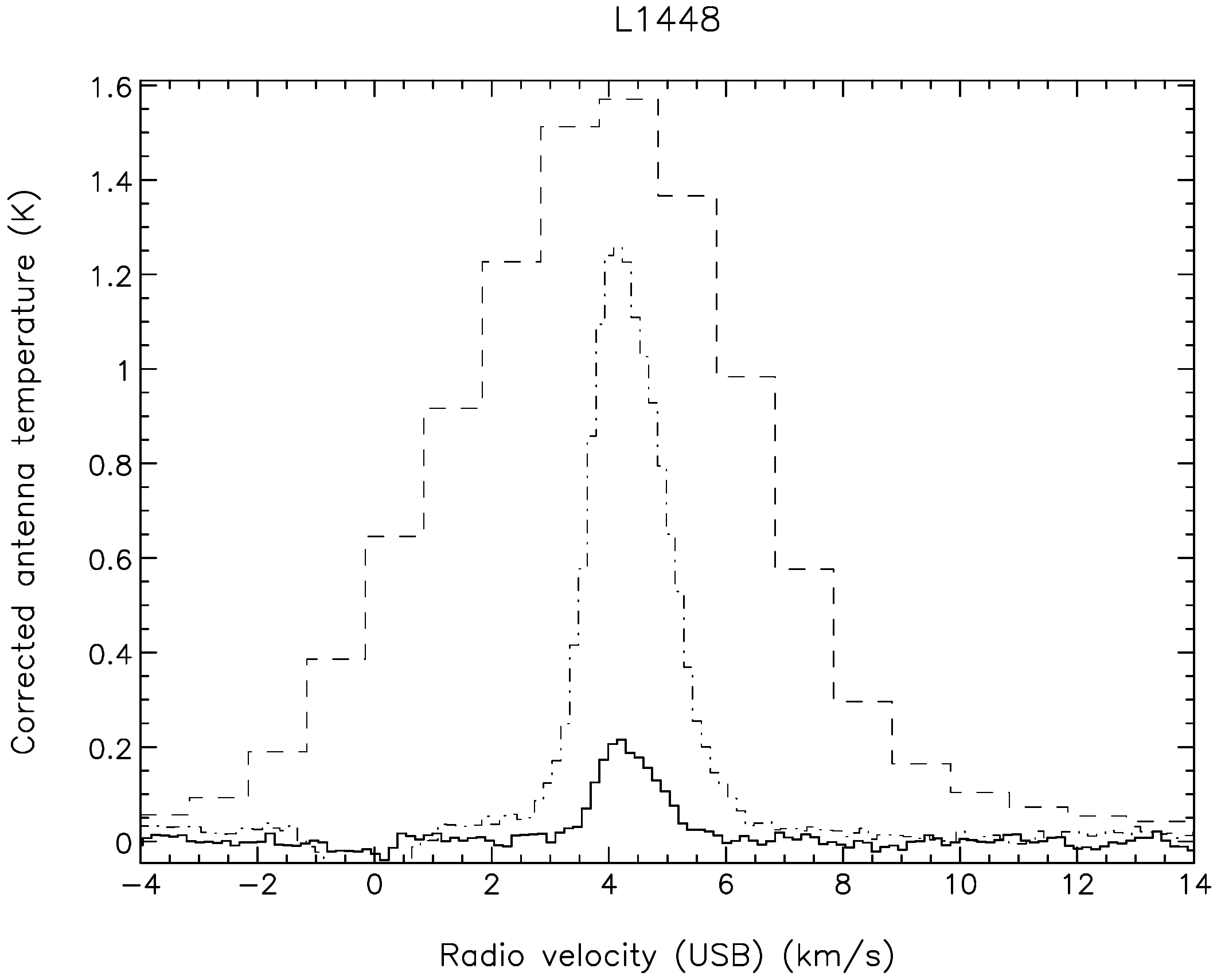}
\includegraphics[angle=270,width=0.235\textwidth]{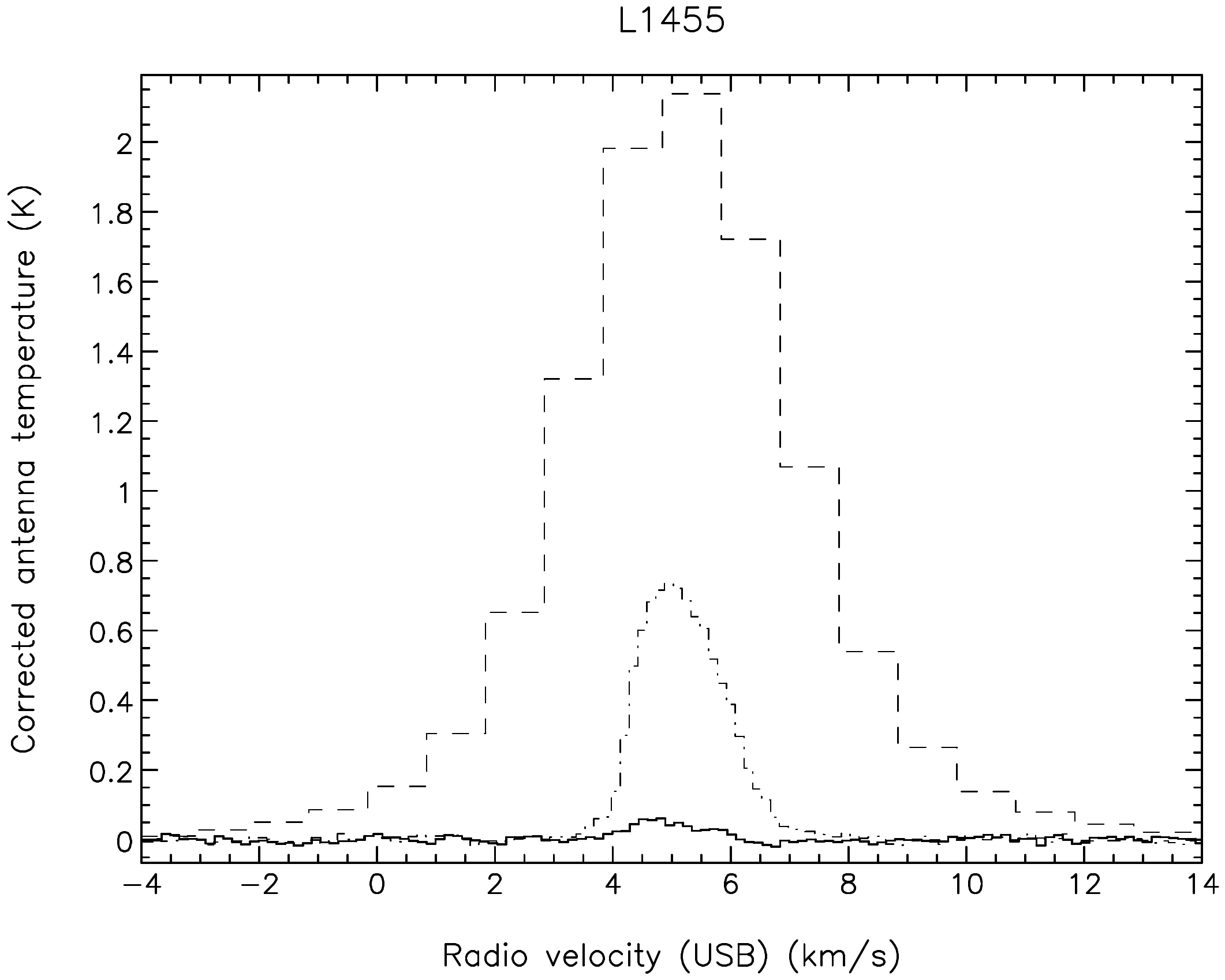}
\caption{Average
  cloud spectra for the different regions. Each plot contains the averages of every spectrum in the HARP \threetotwo\ cubes: \twelveco\ (dashed), \thirteenco\
  (dot-dashed) and \ceighteeno\ (solid).}
\label{fig:averagespectra}
\end{center}
\end{figure}

Average spectra in the different regions are plotted in
Fig. \ref{fig:averagespectra}. All three isotopologues are centred at similar velocities, approximately
7.8, 8.8, 4.3 and 5.0\,\kms\ for NGC1333, IC348, L1448 and L1455
respectively. The strength of the lines are in order of their
abundance as we would expect and these averages show no \emph{global}
evidence for multiple components or self-absorption. However, individual \twelveco\
spectra \emph{do} show self-absorption and/or multiple components. In
L1448, the \thirteenco\ line is nearly the same strength as the
\twelveco, implying the \twelveco\ line is either saturated (and therefore the
gas has a low physical temperature) or there are significant optical
depths in the \twelveco\ gas due to density or temperature
effects. Finally, even an average \twelveco\ spectrum contains
significant high-velocity linewings and there are weaker linewings
clear in the \thirteenco\ spectra towards L1448 and IC348.  

\begin{figure*}
\begin{minipage}{0.49\textwidth}
\includegraphics[angle=270,width=\textwidth]{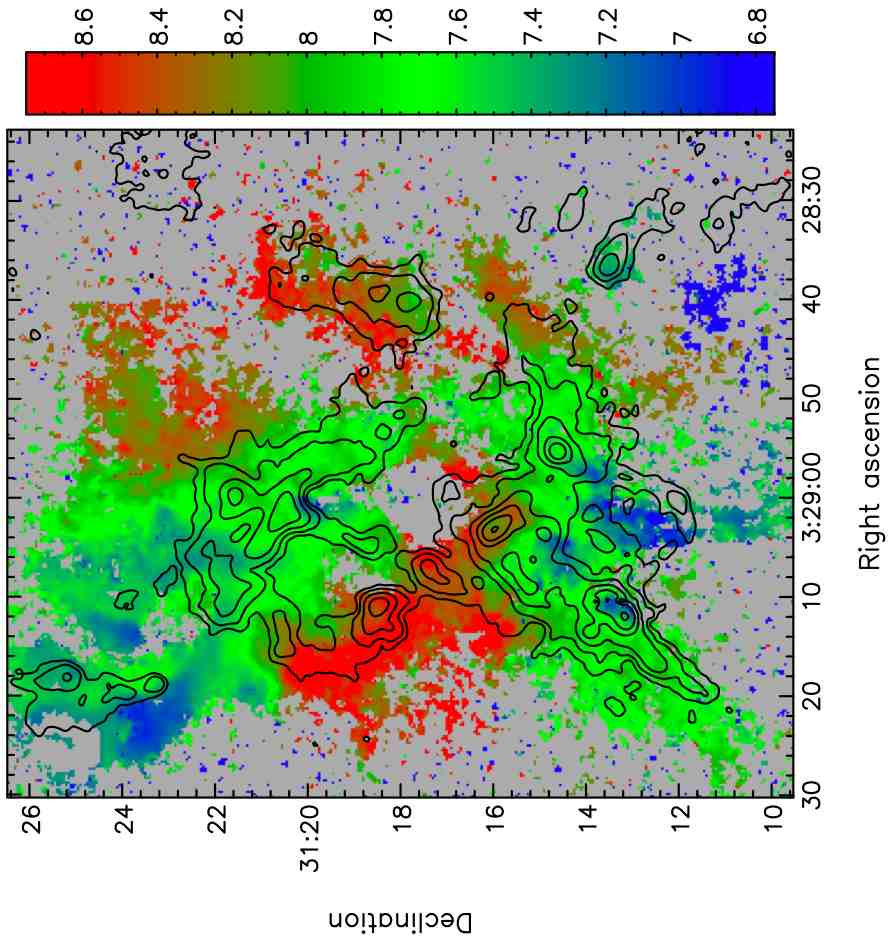}
\end{minipage}\\
\vspace{0.2cm}

\begin{minipage}{0.49\textwidth}
\includegraphics[angle=270,width=\textwidth]{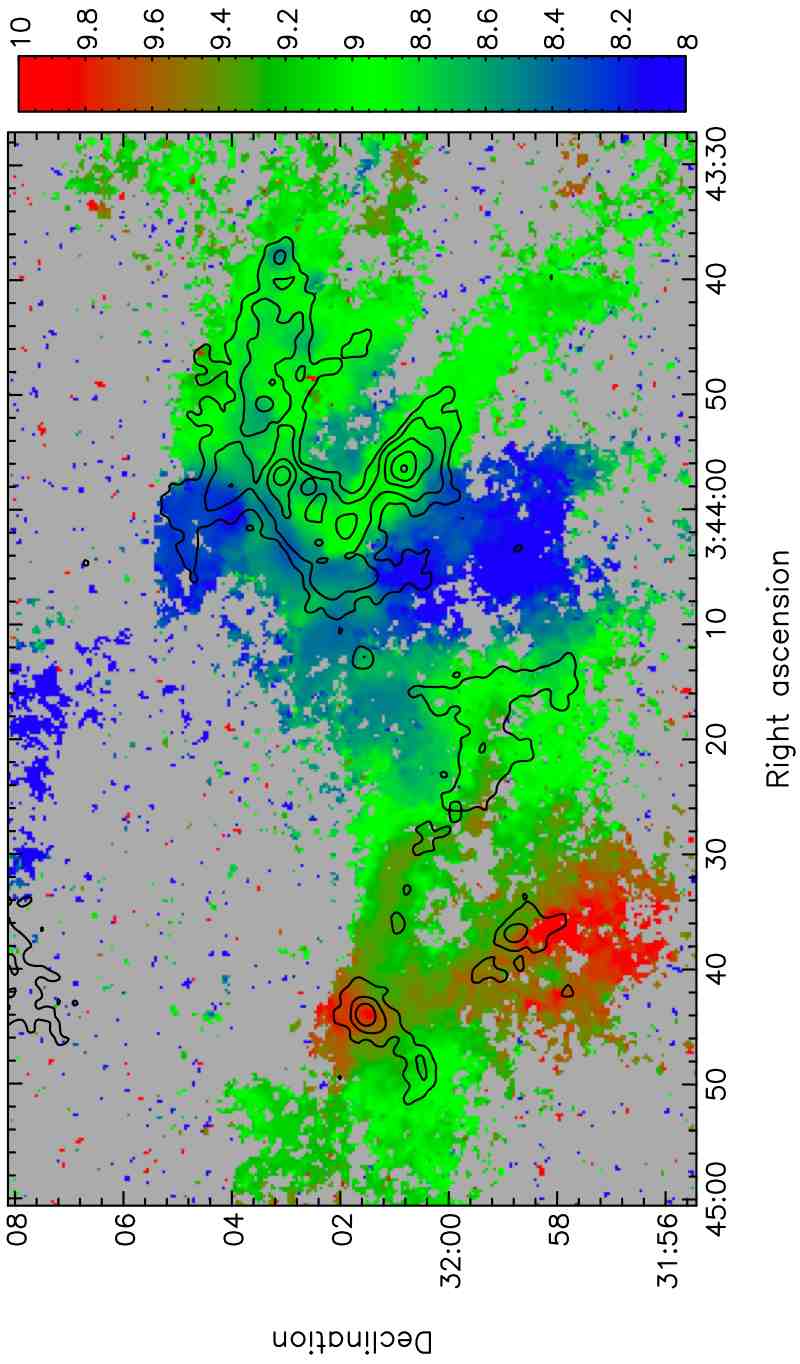}
\end{minipage}
\begin{minipage}{0.49\textwidth}
\includegraphics[angle=270,width=\textwidth]{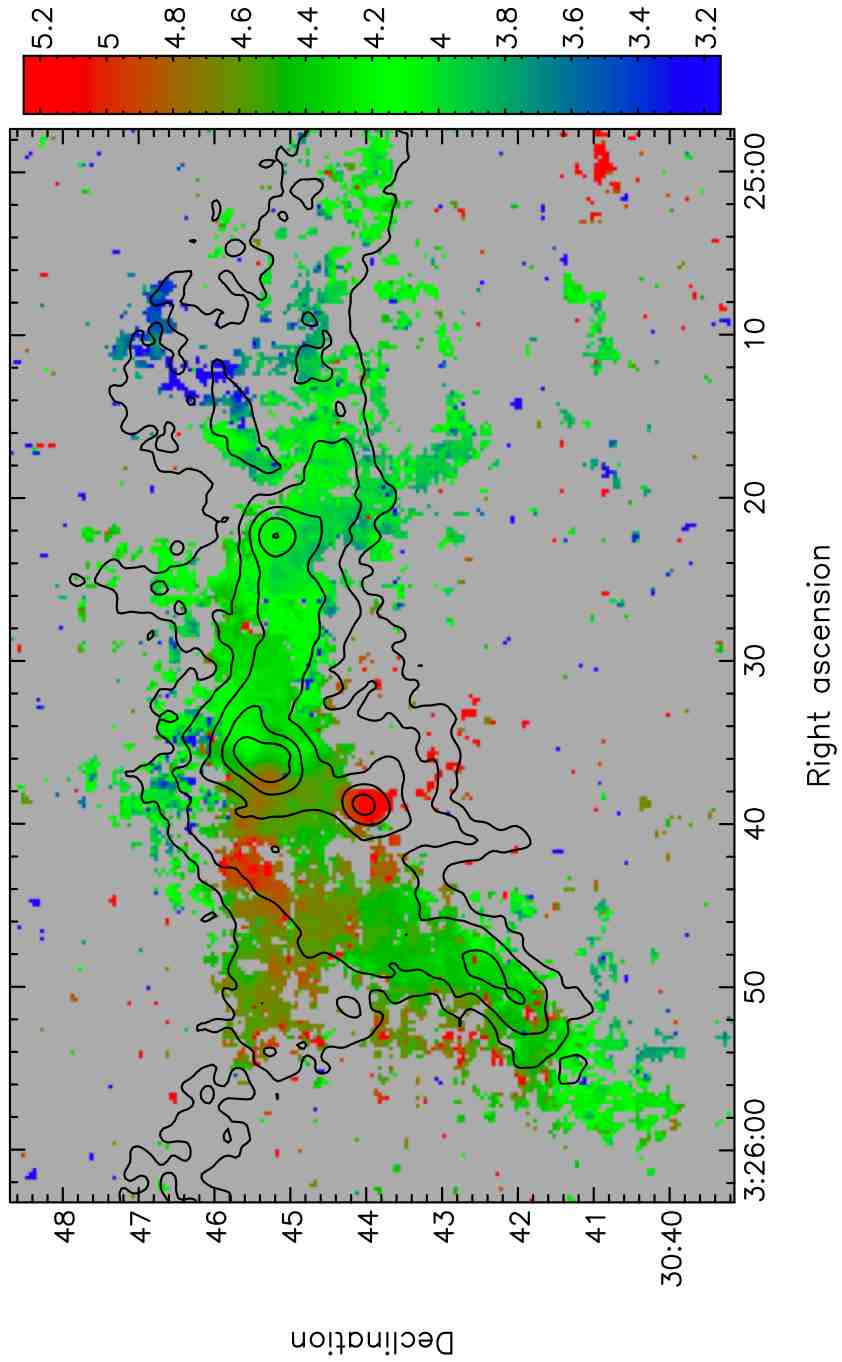}
\end{minipage}
\caption{Motions along the line of sight in our target fields, NGC1333
  (top), IC348 (bottom left) and L1448 (bottom right). At each
  point where the \ceighteeno\ \threetotwo\ line has a peak greater
  than \ceighteeno\ $>3\sigma_\mathrm{RMS}$ the line centre velocity
  (in \kms) is plotted. The line centres have been computed by fitting
  a single Gaussian to the spectrum at each point. L1455 did not have
  strong enough \ceighteeno\ emission for such fitting.}
\label{fig:c18o_centres}
\end{figure*}

\ceighteeno\ is an ideal tracer of the bulk
motions of the molecular gas, given that it is nearly optically thin
throughout and detectable over larger areas than denser-gas tracers.  
We plot \ceighteeno\ line centres across the regions in Fig. \ref{fig:c18o_centres}. The centres are from
non-linear least squares fitting of Gaussian functions to every spectrum with a
peak $>3\sigma_\mathrm{RMS}$ (the noise calculated on each individual
spectrum), performed using software from S. Graves. There are many plausible explanations for variations in
velocity across a cloud region: 
\begin{enumerate}
\item \emph{Rotation}. A systematic shift in the line velocity
  along a particular direction is often suggested to be a signature of
  bulk rotation. For instance the Serpens cloud core is
thought to be undergoing a global east to west rotation
(e.g.\ \citealp{olmi02}) which manifests itself in a very ordered
change in the \ceighteeno\ \threetotwo\ velocity observed with HARP by
the GBS (Graves et al., in preparation). In the Perseus regions there
  are few such global gradients. \citet{ho80} inferred a south to north
  rotation of NGC1333 from \ammonia\ observations, although
  \citet{walsh07} (and this work) find no such \emph{global} gradient
  in \ntwohplus, suggesting \citeauthor{ho80} were biased by the
red-shifted gas around IRAS7 (in the centre-left of our map).
\item \emph{Outflows}. On large scales in our fields, CO outflows do
  not seem to correlate with the \ceighteeno\ centres (although there
  is some link on the scale of individual cores; Curtis \& Richer, in preparation). 
\item \emph{Constituent clumps moving with respect to the
  bulk cloud}. Example SCUBA cores seem to enclose regions
of very different velocity, which implies some of the variations trace
clumps of gas that move at distinct velocities with respect to the ambient
cloud.
\item \emph{Gas Flows}. In a similar fashion to (iii) the
  variations in velocity could reflect larger flows of gas. In
  gravoturbulent models we expect cores to form at stagnation
  points, where convergent streams of material collide (e.g.\ \citealp{padoan01b}). In these data,
  there are large areas of coherent motion but the cores do not seem
  to occur solely where they intersect. 
\end{enumerate}

\subsection{$^{\boldmath{13}}$CO/C$^{\boldmath{18}}$O ratio} \label{sec:lineratio}

The optical depth of a spectral line provides important evidence as to
where in the cloud it has been emitted. Optically-thin molecular lines
can provide a trustworthy measure of the column density and conditions
throughout the cloud, whereas
optically-thick lines saturate, only tracing the outer layers. From
the peak \thirteenco/\ceighteeno\ ratio, we can get a measure of the
optical depth of the \ceighteeno\ \threetotwo\ line in the densest
region along the line of sight and thereby test its
fidelity as a total mass tracer.

Using the radiative transfer equation
  for an isothermal slab (see e.g.\ \citealp{rohlfswilson}), we can relate the \thirteenco/\ceighteeno\ intensity ratio at
  the velocity of the peak of the \ceighteeno\ line, $R$, to
  the optical depths of both species ($\tau_{13}$ and $\tau_{18}$ respectively), assuming both transitions are at the same
  frequency, emanating from the same volume of material and have the
  same $\eta_\mathrm{mb}$ and beam-filling factors:
  \begin{equation} R = \frac{T_\mathrm{A}^*(\mathrm{^{13}CO})}{T_\mathrm{A}^*(\mathrm{C^{18}O})} = \frac{1-\exp(-\tau_{13})}{1-\exp(-\tau_{18})}.\end{equation}
The opacities are expected to be linked through the abundance of
  \thirteenco\ relative to \ceighteeno, which we denote $\zeta$, as
  $\tau_{13}=\zeta\tau_{18}$
  (e.g.\ \citealp*{myerslinkebenson83}). Where the lines (particularly
  of \thirteenco) are optically
  thin, $\tau \to 0$, and $R\to \zeta$
  asymptotically. The absolute values of the \thirteenco\ and
  \ceighteeno\ abundances are not known to high precision and may
  differ across regions or between clouds, due to statistical
  variations across the Galaxy. For
  instance in photon dominated regions, $R$ can be very high where there is
  little self-shielding of the cloud from incident radiation, 
  causing the rarer isotopologue to be almost
  completely destroyed \citep{stoerzer00}. \citet{wilson94} collated various studies
  together and tracked the changes in relative abundances with distance
  from the Galactic centre. In the local interstellar medium they found $\mathrm{[^{12}C]/[^{13}C]}=77 \pm 7$ and
  $\mathrm{[^{16}O]/[^{18}O]}=560 \pm 25$, indicating $\zeta = 7.3 \pm
  0.7$ if there is no fractionation. 

\begin{figure*}
\begin{minipage}{0.49\textwidth}
\includegraphics[width=\textwidth]{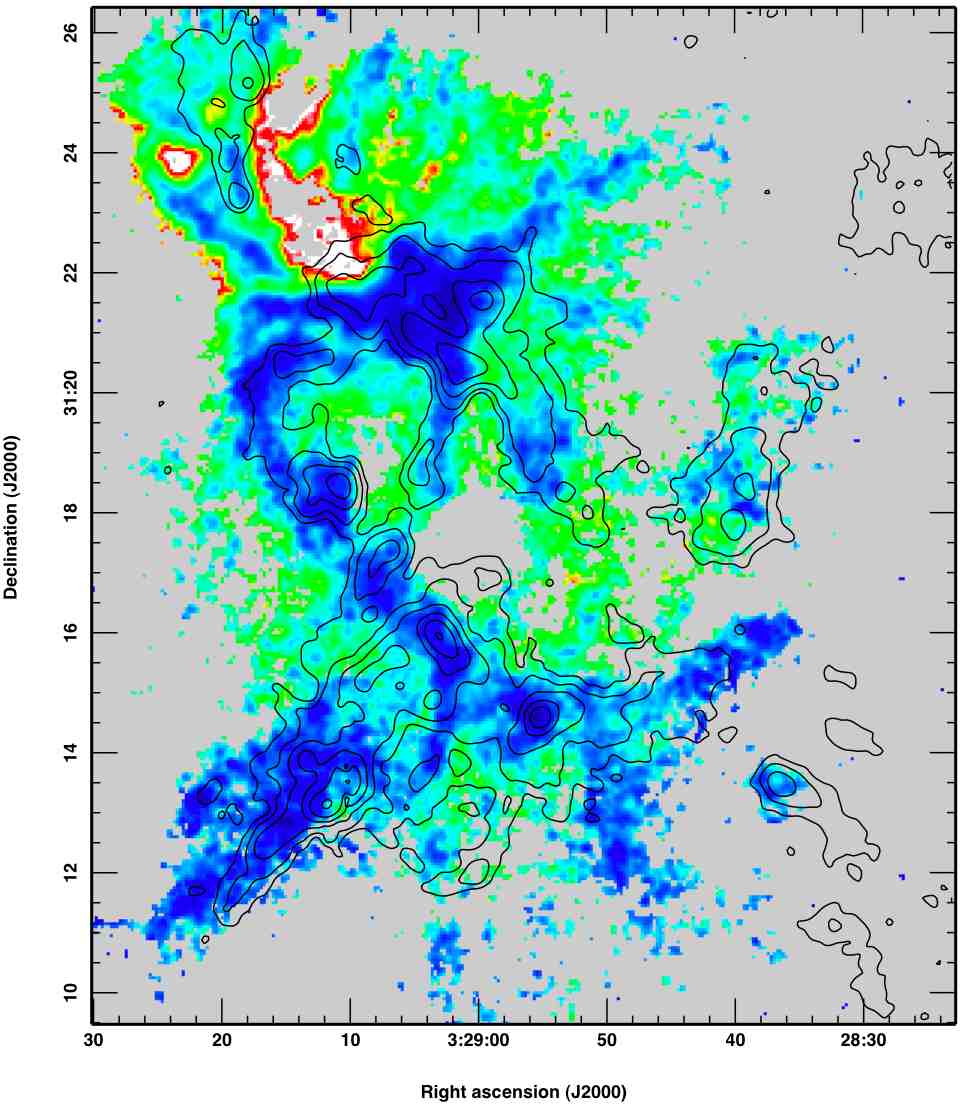}
\end{minipage}
\begin{minipage}{0.49\textwidth}
\includegraphics[width=\textwidth]{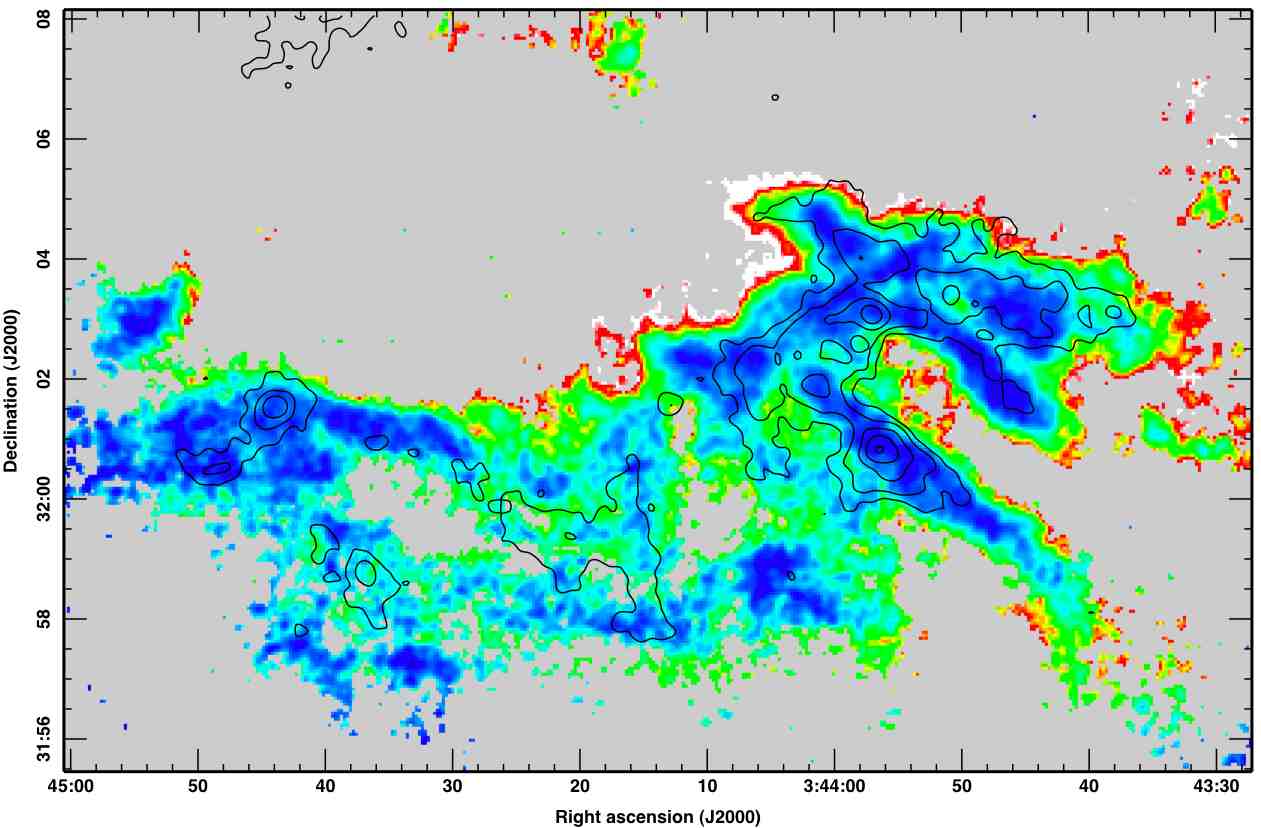}
\end{minipage}
\vspace{0.2cm}
\begin{minipage}{0.49\textwidth}
\includegraphics[width=\textwidth]{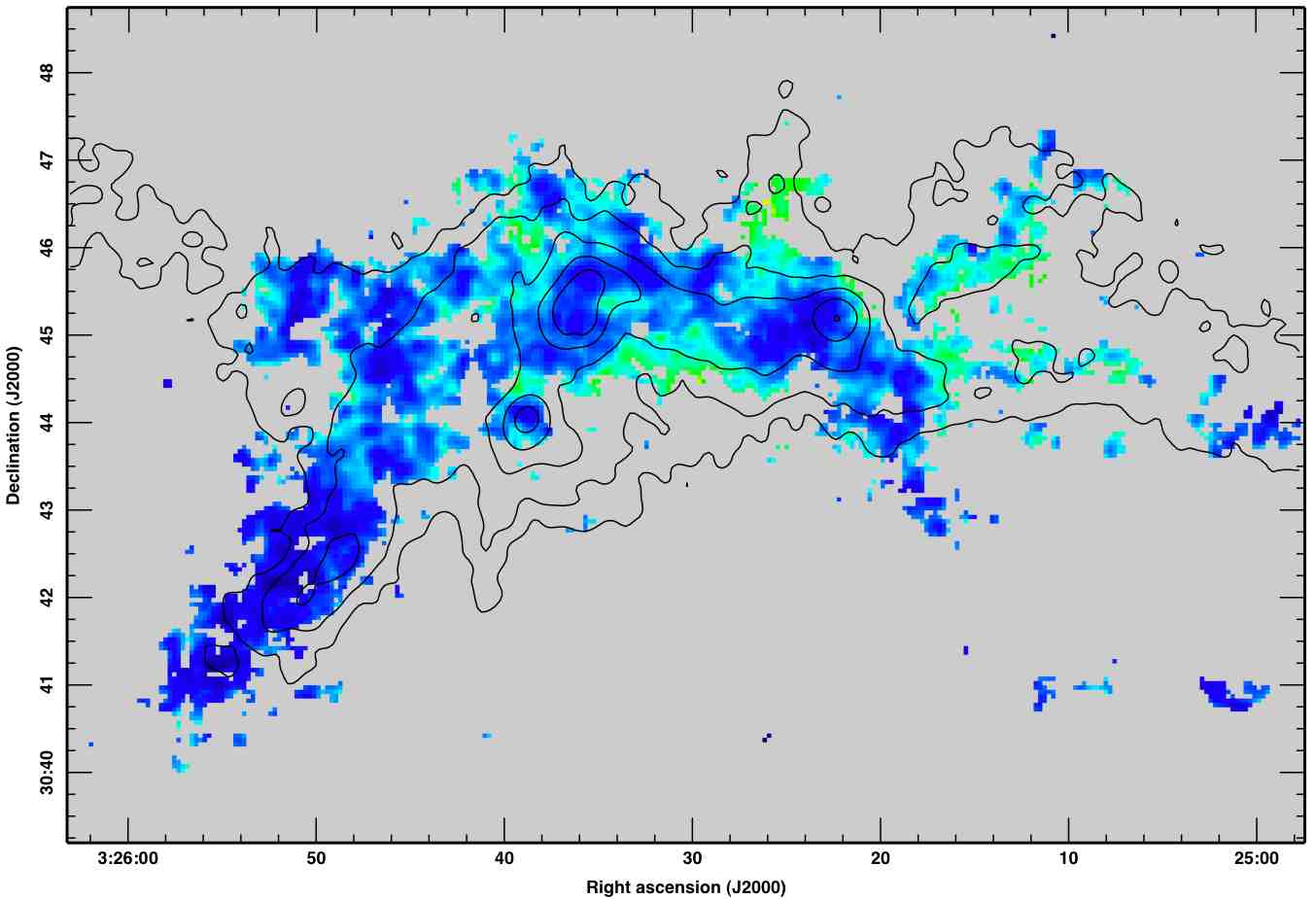}
\end{minipage}
\begin{minipage}{0.49\textwidth}
\includegraphics[width=\textwidth]{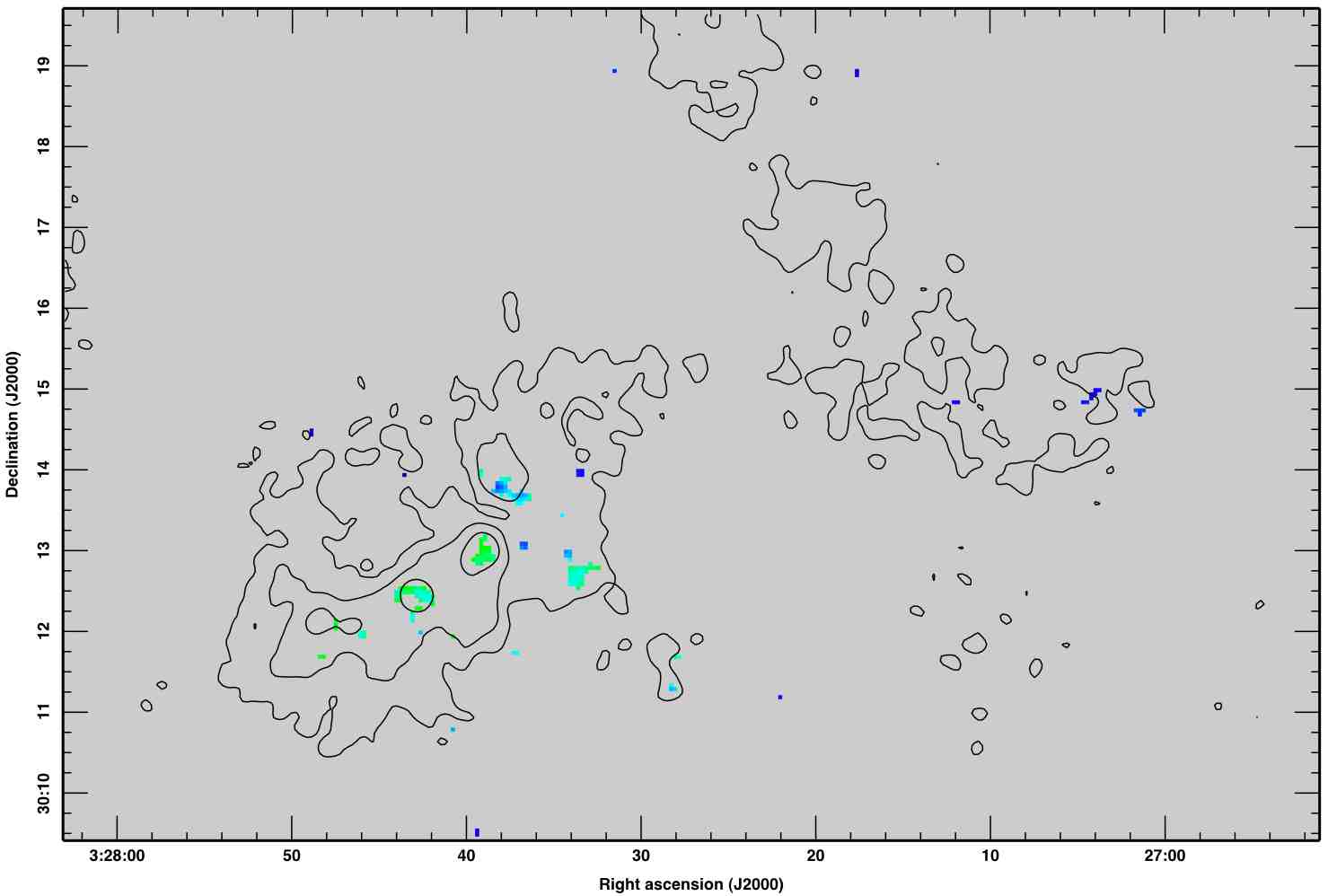}
\end{minipage}
\vspace{0.2cm}
\includegraphics[width=0.8\textwidth]{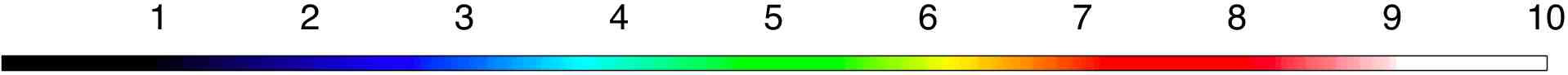}
\caption{Ratio of the \thirteenco\ to \ceighteeno\ antenna
  temperatures ($R$) measured at the \ceighteeno\ line peak in NGC1333 (top left), IC348 (top right),
  L1448 (bottom left) and L1455 (bottom right). Pixels
  are only displayed where both isotopologues have peak intensities
  $>5\sigma_\mathrm{RMS}$. Contours as in Fig.
  \ref{fig:c18o_centres}.}
\label{fig:13cotoc18o}
\end{figure*}

In calculations of $R$, we only include points that have a good
detection of both the \ceighteeno\ and \thirteenco\ lines (peak
line brightness $>5$ times the RMS spectral noise). We plot $R$ in
Fig. \ref{fig:13cotoc18o}. Where the gas becomes less dense $R$ tends to $\zeta$. In Fig.
\ref{fig:13cotoc18o}, outside of the SCUBA contours, so presumably at
lower space densities, the ratio is $\sim$5--6. At the very edges of the
measured values in NGC1333 and IC348, $R$ reaches $\sim$7--8, the
value from \citet{wilson94} and becomes larger still. These larger ratios
could either be a result of actual deviations in $\zeta$ or from noise
effects. Most of the largest $R$ are in the northeast of
NGC1333 and northeast of the main horseshoe of SCUBA emission in
IC348. These two areas are not representative of the two regions. In
NGC1333, this is where a bubble of bright 8.0\,\micron\ \emph{Spitzer}
flux resides (see fig. 1 of \citealp{gutermuth08}), commonly
thought to be emission from polycyclic aromatic hydrocarbon features
excited by UV radiation. The UV source is likely to heat up the
molecular gas in the region and perhaps change the isotopologues' abundances. In the IC348 area, there is a mass of
blue-shifted \twelveco\ gas, expanding into a dust cavity
(\citealp{tafalla06} and Curtis et al., in preparation). Thus, disregarding these areas, $R\to
\sim 7$ at the edges of the map, which we take to be $\zeta$ in the
following analysis. 

\begin{figure}
\begin{center}
\includegraphics[width=0.4\textwidth]{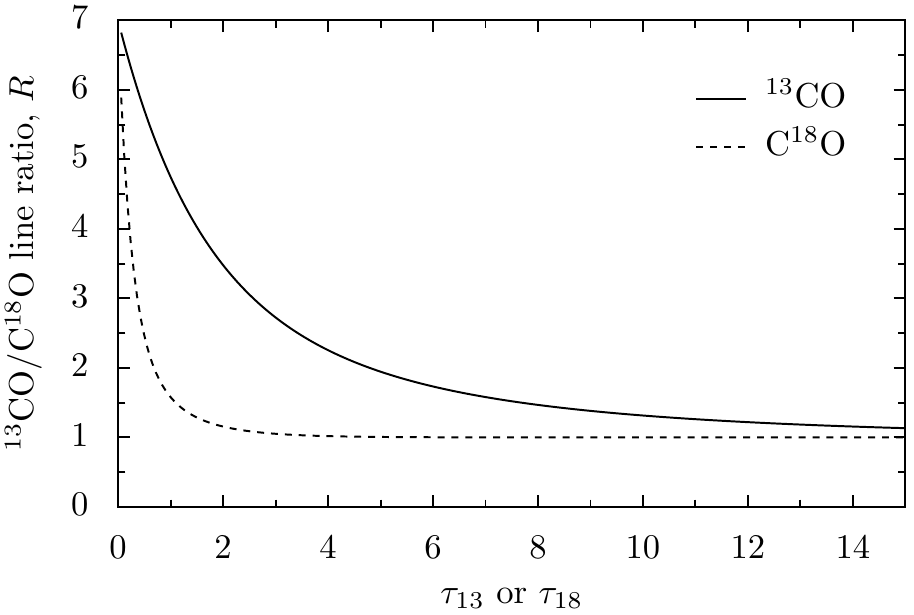}
\caption{Variation of $R$ with $\tau$ for $\zeta=7$.}
\label{fig:ratiowithtau}
\end{center}
\end{figure}

For $\zeta=7$, $\tau_{13} =7\tau_{18}$ and $R$
varies with $\tau$ as in Fig. \ref{fig:ratiowithtau}. $R$
clearly falls towards the centre of the dust filaments and cores, where
we expect the gas to be more dense. The average $R$ in each region is
unlikely to relate convincingly the properties of the whole area as it
is measured over such a small portion of the maps, particularly in
L1448 and L1455. The typical ratios can be measured in the majority of
the star-forming SCUBA cores, which may prove more informative. We measured $R$ at the peak of each
core identified by \citet{hatchell07a} and present their
averages with those in the entire regions for comparison in Tables
\ref{table:r_bysource} and \ref{table:r_byregion}.

\begin{table}
\caption{\thirteenco/\ceighteeno\ ratios, $R$, at the peak of the
  SCUBA cores in the \citet{hatchell07a} catalogue.  $\sigma_R$ is
  the standard deviation of $R$ across the cores.}
\begin{tabular}{lcccc}
\hline
Region or & Number & $\langle R \rangle$ & $\sigma_R$ \\
Core Class \\
\hline
NGC1333 & 28 & 2.8 & 0.8 \\ 
IC348 & 16 & 3.0 & 0.8 \\
L1448 & 7 & 2.2 & 0.3 \\
L1455 & 1 & 3.6 & -- \\
\hline
Starless & 21 & 2.8 & 0.6\\
Class 0 & 21 & 2.6 & 0.8 \\
Class I& 10 & 3.1 & 1.0 \\
\hline
\end{tabular}
\label{table:r_bysource}
\end{table}

\begin{table}
\caption{\thirteenco/\ceighteeno\ ratios, $R$, for every map pixel in Fig. \ref{fig:13cotoc18o} $\sigma_R$ is the standard deviation
  of $R$ across the pixels.}
\begin{tabular}{lccc}
\hline
Region & $\langle R \rangle$ & $\sigma_R$ & Median $R$\\
\hline
NGC1333 & 3.7 & 1.3 & 3.5 \\
IC348 & 4.1 & 1.8 & 3.6 \\
L1448 & 2.8 & 0.7 & 2.7 \\
L1455 & 3.6 & 0.9 & 3.8 \\ 
\hline
\end{tabular}
\label{table:r_byregion}
\end{table}

The typical ratios across the four regions, $R\approx2-4$, translate into
$\tau_{13}\approx 1.52-0.15$ and $\tau_{18}\approx 0.22-0.02$. This suggests
that the \ceighteeno\ line is typically optically thin but the
\thirteenco\ can be classed neither as optically thin nor thick. The pattern
of $R$ suggests that on average L1448 has the highest column density,
followed by NGC1333, whilst IC348 has the smallest. The value
for L1455 is not representative of the region as it is
only measured in the dense cores. When just the cores
are examined the pattern is the same IC348 has the largest $R$ followed by
NGC1333 and then L1448. As we expect, the cores have larger
optical depths than their parent clouds on average. However, the reduction in $R$ is not
enough to render the core \ceighteeno\ gas optically thick. The \emph{smallest}
core ratio is 1.62 in NGC1333, which implies $\tau_{18}\approx
1.0$ whereas must cores have much smaller
$\tau_{18}$. The mean core ratios are smaller (and thus the
  column densities are higher) for Class 0 than I protostars, although there is significant spread in both populations. As collapse progresses in
protostars, they accrete an increasing fraction of
their envelopes on to the central object so reducing the column density seen towards Class I
over 0 cores. The column density seen towards starless cores
presumably depends on the exact age of the core so could be comparable
to the protostars' or entirely different. 

\subsection{Temperature} \label{sec:12co_temp}

In this section, we derive $T_\mathrm{ex}$ for the \twelveco\
\threetotwo\ line assuming it is optically
thick. We include \twelveco\ HARP data for NGC1333 over the region
observed in \thirteenco/\ceighteeno\ (J. Swift, personal communication)
which are distributed on a coarser grid of 6\,arcsec pixels using a
nearest-neighbour scheme. 

\begin{figure*}
\begin{center}
\begin{minipage}{0.49\textwidth}
\begin{flushright}
\includegraphics[width=0.94\textwidth]{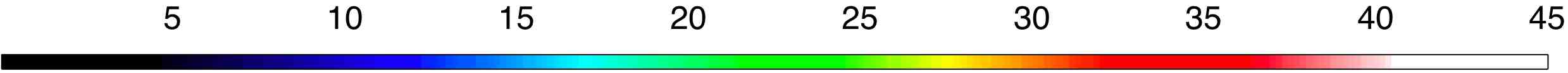}
\includegraphics[width=\textwidth]{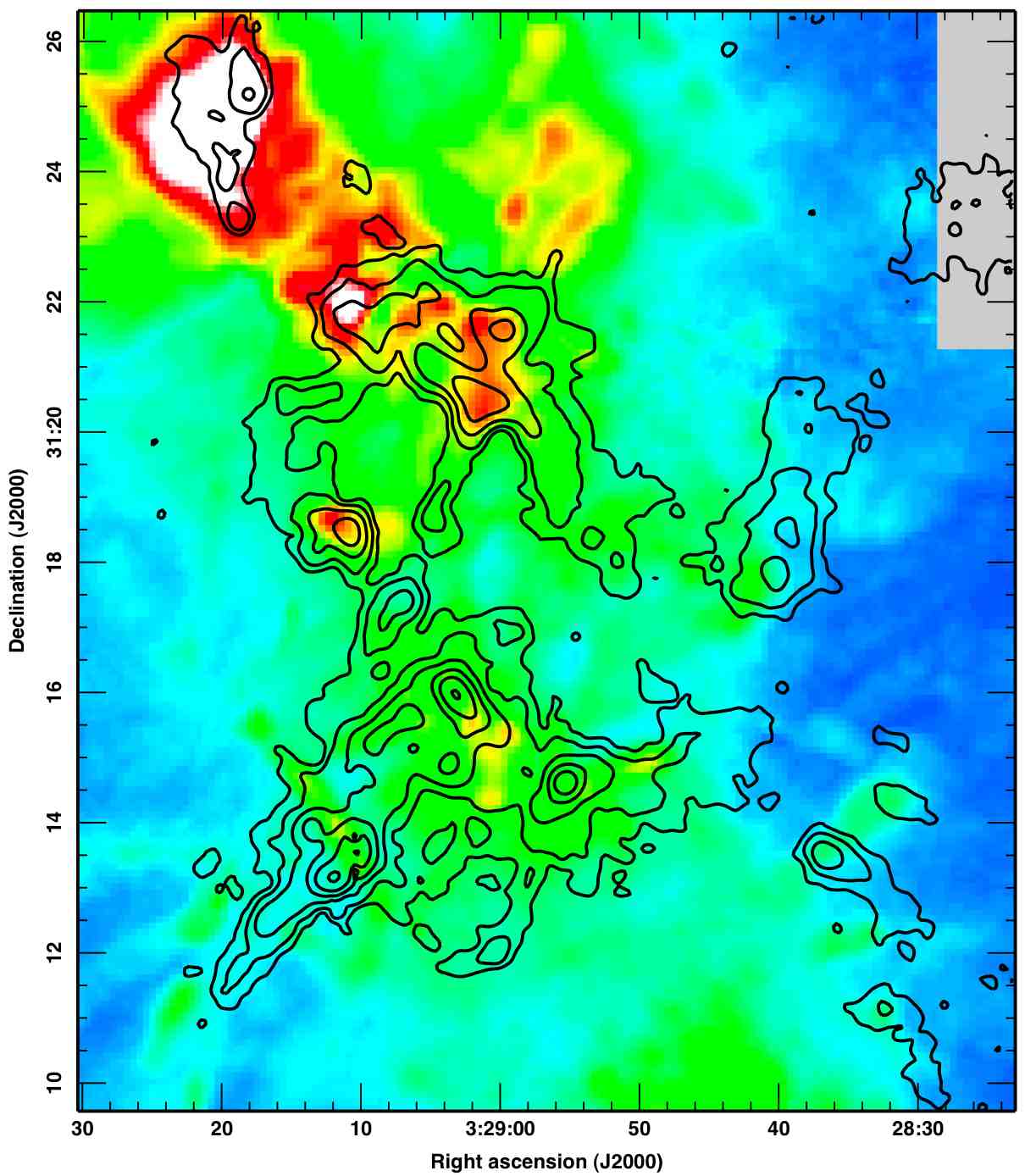}
\end{flushright}
\end{minipage}
\begin{minipage}{0.49\textwidth}
\begin{flushright}
\includegraphics[width=0.95\textwidth]{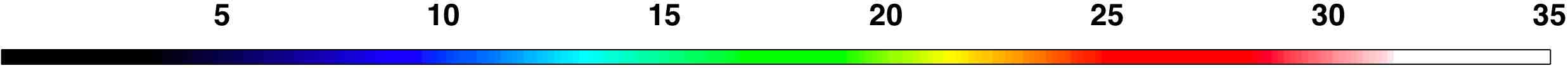}
\includegraphics[width=\textwidth]{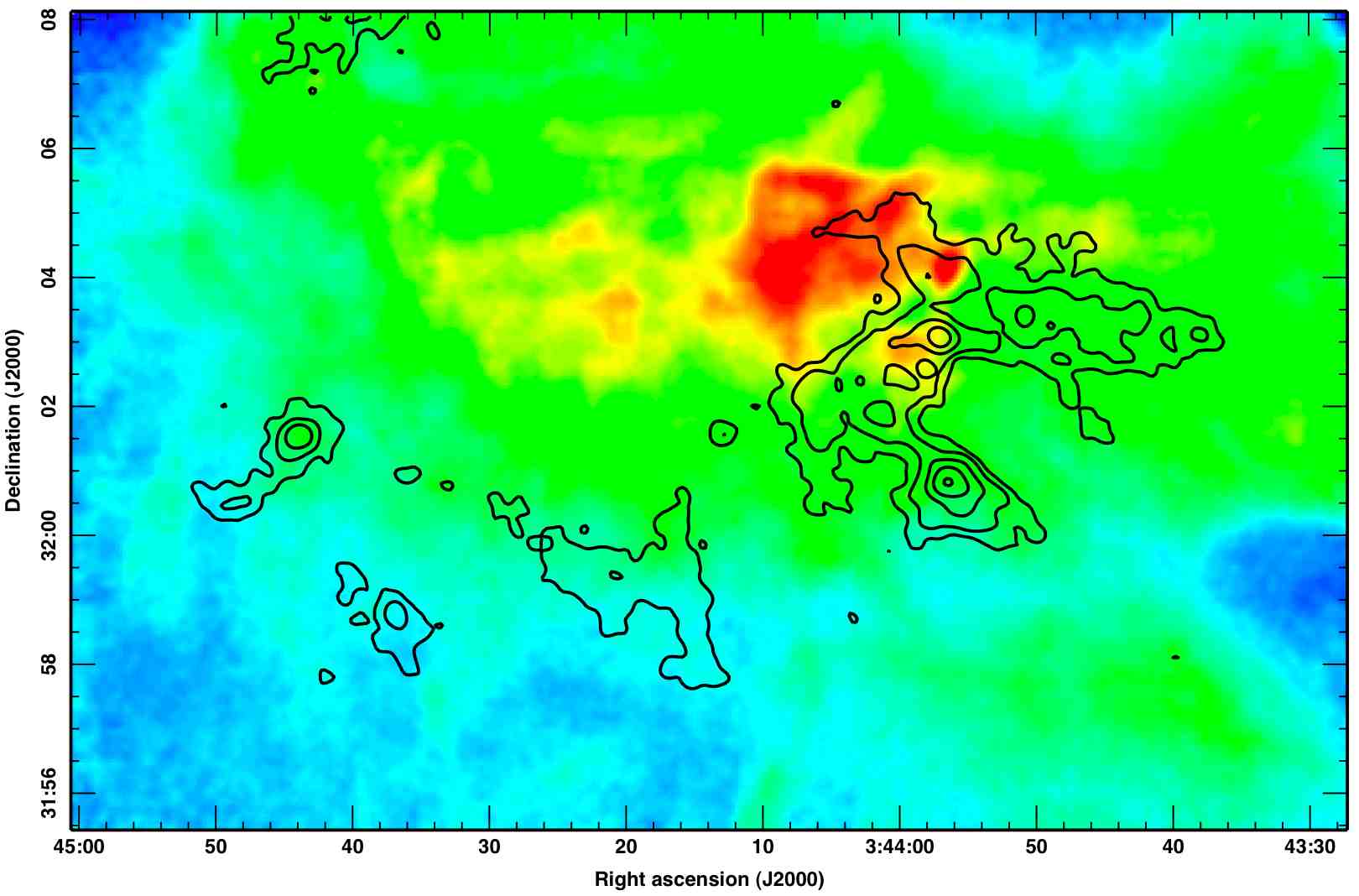}
\end{flushright}
\end{minipage}
\vspace{0.2cm}

\begin{minipage}{0.49\textwidth}
\begin{flushright}
\includegraphics[width=0.94\textwidth]{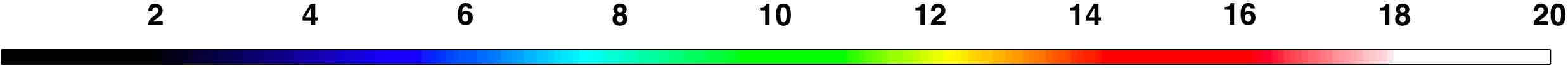}
\includegraphics[width=\textwidth]{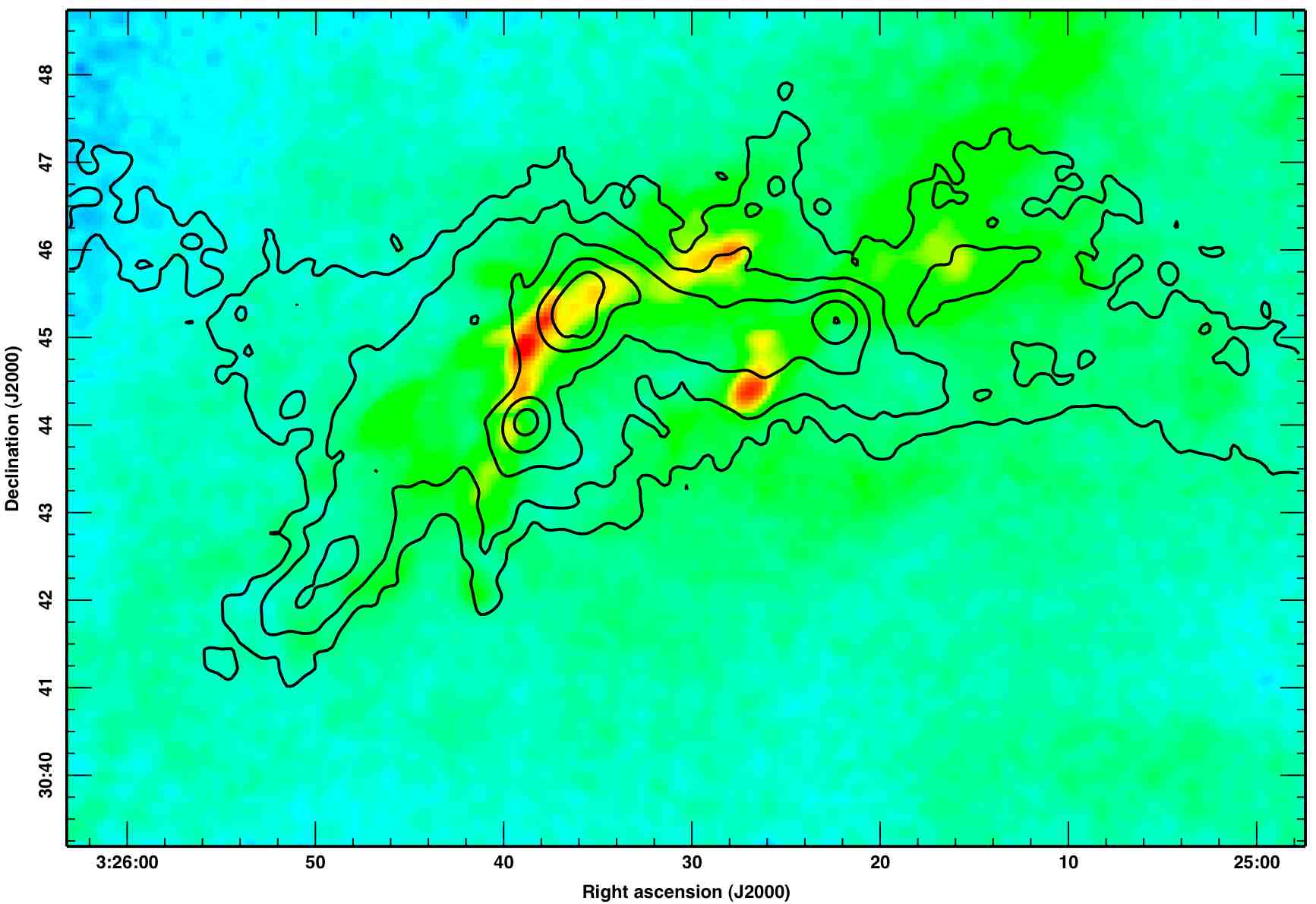}
\end{flushright}
\end{minipage}
\begin{minipage}{0.49\textwidth}
\begin{flushright}
\includegraphics[width=0.96\textwidth]{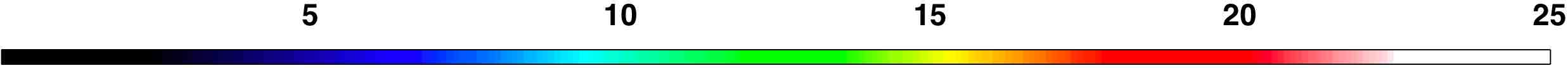}
\includegraphics[width=\textwidth]{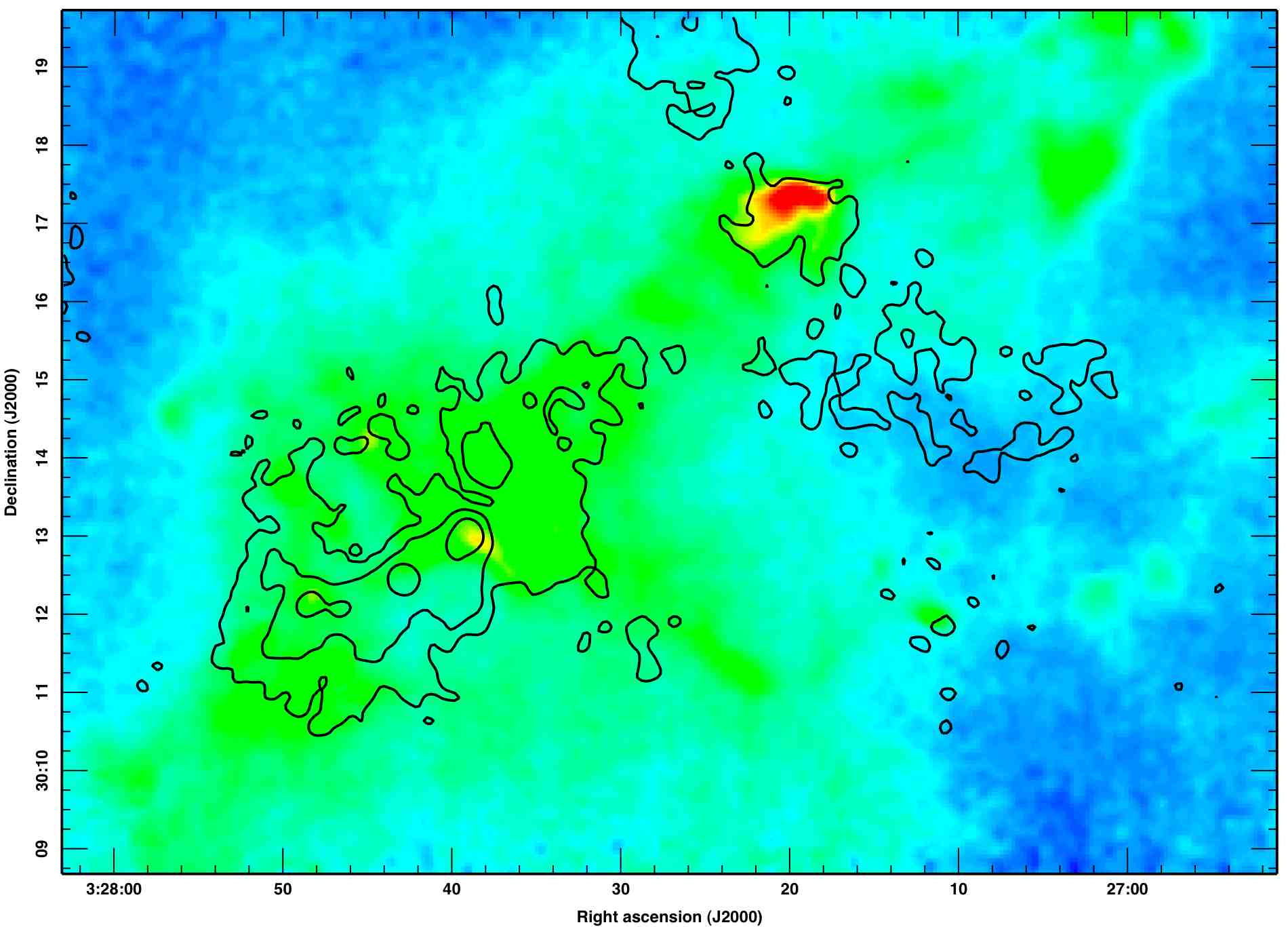}
\end{flushright}
\end{minipage}
\caption{$T_\mathrm{ex}$ in K derived from the peak \twelveco\
  main beam brightness temperature with panels as arrange in Fig.\ \ref{fig:13cotoc18o}. Contours as in Fig.
  \ref{fig:c18o_centres}.}
\label{fig:12co_temps}
\end{center}
\end{figure*}

\begin{figure}
\begin{center}
\begin{minipage}{0.47\textwidth}
\includegraphics[width=\textwidth]{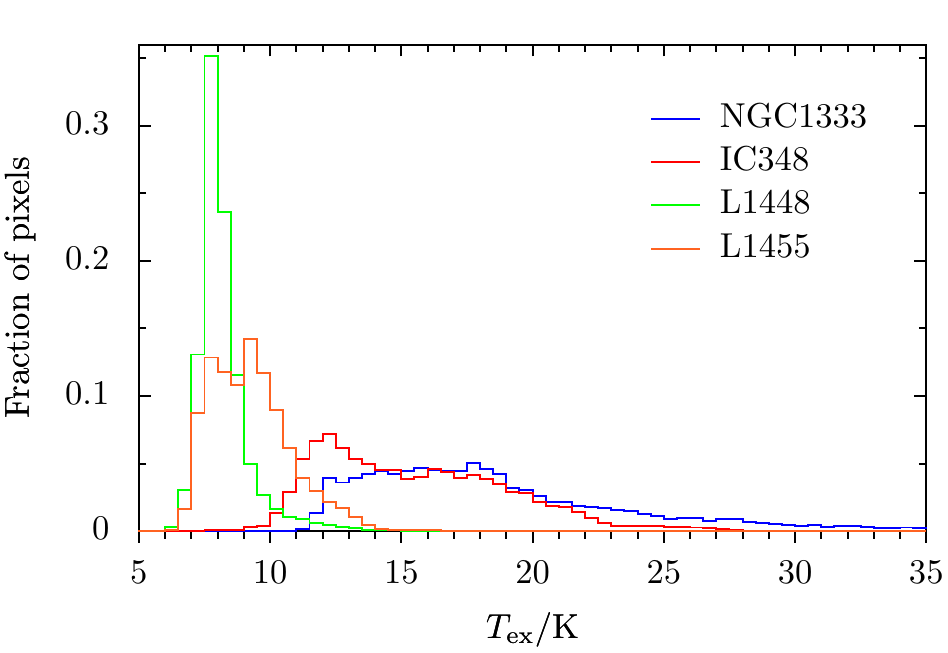}
\end{minipage}
\begin{minipage}{0.49\textwidth}
\begin{tabular}{lccc}
\hline
Region & $\langle T_\mathrm{ex} \rangle$ (K) & $\sigma_T$ (K)& Median
$T_\mathrm{ex}$ (K)\\ 
\hline
NGC1333 & 19.3 & 7.0 & 17.7 \\  
IC348 & 15.5 & 3.6 & 15.0 \\
L1448 & 8.2 & 1.0 & 8.0 \\
L1455 & 9.3 & 1.6 & 9.2 \\
\hline
\end{tabular}
\end{minipage}
\caption[]{$T_\mathrm{ex}$ derived from peak \twelveco\
  values. $\sigma_T$ is the standard deviation of $T_\mathrm{ex}$.}
\label{fig:12coT_histogram}
\end{center}
\end{figure}

In LTE, $T_\mathrm{ex}$ \emph{is} the physical temperature of the gas,
although this will of course not be the case in general. We have noted typical \thirteenco\ optical depths of $\tau_{13}\approx 1-2$, which provided the abundance
$\mathrm{[^{12}C]/[^{13}C]}=77$ \citep{wilson94} should ensure the
opacity of the \twelveco\ gas is $\tau_{12}\approx 77-154$. Ratio maps with
other isotopologues for \twelveco\ are unlikely to produce accurate
$\tau_{12}$ values as \twelveco\ traces outer cloud layers and not the
volumes where \thirteenco/\ceighteeno\ are emitted. However, provided
the line is optically thick, i.e.\ $\tau \to \infty$, and not
self-absorbed, we can derive
$T_\mathrm{ex}$ thus (see e.g.\ \citealt{pineda08}): \begin{equation} T_\mathrm{ex}(\mathrm{^{12}CO}) = \frac{16.59\,\mathrm{K}}{\ln \left\{ 1 +
      16.6\,\rmn{K}/[T_\mathrm{max}(\mathrm{^{12}CO})+0.036\,\rmn{K}]
    \right\}} \label{eqn:12co_temps} \end{equation}
where $T_0=16.59$\,K and $T_\mathrm{max}(\mathrm{^{12}CO})$ is the peak
\twelveco\ main beam brightness temperature. We plot $T_\mathrm{ex}$ in Figs.
\ref{fig:12co_temps} and \ref{fig:12coT_histogram}. $T_\mathrm{ex}$
is typically 5--25\,K, generally
increasing towards the centre of each region where it can reach 30\,K. 

There are a number of caveats to this analysis that should be noted. First, in some respects the most interesting regions of
molecular clouds are the cores in which stars are formed. The
temperatures of such high-density regions will not be accurately
probed by \twelveco, whose emission is restricted to outer cloud
material due to its higher opacity. However, we might naturally expect
the temperature of starless cores (without an internal heating
source) and possibly protostars to reflect that of their natal
environment so the \twelveco\ temperature should reflect the pattern of core conditions. Second, \twelveco\ may not be
optically thick everywhere, although the typical \thirteenco\
optical depths in regions with bright SCUBA emission do suggest it
will be for most of the areas. Furthermore, we may underestimate the
true \twelveco\ peak temperature if the line profile is complicated by
effects such as infall and self-absorption.

In L1455, L1448 and IC348, the $T_\mathrm{ex}$ maps follow the integrated
\twelveco\ intensities closely. Any strong outflows are prominent areas of higher $T_\mathrm{ex}$, as suggested by
other authors (\citealp*{hatchell99}; \citealp{nisini00}). The dominant feature in
NGC1333's $T_\mathrm{ex}$ map is the blue-shifted bubble of gas in the
northeast corner. We have already mentioned this is an area likely to
be illuminated by a
UV source. By contrast, the most pronounced feature in the \twelveco\
integrated intensity, the bright outflow from SVS13, is almost
absent. The bubble's $T_\mathrm{ex}$ is some 40\,K higher than the maximum
in the other regions. In this part of the map, \twelveco\ is not optically
thick and the gas probably has a higher physical temperature -- as we noted the \thirteenco/\ceighteeno\ ratio is
larger in this region. Additionally, it
only has single-peaked lines, devoid of some of the more complicated
profiles seen in NGC1333, suggesting its peak temperature is
over-estimated compared to the rest of the cloud. This all points to a
warm, low-density region with material perhaps dispersed by nearby
luminous stars, which \citet*{hatchell07b} suggest are the cause of a
lack of outflow detections in this area. 

It is interesting
that the largest temperatures in NGC1333 are to the northeast, in the
direction of 40 Persei
(03$^\mathrm{h}$42$^\mathrm{m}$22.6$^\mathrm{s}$,$+33^\circ$57$'$54.1$''$),
a B0.5 star, part of the Perseus OB association, suggested to be triggering the
star formation in these clouds \citep{walawender04,kirk06}. Indeed,
the highest temperatures are to the north in IC348, also in the
direction of 40 Per, although no such gradients are clear in L1448
and L1455, possibly as they are further away.  

On average, L1448 is the coldest, followed in order of increasing
temperature by
L1455 then IC348, with NGC1333 the hottest. The histograms of
Fig. \ref{fig:12coT_histogram} show two distinct temperature environments:
one colder in L1455 and L1448 and the other hotter in IC348 and
NGC1333. The latter two regions show a large range of
temperatures. NGC1333 and IC348 are probably heated more by their
namesake IR clusters and the
triggering radiation from 40 Per than L1448 and L1455. Protostars
may also heat their surroundings to a limited extent via radiation and
outflows. NGC1333 with its numerous energetic flows and protostars is
most likely to be warmed up in this way. L1448 has a high protostellar
fraction as well and powerful outflows but less ambient gas with which to
interact -- most of the outflows have broken out of the region.  

As we have already mentioned, temperatures derived from \twelveco\ are
unlikely to be best probes of conditions inside dense
cores. Nevertheless we might expect the core temperatures to be affected
to some extent by the temperature in the bulk of the gas. From Table
\ref{table:12coT_cores}, it is clear the temperatures towards the
cores are larger on average than within their parent region as a whole but
exhibit the same trends. When comparing cores of different ages it is
worth noting that we expect fundamentally different relations
between $T_\mathrm{ex}$ and the true temperatures for
starless and protostellar cores, $T_\mathrm{core}$. Towards the centres of starless cores,
temperatures fall whereas for protostars temperatures rise. Thus, for
starless cores $T_\mathrm{ex}>T_\mathrm{core}$ and for protostars
$T_\mathrm{ex}<T_\mathrm{core}$. In Table \ref{table:12coT_cores}
the \emph{environments} of Class I protostars appear slightly
warmer than both starless cores and Class 0 protostars on
average. As a
source ages we expect its core temperature to increase, resulting from
the increase in brightness of its central object. These data would
exhibit this trend except the Class 0 and starless cores have similar
temperatures, probably illustrating that the \twelveco\ line probes an environment outside the
high-density core, where $T_\mathrm{ex}>T_\mathrm{core}$ for the
starless cores and $T_\mathrm{ex}<T_\mathrm{core}$ for the protostars. 

\begin{table}
\caption[]{$T_\mathrm{ex}$ from \twelveco\ at the peak of the
  SCUBA cores in the \citet{hatchell07a} catalogue by region or
  source age.}
\begin{tabular}{lcccc}
\hline
Region or Core Class & Number & $\langle T_\mathrm{ex} \rangle$/K & $\sigma_T$/K \\
\hline
NGC1333	& 29 & 25.2 & 10.1 \\
IC348 & 17 & 17.6 & 3.2 \\
L1448 & 7 & 10.6 & 2.0\\
L1455 & 5 & 12.6 & 2.2\\
\hline			
Starless & 23 & 18.8 & 9.7\\
Class 0	& 21 & 18.3 & 6.2\\
Class I	& 14 & 24.9 & 11.2\\
\hline
\end{tabular}
\label{table:12coT_cores}
\end{table}

\begin{figure}
\begin{center}
\includegraphics[width=0.4\textwidth]{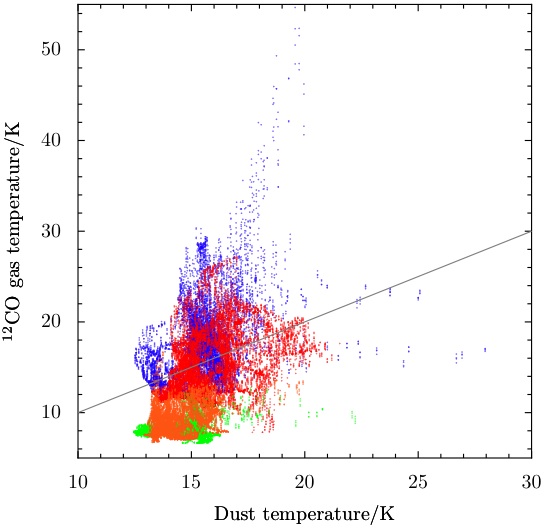}
\includegraphics[width=0.4\textwidth]{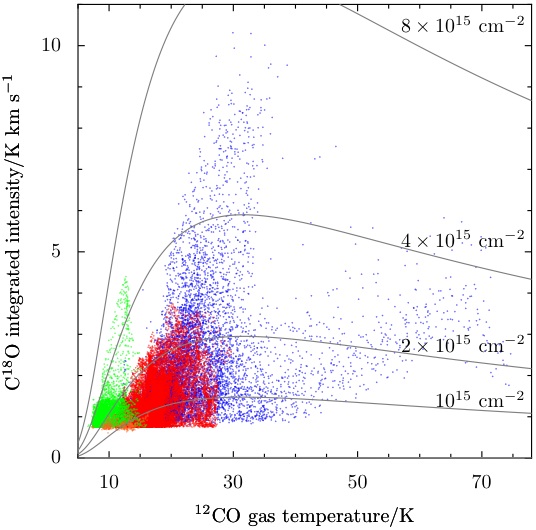}
\caption{Points mark pixels in the various regions' maps: NGC1333
  (blue), IC348 (red), L1448 (green) and L1455
  (orange). Top: $T_\mathrm{ex}$ derived from the
  peak \twelveco\ $T_\mathrm{A}^*$ versus the dust temperature derived
  from \emph{Spitzer} MIPS data at 70 and 160~\micron\ with
  40~arcsec resolution \citep{schnee08b}. The line marks where both
  temperatures are equal. Bottom: \ceighteeno\
  integrated intensity versus \twelveco\ derived $T_\mathrm{ex}$. The
  lines denoted the anticipated dependence for constant column
  densities of \ceighteeno\ as labelled.  }
\label{fig:12coT_scatter}
\end{center}
\end{figure}

We can make an interesting comparison between this gas
temperature and that of the dust. At high densities, $n\ga 2\times
10^4$\,cm$^{-3}$, conditions likely to be probed in CO with HARP, the
gas and dust temperatures are coupled \citep*{galli02}. Recently,
\citet{schnee08b} computed the dust temperature at 40\,arcsec
resolution across Perseus using
\emph{Spitzer} and \emph{IRAS} measurements. We plot $T_\mathrm{ex}$ degraded to 40\,arcsec resolution versus
\citeauthor{schnee08b}'s dust temperature in Fig.
\ref{fig:12coT_scatter}. The dust
temperature, $T_\mathrm{D}$, was calculated from \emph{Spitzer} 70 and 160\,\micron\
fluxes, which is dominated by warm dust along the line of sight rather
than the denser, colder material in say starless cores and is more
likely to match the material traced in \twelveco. There is
actually little
correlation between the two temperatures. Typically $T_\mathrm{D}$ is
slightly lower than $T_\mathrm{ex}$: 12--20\,K
compared to 8--30\,K. In NGC1333 and IC348, $T_\mathrm{ex}$ is mostly
larger than $T_\mathrm{D}$ although there is a significant proportion
of pixels where the opposite is true. However, in L1448 and L1455,
nearly everywhere $T_\mathrm{D}>T_\mathrm{ex}$. This may be explained
if the \twelveco\ gas is not tracing the same areas as the dust
emission, with the \twelveco\ emitted in a colder layer whilst the dust
comes from a hotter inner cloud region. Additionally, if the gas is
not optically thick, then we would over-estimate the \twelveco\
excitation temperature and $T_\mathrm{ex} > T_\mathrm{D}$. 

The \twelveco\ excitation temperature may be correlated
with the cloud column density. As the rotational transition is
collisionally excited, on moving to denser and higher extinction
portions of the cloud with densities above the
critical density, the gas $T_\mathrm{ex}$ should increase. We use
the integrated \ceighteeno\ \threetotwo\ intensity as a proxy for the
visual extinction since typically there is a linear relation between
them. Using \ceighteeno\ \onetozero\ data across the Perseus molecular cloud \citet{pineda08} measured: \begin{equation} A_\mathrm{v}/\mathrm{mag} = (2.4\pm 0.1) \left(\frac{\int
  T_\mathrm{mb}(\mathrm{C^{18}O})\mathrm{d}v}{\mathrm{K\,km\,s^{-1}}}\right) +(2.9\pm0.9)\mathrm{.}\end{equation}
In Fig. \ref{fig:12coT_scatter}, we plot the \ceighteeno\ integrated intensity, $\int
T_\mathrm{A}^*~\mathrm{d}v$ versus $T_\mathrm{ex}$, with lines
denoting the expected variation for various column densities of \ceighteeno\ assuming
LTE. No single line is a good fit for all the points as in each region
there is a range of column densities. However, a moderate range of
column densities will span the whole parameter space. At low temperatures with a good deal of
scatter the integrated intensity increases linearly with $T_\mathrm{ex}$. L1448
again requires the highest column densities with IC348 and NGC1333 having a large spread
in conditions.

\section{Summary} \label{sec:summary}

This paper presents the technical details and preliminary analysis of
a large-scale survey of the kinematics of
molecular gas in the Perseus molecular cloud. Observations of the \threetotwo\ rotational
transitions of \twelveco, \thirteenco\ and \ceighteeno\ in over
600\,arcmin$^2$ of NGC1333, IC348, L1448 and L1455 were undertaken
with HARP on the JCMT. We introduce a new
`flatfield' procedure to account for striping artefacts in HARP scan maps
of molecular clouds, apparently resulting from differential
performance across the detectors and/or their samplers. The data from each
working detector is multiplied by a conversion factor which scales its intensity
to match the nominated reference detector. The factors are computed
from the scientific observations themselves by calculating the total
intensity received by each detector across the whole map. 

We compare integrated intensity maps of the three tracers to SCUBA
 850\,\micron\ emission \citep{hatchell05} and the position of
 protostars and starless cores \citep{hatchell07a} in each
 field. There is a striking similarity between the SCUBA maps and
 \ceighteeno\ emission, hinting that similar densities of material are
 traced with the gas and dust strongly coupled. However, the detailed structure of
 individual star-forming cores is not always so simple, for example very
 bright SCUBA cores sometimes have only weak \ceighteeno\
 emission and vice versa. Many outflows are obvious in the \twelveco\ data, which we will examine in
detail in a subsequent study (Curtis et al. in preparation). The
 \thirteenco\ maps are somewhat intermediate between the other two
 isotopologues and average spectra across the maps do show weak
 linewings from outflows. Intriguingly, masses derived from the HARP
 and SCUBA data exhibit different trends across the four
 regions, emphasizing the need to examine the detailed excitation
 conditions across regions rather than simple constant assumptions.
  
From the \thirteenco/\ceighteeno\ integrated line ratio, $R$, we explore
variations in the gas opacity and estimate the two species' relative
abundance ($\rmn{[^{13}CO]/[C^{18}O]}\sim 7$). Across most of the
regions the \ceighteeno\ gas is optically thin ($\tau_{18}=0.02-0.22$)
and therefore is a
reliable total mass tracer. Indeed, inside the denser parts of the clouds, the
star-forming cores, the opacity does not increase beyond a maximum,
$\tau_{18}=0.9$. When we also consider that the critical density of
the transition is some $10^4$\,\cmthree, we expect to probe an
intermediate region between a dense core and its envelope with the
\ceighteeno\ line. The \thirteenco\ optical depths are neither in the optically thin nor thick regimes ($\tau_{13}$ is
typically 0.15--1.52). Class 0 protostars have smaller ratios than Class
Is as expected as more material accretes on to the central object over
time. 

If we assume the \twelveco\ line is optically thick, an estimate of
the excitation temperature can be gathered from the peak line
temperature, assuming LTE (in which case $T_\rmn{ex}$ is the
physical gas temperature). In general we derive temperature of
5--25\,K, increasing towards the centres of the individual regions and
in outflow lobes. An area in the northwest of NGC1333 has
temperatures of over 45\,K, probably as it is heated by a UV source
and/or the region has lower density so is not optically thick. IC348
and NGC1333 are on average much warmer than L1448 and L1455. This is partly as the averages for NGC1333 are skewed upwards because of
the warm gas in the north and perhaps since NGC1333 and
IC348 are closer to 40 Per (thought to be triggering star formation
in Perseus). There
is little correlation between $T_\rmn{ex}$ and the dust temperature, $T_\rmn{D}$,
derived from \emph{Spitzer} observations \citep{schnee08b}. Typically
$T_\rmn{D}$ is slightly lower than $T_\rmn{ex}$: 12--20\,K compared to
8--30\,K. However, in L1448 and L1445, typically
$T_\rmn{ex}<T_\rmn{D}$, this may imply we are tracing completely
different regions with the \twelveco\ gas and warm dust or that the gas
is not optically thick, so we are over-estimating $T_\rmn{ex}$.    

This work demonstrates the utility of HARP for large-scale surveys
of gas kinematics in nearby molecular clouds. The \threetotwo\ transitions of CO
and its isotopologues are powerful probes of the
conditions of star formation when used in combination, examining moderately
high densities comparable to the dust densities seen by SCUBA. 

\section{Acknowledgments}

EIC thanks the Science and Technology Facilities Council (STFC) for studentship support while carrying out this
work. The authors thank Jonathan Swift for use of the \twelveco\ data
towards NGC1333 in advance of publication. We are also
  grateful to the referee, whose useful comments and suggestions significantly
  improved the clarity of this paper. The JCMT is operated by The Joint Astronomy
Centre (JAC) on behalf of the STFC of
the United Kingdom, the Netherlands Organisation for Scientific
Research and the National Research Council (NRC) of Canada. We have also
made extensive use of the SIMBAD data base, operated at CDS,
Strasbourg, France. We acknowledge the data analysis
facilities provided by the Starlink Project which is maintained by JAC
with support from STFC. This research used the facilities of the Canadian
Astronomy Data Centre operated by the NRC with the support of the Canadian Space Agency. 

\renewcommand{\bibname}{References}

\bsp

\label{lastpage}

\end{document}